\documentclass[12pt]{iopart}
\usepackage{import}
\eqnobysec 
\usepackage{iopams} 
\usepackage{dcolumn}
\usepackage{amsmath}  
\usepackage{amsfonts} 
\usepackage{textcomp}
\usepackage[utf8]{inputenc} 
\usepackage[T1]{fontenc}    
\usepackage{hyperref}       
\usepackage{url}            
\usepackage{booktabs}       
\usepackage{nicefrac}       
\usepackage{microtype}      
\usepackage{lipsum}
\usepackage[pdftex]{graphicx}
\usepackage{caption}
\usepackage{amssymb, mathtools, amsthm}
\usepackage{xcolor}
\usepackage{mathrsfs}
\usepackage{tensor}
\usepackage{todonotes}
\usepackage{float} 
\usepackage{ragged2e}
\usepackage[normalem]{ulem}
\usepackage[english]{babel}
\usepackage{graphicx}
\usepackage{dcolumn}
\usepackage{bm}
\usepackage{blindtext}
\usepackage{verbatim}
\usepackage{relsize}
\usepackage{blindtext}
\usepackage{cancel}
\usepackage{epstopdf}
\usepackage{mathtools}
\usepackage{blindtext}
\usepackage{color}
\usepackage[usenames,dvipsnames]{pstricks}
\usepackage{epsfig}
\usepackage{pst-grad} 
\usepackage{pst-plot} 
\usepackage{hyperref}
\usepackage{verbatim}
\usepackage{slashed}
\usepackage{dsfont}
\usepackage[english]{babel}
\usepackage{cleveref}
\usepackage{subcaption}
\usepackage{xcolor}
\usepackage{upgreek}

\newcommand{\p}{\partial}

\newcommand{\RR}{\mathbb{R}}
\newcommand{\SSS}{\mathbb{S}}

\newcommand{\OO}{\mathcal{O}}

\newcommand{\ee}{\mathrm{e}}

\usepackage{float}
\usepackage{tikz}
\usetikzlibrary{decorations.markings}

\tikzset{double arrow/.style={
    decoration={
      markings,
      mark=at position 0.45 with {
        \node[transform shape, rotate=90, anchor=center] at (0,0) {\tikz{\draw[line width=2.5pt, -{>>}] (0,-0.1) -- (0,0.1);}};
      }
    },
    postaction={decorate}
  }
}

\tikzset{double arrow other/.style={
    decoration={
      markings,
      mark=at position 0.55 with {
        \node[transform shape, rotate=90, anchor=center] at (0,0) {\tikz{\draw[line width=2.5pt, -{>>}] (0,-0.1) -- (0,0.1);}};
      }
    },
    postaction={decorate}
  }
}

\usepackage{amsthm}
\usepackage{geometry}
\usepackage{braket}

\DeclareMathOperator{\arctanh}{arctanh}
\DeclareMathOperator{\arccosh}{arccosh}

\theoremstyle{plain}

\theoremstyle{definition}

\newcommand{\mrm}[1]{\mathrm{#1}}
\newcommand{\mc}[1]{\mathcal{#1}}
\newcommand{\wt}[1]{\widetilde{#1}}

\renewcommand{\ms}[1]{\mathscr{#1}}
\newcommand{\lr}[1]{\left(#1\right)}
\newcommand{\lrc}[1]{\left\{#1\right\}}
\newcommand{\lrb}[1]{\left[#1\right]}

\newcommand{\lrl}[2]{\!\left.#1\right|_{#2}}

\newcommand{\Mmin}{M_{\mathrm{min}}}

\newcommand{\rc}{r_{\mathrm{conf}}}

\newcommand{\dd}{\mathrm{d}}

\usepackage[final]{showlabels}

\newcommand{\td}[2]{\ensuremath{\frac{\dd #1}{\dd #2}}}

\newcommand{\beq}{\begin{equation}}
\newcommand{\eeq}{\end{equation}}

\newcommand{\ba}{\begin{align}}
\newcommand{\ea}{\end{align}}

\allowdisplaybreaks[1] 

\begin{document}
\maketitle

\title{C-metric in a (nut)shell}

\author{Cameron R D Bunney$^{1}$\footnote[1]{Author to whom any correspondence should be addressed.} and Robert B Mann$^{2,3}$}

\address{$^1$School of Physics and Astronomy, University of Nottingham, Nottingham NG$7$ $2$RD, UK}
\address{$^2$Department of Physics and Astronomy, University of Waterloo, Waterloo, Ontario, N$2$L $3$G$1$, Canada}
\address{$^3$Perimeter Institute for Theoretical Physics, $31$ Caroline St., Waterloo, Ontario, N$2$L $2$Y$5$, Canada}
\ead{\mailto{cameron.bunney@nottingham.ac.uk}, \mailto{rbmann@uwaterloo.ca}}

\vspace{10pt}
\begin{indented}
    \item[]{October 2024. Revised February 2025.\footnote[3]{Published in Classical and Quantum Gravity \textbf{42}, 075001 (2025), 
doi:10.1088/1361-6382/adbc3f. 
For Open Access purposes, 
this Author Accepted Manuscript is made available under CC BY public copyright.}}
\end{indented}

\begin{abstract}
    We present a comprehensive study of the C-metric in $2+1$ dimensions, placing it within a shell of stress energy and matching it to an exterior vacuum AdS metric. The $2+1$ C-metric is not circularly symmetric and hence neither are the constructed shells, which instead take on a cuspoidal or teardrop shape. We interpret the stress energy of the shells as a perfect fluid, calculating the energy density and pressure. For accelerating particles (Class I), we find the stress energy is concentrated on the part of shell farthest from the direction of acceleration and always respects the strong and weak energy conditions. For accelerating black holes (Class I$_{\mrm{C}}$, II, and III), the shell stress energy may either respect or violate the energy conditions depending on the parameter of the exterior metric --- between the two regimes lies a critical value of the external parameter for which the shell stress energy vanishes, leading to new solutions of Einstein's field equations, which fall into three categories: an accelerated black hole pulled by a finite-length string with a point particle at the other end, an accelerated black hole pushed by a finite-length strut with a point particle at the other end, and an accelerated black hole pushed from one side by a finite-length strut and pulled from the other by a finite-length string, each with a point particle at the other end.
\end{abstract}

\section{Introduction}

Accelerating black holes are garnering increased interest in the physics community, in large part because historically the effects of acceleration on the properties of the black hole have not been understood. 
In four spacetime dimensions, 
these objects are described by
the C-metric \cite{LeviCivita:1917,Weyl:1919}, whose maximal extension is interpreted as a black hole undergoing uniform acceleration, isolated from a second black hole by a non-compact acceleration horizon 
\cite{KinnersleyWalker:1970,Bonnor:1982,InterpretingCMetric}. The acceleration 
is due to the force from a
string of positive tension
stretching from infinity to the horizon
along the axis of symmetry,
pulling the black hole; alternatively it can be due to  a strut (of negative tension)  pushing the black hole and connecting its horizon to the second black hole. The Plebański-Demiański metric~\cite{PlebanskiDemianski:1976} forms the complete family of Petrov type-D spacetimes, which includes as a special case the C-metric. The form of the Plebański-Demiański metric given in~\cite{PlebanskiDemianski:1976} is not the most accessible; hence, the C-metric and its generalisations have been studied in their own right, including a cosmological constant $\Lambda$ \cite{MannRoss:1995,Podolsky:2002,Dias:2003xp, Batic:2021tjh}, rotation and electromagnetic charge \cite{PlebanskiDemianski:1976, BHinElectricField}, and
dilatons \cite{Dowker:1993bt}. {Furthermore, the physics of these spacetimes has also been thoroughly investigated, including aspects of geodesic motion~\cite{Pravda:2000zm, Bini:2005qyt, Bini:2007kvf,Grenzebach:2015oea,Alawadi:2020qdz}.}

The anti-de Sitter (AdS) case, with $\Lambda < 0$, has been of considerable recent attention. In this case, if the acceleration is sufficiently small,
the black hole is  suspended at a fixed distance from the centre of AdS, and there is neither an acceleration horizon nor a second black hole. In such situations the laws of thermodynamics can be consistently formulated 
\cite{Anabalon:2018ydc,Anabalon:2018qfv,Gregory:2019dtq,CmetricThermoSuSY,CmetricThermoPhaseSpace}, and applied to uncover
new and interesting thermodynamic behaviour
\cite{Abbasvandi:2018vsh,Abbasvandi:2019vfz,Ahmed:2019yci}. Extensions of the first law to  black holes that possess an acceleration horizon
have been carried out
\cite{Gregory:2020mmi,Ball:2020vzo}.

Accelerating black holes have proven to be very useful in understanding the quantum properties of black holes 
\cite{Emparan:1999wa,Emparan:1999fd} and, for non-rotating C-metrics, an equation analogous to the Teukolsky equation has been derived~\cite{Kofron:2015gli}. 
By placing a brane, a defect of codimension one, in AdS that slices an accelerating black hole, excising the part of the spacetime containing the string, and then gluing it to a copy of itself, a static black hole
solution in (2+1) dimensions (i.e. on the brane) is obtained that takes into account  quantum backreaction \cite{Emparan:2020znc,Feng:2024uia,Climent:2024nuj}. This construction  is a realisation of the  braneworld holography principle: gravity emerges from the brane.  

For these reasons, amongst others, it is of interest to consider accelerating black holes in (2+1) dimensions. First discovered in~\cite{CMetricOriginalAstorino}, an investigation of such solutions was recently carried out
\cite{Accin3D,RuthAccBH} and a rich panoply of solutions obtained.  Three classes of (2+1)-dimensional versions of the C-metric were obtained, with different interpretations for each. Uniting each solution was the presence of a domain wall: a thread of stress energy that pulled or pushed the black hole, thereby accelerating it. These solutions straightforwardly generalise to include accelerating point masses \cite{Accin3D,RuthAccBH}. These solutions may be extended to the real world (($3+1$) dimensions) to describe an accelerating black string by the addition of a translation-invariant spatial dimension or a black ring by introducing a periodic coordinate and forming a warped product~\cite{CMetricOriginalAstorino}, albeit perhaps with additional stress-energy. 
As the wall is one-dimensional, we shall henceforth refer to it as a string (or strut), recognising that in two spatial dimensions this object is actually a domain wall.

Including rotation in such solutions is not straightforward, since 
a portion of the domain wall 
beyond a sufficiently large distance from the horizon would have a superluminal rotation velocity. One way of addressing this problem would be to place a rotating black hole inside a shell whose size is small enough to ensure such superluminal motion does not occur.

Motivated by this we consider here
the problem of placing a non-rotating black hole inside a shell of stress-energy. We find a range of solutions that is as rich -- or even richer -- as those for the C-metric. Shells in general take the form of cardioids or teardrops, depending on the parameters of the interior metric.
The metric outside the shell is that of a static AdS spacetime for either a black hole or a point mass; which of these is realised likewise depends on the parameters of the solution. 

Our approach is in some sense complementary to the  problem
of  Oppenheimer-Snyder collapse
\cite{Oppenheimer_Snyder} 
in (2+1) dimensions 
\cite{Ross:1992ba}.
Rather than assume circular symmetry
and a dynamically collapsing shell
\cite{Mann:2006yu}, we  assume a static shell but not  circular symmetry. 

The outline of our paper is as follows. We begin in Section~\ref{Sec2} by developing the general formalism for placing the
 C-metric inside a shell, computing
 the junction conditions and the
 resultant density and pressure of the shell.  Both of these quantities depend on the angular location of a Section of the shell relative to the domain wall axis: essentially the stress energy collects away from the direction of acceleration.  In Section~\ref{Sec3}, we construct shells for the first class of black holes pulled by a string and in Section~\ref{Sec4} for those pulled by a strut.  The two situations are similar, but not symmetric. In the next two sections, we carry out the shell construction for the second class of accelerating black holes
that are either pushed (Section~\ref{Sec5}) or pulled (Section~\ref{Sec6}).  In Section~\ref{Sec7}, we consider shell construction for 
the third class, which hitherto has received relatively little attention.
We summarise our work in a concluding section.

{Before proceeding, we note that the primary relevance of $(2+1)$-dimensional gravity is as a `theoretical laboratory' ground for understanding complex gravitational phenomena.  The $(2+1)$-dimensional setting furnishes in many cases mathematically tractable models for studying concepts in classical and quantum gravity, in large part due to the lack of local gravitational degrees of freedom.  We expect that for the solutions we study, the greatest physical relevance will be those of the point particle solutions in Sections~\ref{Sec3}
and~\ref{Sec4}, which have straightforward extensions to accelerating cosmic strings attached to domain walls. The physics of these objects remains an interesting subject for future investigation. }

\section{C-metric in a shell}
\label{Sec2}

There are three classes of C-metric, each with its own interpretation. Class I represents an accelerated particle and Class II an accelerated BTZ black hole, pushed or pulled by a strut. Class III does not have an interpretation in terms of particle-like or black-hole-like solutions. We confine the C-metric within a shell with an asymptotically-AdS spacetime outside the shell absent of any topological defects.

\subsection{C-metric}\label{sec:C metric}

\noindent We present first the C-metric in prolate coordinates~\cite{Accin3D,RuthAccBH, CisternaMannHairyBTZ},
\begin{align}
    \dd s^2&~=~\frac{1}{\Omega^2(x,y)}\lr{-P(y)\dd\tau^2+\frac{\dd y^2}{P(y)}+\frac{\dd x^2}{Q(x)}}\,,\\
    \Omega(x,y)&~=~A(x-y)\,,
\end{align}where $A$ is an acceleration parameter and $P$ and $Q$ are polynomials. There are three distinct classes of C-metric, summarised in~\autoref{table:: Classes Prolate}. Class I represents accelerating particle-like solutions, and has a  subclass dubbed $\text{I}_{\text{C}}$, representing an accelerated black-hole solution~\cite{Accin3D}. 
Class II represents an accelerated black hole and is a one-parameter extension of the BTZ geometry but is disconnected from the Class $\text{I}_{\text{C}}$ black-hole solution. Class III has neither a particle-like nor a black-hole-like interpretation.
\begin{table}[t!]
\centering
\caption{\justifying The three classes of C-metric solutions in prolate coordinates $(x,y)$, each  with its maximal range of $x$. AdS length scale $\ell$ and acceleration parameter $A$.}
\begin{tabular}{|c|c|c|c|} 
 \hline
 Class & $Q(x)$ & $P(y)$ & Maximal range of $x$ \\ [0.5ex] 
 \hline
 I & $1-x^2$ & $\frac{1}{A^2\ell^2}+(y^2-1)$ & $|x|<1$ \\ [0.5ex]
 II & $x^2-1$ & $\frac{1}{A^2\ell^2}+(1-y^2)$ & $x>1$ or $x<-1$ \\[0.5ex]
 III & $1+x^2$ & $\frac{1}{A^2\ell^2}-(1+y^2)$ & $\RR$ \\ [0.5ex] 
 \hline
\end{tabular}
\label{table:: Classes Prolate}
\end{table}

It is more intuitive to work in polar coordinates $(\sigma,r,\phi)$,
\begin{subequations}
\label{eqn:: polars}
\begin{align}
    \tau&~=~\frac{m^2\mc{A} \sigma}{\alpha}\,,\\
    y&~=~-\frac{1}{\mc{A}r}\,,\\
    x&~=~\begin{cases}
        \cos(m\phi)&\text{Class I~}(0\leq m<1)\,,\\
        \cosh(m\phi)&\text{Class II~}(m>0)\,,\\
        \sinh(m\phi)&\text{Class III~}(m\in\RR)\,,
    \end{cases}
\end{align}    
\end{subequations}
where $\mc{A}=A/m$, $\alpha$, and $m$ are dimensionless parameters. 
To ensure that the C-metric~\cite{Accin3D}
solves the Einstein equations
everywhere, a domain wall is inserted at $\phi=\pm\pi$. This ensures the necessary force for acceleration. We refer to the domain wall as a string if it has positive tension and a strut if it has negative tension. In these coordinates, the C-metric reads
\begin{equation}\label{eqn:: general c metric}
    \dd s^2~=~\frac{1}{\Omega^2(r,\phi)}\lr{-\frac{f(r)}{\alpha^2}\dd\sigma^2+\frac{\dd r^2}{f(r)}+r^2\dd\phi^2}\,,
\end{equation}where $f(r)$ and $\Omega(r,\phi)$ are given in~\autoref{table:: Classes Polar} for the three classes of C-metric. Conformal infinity is found at the zeros of $\Omega$.

\begin{table}[t!]
\centering
\caption{\justifying Three classes of C-metric solutions in polar coordinates $(r,\phi)$.}
\begin{tabular}{|c|c|c|} 
 \hline
 Class & $f(r)$ & $\Omega(r,\phi)$\\ [0.5ex] 
 \hline
 I & $\frac{r^2}{\ell^2}+m^2(1-\mc{A}^2r^2)$ & $1+\mc{A}r\cos(m\phi)$ \\ [0.5ex]
 II & $\frac{r^2}{\ell^2}-m^2(1-\mc{A}^2r^2)$ & $1+\mc{A}r\cosh(m\phi)$\\[0.5ex]
 III & $\frac{r^2}{\ell^2}-m^2(1+\mc{A}^2r^2)$ & $1+\mc{A}r\sinh(m\phi)$ \\ [0.5ex] 
 \hline
\end{tabular}
\label{table:: Classes Polar}
\end{table}

\subsection{BTZ black holes and point particles}\label{sec:BTZ particle}

In Section~\ref{sec:shell construct}, we will construct a shell of stress energy around the C-metric described in~\Sref{sec:C metric}. In this Section, we describe the geometry exterior to the shell, which we require to be the spacetime of constant negative curvature with mass parameter $M$~\cite{BTZoriginal,BTZgeometry,CarlipBTZ}. In Schwarzschild-like coordinates, this metric reads
\begin{subequations}\label{eqn:: btz metric}
    \begin{align}
         \dd s_+^2&~=~-g(r_+)\dd t^2+\frac{\dd r_+^2}{g(r_+)}+r_+^2\dd\theta^2\,,\\
         g(r_+)&~=~\frac{r_+^2}{\ell^2}-M\,,
    \end{align}
\end{subequations}where $\ell$ is the AdS length scale and $M$ is a dimensionless mass parameter. When $\theta$ is $2\pi$ periodic, $M/(8G)$ is the ADM mass of the spacetime~\cite{BTZoriginal,CarlipBTZ}. In this paper, we set $G=1$.

There are four distinct physical interpretations of the metric~\eqref{eqn:: btz metric}, depending on the value of $M$~\cite{QuantumCorrectedBTZ}: for $M\geq0$, the metric~\eqref{eqn:: btz metric} describes a spinless black hole; for $-1<M<0$, the metric~\eqref{eqn:: btz metric} has an angular defect, interpreted as a particle of positive mass at the centre of the spacetime; for $M=-1$, the metric~\eqref{eqn:: btz metric} describes global anti-de Sitter spacetime; and for $M<-1$, the metric~\eqref{eqn:: btz metric} has an angular excess, interpreted as a particle of negative mass at the centre of the spacetime.

We comment briefly now on the transition between the black-hole-like and particle-like solutions at $M=0$. For $M=0$, the metric~\eqref{eqn:: btz metric} reads
\begin{equation}\label{eqn:M0 metric}
    \dd s_+^2~=~-\frac{r_+^2}{\ell^2}\dd t^2+\ell^2\frac{\dd r_+^2}{r_+^2}+r_+^2\dd\theta^2\,.
\end{equation}Introducing $t'=t/\ell, z=\ell/r_+$, the metric~\eqref{eqn:M0 metric} reduces
\begin{equation}\label{eqn:cylinder}
    \dd s_+^2~=~\frac{\ell^2}{z^2}\lr{-\dd{t'}^2+\dd z^2+\dd\theta^2}\,,
\end{equation}where $z>0$. As $\theta$ is identified, the geometry described by~\eqref{eqn:cylinder} is that of Torricelli's trumpet or Gabriel's horn and is conformally a half-cylinder (as $z>0$). If $\theta$, however, were not identified ($-\infty<\theta<\infty$), the geometry would instead be that of the Poincaré patch of AdS.

\subsection{Shell construction}\label{sec:shell construct}

We now consider a spacetime separated into two regions $\ms{V}^+$ and $\ms{V}^-$ by a hypersurface $\Sigma$. Throughout this Section, we denote spacetime indices using Greek indices and hypersurface indices with Latin indices. In the region $\ms{V}^-$ interior to the shell, we consider in turn each class of the C-metric~\eqref{eqn:: general c metric}. In the exterior region $\ms{V}^+$, we take the geometry described in~\Sref{sec:BTZ particle}. Coordinates within $\ms{V}^-$ are the $(\sigma,r,\phi)$, coordinates of \eqref{eqn:: general c metric}, within $\ms{V}^+$ are the $(t,r_+,\theta)$  coordinates of \eqref{eqn:: btz metric}, and on $\Sigma$ are $(\sigma,\phi)$.

The hypersurface $\Sigma$ is placed at $(\sigma,r,\phi)=(\sigma,r_0,\phi)$ 
in the region $\ms{V}^-$,
and  in the region $\ms{V}^+$ at $(t,r_+,\theta)=(\sigma,R_M(\phi),\Theta_M(\phi))$, where $\phi$ is the inner angular coordinate $\phi\in(-\pi,\pi)$ and the subscript $M$ makes the $M$-dependence explicit. 
The induced metric on $\Sigma$  is then
\begin{subequations}\label{eqn:induced metrics}
    \begin{align}
        \dd s_+^2|_\Sigma&~=~\left.g^+_{\mu\nu}\dd x^\mu\dd x^\nu\right|_\Sigma~=~-\lr{\frac{R_M^2}{\ell^2}-M}\dd\sigma^2+\lr{\frac{\ell^2\dot{R}_M^2}{R_M^2-M\ell^2}+R_M^2\dot{\Theta}_M^2}\dd\phi^2\,,\\
        \dd s_-^2|_\Sigma&~=~\left.g^-_{\mu\nu}\dd x^\mu\dd x^\nu\right|_\Sigma~=~\frac{1}{\Omega^2(r_0,\phi)}\lr{-\frac{f(r_0)}{\alpha^2}\dd\sigma^2+r_0^2\dd\phi^2}\,,
    \end{align}
\end{subequations}
in the respective $\ms{V}^+$ and $\ms{V}^-$ coordinates,
where $f(r)$ and $\Omega(r,\phi)$ are given in~\autoref{table:: Classes Polar} for the three classes of C-metric and an overdot denotes a derivative with respect to $\phi$.

The metrics in \eqref{eqn:induced metrics} must match, as required by
the first junction condition~\cite{Israel_thin_shell}. By comparing the coefficients of $\dd\sigma^2$ and $\dd\phi^2$, this yields expressions for $R_M$ and $\Theta_M$ 
\begin{subequations}\label{eqn:: Matching Condition}
    \begin{align}\label{eqn:: Radial matching}
        R_M(\phi)&~=~\ell\sqrt{\frac{f(r_0)}{\alpha^2}\frac{1}{\Omega^2(r_0,\phi)}+M}\,,\\
        \Theta_M(\phi)&~=~\int_0^\phi\dd\phi'\,\frac{1}{R_M(\phi')}\sqrt{\frac{r_0^2}{\Omega^2(r_0,\phi')}-\frac{\ell^2\dot{R}_M^2(\phi')}{R_M^2(\phi')-M\ell^2}}\,.\label{eqn:: Theta matching}
    \end{align}
\end{subequations}
in terms of the interior coordinate $\phi$.

For a well-defined shell, the arguments of the square roots in~\eqref{eqn:: Matching Condition} must be positive. A necessary and sufficient condition for this is the non negativity of the argument in the square root of~\eqref{eqn:: Theta matching}, which imposes a constraint on the possible values of $M$
\begin{equation}\label{eqn:: General min M}
    M~~\geq~M_{\mrm{min}}~=~\frac{f(r_0)}{\alpha^2}\frac{1}{\Omega^2(r_0,\phi_*)}\lr{\frac{\ell^2}{r_0^2}\dot{\Omega}^2(r_0,\phi_*)-1}\,,
\end{equation}
for a given pair $(r_0,\alpha)$, 
where $\phi_*\in\lrb{-\pi,\pi}$ is the value of $\phi$ maximising the right-hand side of~\eqref{eqn:: General min M}.

Using the second junction condition,  
we interpret the hypersurface as a thin shell of stress energy, governed by the Lanczos equation~\cite{Israel_thin_shell},
\begin{equation}\label{eqn:: Lanczos equation}
    S_{ab}~=~-\frac{1}{8\pi}\lr{\lrb{K_{ab}}-\lrb{K}h_{ab}}\,,
\end{equation}where $K_{ab}$ is the extrinsic curvature, $h_{ab}$ is the induced metric on $\Sigma$, $K=h^{ab}K_{ab}$, and  the square bracket notation means $\lrb{A}=A(\ms{V}^+)|_\Sigma-A(\ms{V}^-)|_\Sigma=\lrl{A^+}{\Sigma}-\lrl{A^-}{\Sigma}$ for any tensorial quantity $A$.

Projecting from $\ms{V}^\pm$ to $\Sigma$ yields the induced metric
\begin{equation}
    h_{ab}~=~\left.g^\pm_{\mu\nu}\right|_\Sigma{\ee^\pm}^\mu_a{\ee^\pm}^\nu_b\,,
\end{equation}where
\begin{subequations}\label{eqn:: projectors}
    \begin{align}
        &\ee^-_\sigma~=~(1,0,0)\,,&&\ee^-_\phi~=~(0,0,1)\,,\\
        &\ee^+_\sigma~=~(1,0,0)\,,&&\ee^+_\phi~=~(0,\dot{R}_M,\dot{\Theta}_M)~\eqcolon~{v^+}^\mu\,,
    \end{align}
\end{subequations}
and ${v^+}^\mu$ is the tangent vector to $\Sigma$ in $\ms{V}^+$. The extrinsic curvature is then given by
\begin{equation}
    K^\pm_{ab}~=~\left.\lr{\nabla_\mu n^\pm_\nu}\right|_\Sigma{\ee^\pm}^\mu_a{\ee^\pm}^\nu_b\,,
\end{equation}
where $n^\pm_\mu$ is the normal to $\Sigma$ in $\ms{V}^\pm$ such that ${n^\pm}^\mu{n^\pm}_\mu=1$,
\begin{subequations}\label{eqn:: normal vectors}
    \begin{align}
        &n^-_\mu~=~k^-(\phi)(0,1,0)\,, &&k^-(\phi)~=~\frac{1}{\sqrt{f(r_0)}\Omega(r_0,\phi)}\,,\\
        &n^+_\mu~=~k^+(\phi)(0,\dot{\Theta}_M,-\dot{R}_M)\,, &&k^+(\phi)~=~\frac{1}{\sqrt{g(R_M)\dot{\Theta}_M^2+\dot{R}_M^2/R_M^2}} 
    \end{align}
\end{subequations}
and ${v^+}^\mu n^+_\mu=0$.
We note in passing that in the thin-shell approach to Oppenheimer-Snyder collapse~\cite{Oppenheimer_Snyder}, one relaxes the assumption of staticity of the shell but assumes circular symmetry. In the present formulation, we assume staticity without circular symmetry. As such, the calculations in this formulation have strong parallels with those in the collapse of a uniform thin shell.

The nonzero components of the extrinsic curvature are
\begin{subequations}\label{eqn:ext K curvature}
    \begin{align}\label{eqn:: Km}
        &K^-_{\sigma\sigma}~=~-k^-(\phi)\!\left.\Gamma^r_{\sigma\sigma}\right|_\Sigma\,,&&K^-_{\phi\phi}~=~-k^-(\phi)\!\left.\Gamma^r_{\phi\phi}\right|_\Sigma\,,\\
       &K^+_{\sigma\sigma}~=~-k^+(\phi)\dot{\Theta}_M\!\left.\Gamma^{r_+}_{tt}\right|_\Sigma\,,&&K^+_{\phi\phi}~=~-n^+_\mu\!\left.\frac{\mrm{D}{v^+}^\mu}{\mrm{D}\phi~}\right|_\Sigma\,,\label{eqn:: Kp}
    \end{align}
\end{subequations}    
where    
\begin{subequations}
    \begin{align}
    &\lrl{\Gamma^r_{\phi\phi}}{\Sigma}~=~-\frac{r_0f(r_0)}{\Omega(r_0,\phi)}\,,&&\lrl{\Gamma^r_{\sigma\sigma}}{\Sigma}~=~\frac{f(r_0)}{2\alpha^2\Omega(r_0,\phi)}\lr{f'(r_0)\Omega(r_0,\phi)-2f(r_0)\Omega'(r_0,\phi)}\,,\\
    &\lrl{\Gamma^{r_+}_{r_+r_+}}{\Sigma}~=~-\frac{R_M}{\ell^2}\frac{1}{g(R_M)}\,, &&\lrl{\Gamma^{r_+}_{\theta\theta}}{\Sigma}~=~-R_Mg(R_M)\,,\,\,\,\,\lrl{\Gamma^\theta_{\theta r_+}}{\Sigma}~=~\frac{1}{R_M}\,,\,\,\,\,\lrl{\Gamma^{r_+}_{tt}}{\Sigma}~=~\frac{R_M}{\ell^2}g(R_M)\,,
\end{align}
\end{subequations}
are the relevant nonzero Christoffel symbols on $\ms{V}^\pm$, with $g$ given in~\eqref{eqn:: btz metric},
$f'(r_0)=\p_r f(r)|_{r=r_0}$, and $\Omega'(r_0,\phi)=\lrl{\p_r\Omega(r,\phi)}{r=r_0}$; both of these latter quantities can be obtained from~\autoref{table:: Classes Polar}.  
Noting for any vector $A^\mu$ that
 $\frac{\mrm{D}A^\mu}{\mrm{D}\phi}=\frac{\dd A^\mu}{\dd\phi}+{v^+}^\alpha\Gamma^\mu_{\alpha\beta} A^\beta$, we obtain
    \begin{equation}\label{eqn:: Kphiphi}
        K^+_{\phi\phi}~=~-k^+(\phi)\lr{\dot{\Theta}_M\ddot{R}_M-\ddot{\Theta}_M\dot{R}_M+\dot{\Theta}_M\dot{R}^2_M\lrl{\Gamma^{r_+}_{r_+r_+}}{\Sigma}+\dot{\Theta}_M^3\lrl{\Gamma^{r_+}_{\theta\theta}}{\Sigma}-2\dot{\Theta}_M\dot{R}^2_M\lrl{\Gamma^\theta_{\theta r_+}}{\Sigma}}\,.
    \end{equation}
Using  the following relations
    \begin{align}
        &K^\pm_{\sigma\sigma}-K^\pm h_{\sigma\sigma}~=~-h_{\sigma\sigma}h^{\phi\phi}K^\pm_{\phi\phi}\,,
        &&K^\pm_{\phi\phi}-K^\pm h_{\phi\phi}~=~-h_{\phi\phi}h^{\sigma\sigma}K^\pm_{\sigma\sigma}\,,
    \end{align}
which follow from the definition $K^\pm=h^{ab}K_{ab}$, we find the nonzero components of the Lanczos equations~\eqref{eqn:: Lanczos equation},
\begin{subequations}
    \begin{align}\label{eqn:: shell SE}
       S_{\sigma\sigma}&~=~\frac{h_{\sigma\sigma}h^{\phi\phi}}{8\pi}\lr{K^+_{\phi\phi}-K^-_{\phi\phi}}\,,\\
       S_{\phi\phi}&~=~\frac{h_{\phi\phi}h^{\sigma\sigma}}{8\pi}\lr{K^+_{\sigma\sigma}-K^-_{\sigma\sigma}}\,,
    \end{align}
\end{subequations}

We interpret the stress energy of the shell as a perfect fluid,
\begin{equation}\label{eqn:: SE perfect fluid}
    S^{ab}~=~(\rho+p)u^a u^b+p h^{ab}\,,
\end{equation}where $u^a$ is the two-velocity of an observer on $\Sigma$, $\rho$ is the energy density of the shell, and $p$ is the pressure of the shell. Since the shell is static, we have
\begin{equation}
    u^a~=~\frac{\alpha\Omega(r_0,\phi)}{\sqrt{f(r_0)}}(1,0) \,,
\end{equation} 
implying 
\begin{equation}\label{eqn:: density and pressure}
        \rho(\phi)~=~-h^{\sigma\sigma}S_{\sigma\sigma}\,,
        \quad\quad p(\phi)~=~h^{\phi\phi}S_{\phi\phi}\,,
\end{equation}
for the respective density and pressure of the shell.

\subsection{String ends and point particles}\label{sec:string end}

Both the interior and exterior spacetimes are formed by an identification of the angular coordinate, introducing a string in the interior spacetime. We consider now the point at which the string meets the shell. In the Class $\mrm{I}$ and $\mrm{II}$ solutions, there is a single string at the point of identification $\phi=\pm\pi$. We may calculate the presence of an angular deficit or excess by considering the angle made between a tangent to the shell and a line of constant angle, along which the identifications are made.

In the interior spacetime, the tangent vector to $\phi=\,\,$constant is given by ${v^-}^\mu\propto(0,1,0)$, whereas the tangent to the shell at $r=r_0$ is given by ${u^-}^\mu\propto(0,0,1)$. These vectors are clearly orthogonal, and so  the angle between the shell and the line $\phi=\pi$ is $\pi/2$ and the angle between the shell and the line $\phi=-\pi$ is $\pi/2$.

We consider now the same calculation for the exterior spacetime. The corresponding tangent vectors are given by
    \begin{equation}\label{eqn:pi tangent}
     {v^+}^\mu~=~\kappa_1(0,1,0)\,,\quad
        {u^+}^\mu~=~\kappa_2(0,\mp\dot{R}_M(\pm\pi),\dot{\Theta}_M(\pm\pi))\,,
    \end{equation}
    at $\phi=\pm\pi$. The normalisation functions $\kappa_1$ and $\kappa_2$ are given by
    \begin{equation}
        \kappa_1~=~\sqrt{g(R_M(\pm\pi))}\,,\quad\kappa_2~=~\frac{1}{\sqrt{h_{\phi\phi}(\pm\pi)}}\,,
    \end{equation} where $g(R_M)$ and $h_{\phi\phi}(\phi)$ are given by~\eqref{eqn:induced metrics}. We note that $\dot{R}_M$ is an odd function of its argument, whereas $\dot{\Theta}_M$, $R_M$, and $h_{\phi\phi}$ are even functions. As such, we see that the tangent vectors 
  \eqref{eqn:pi tangent}  at $\phi=\pm\pi$ are the same. In particular, the angle between the two tangents at $\phi=\pi$ and $\phi=-\pi$ is the same. The angle between the tangents at $\phi=\pm\pi$ and the tangent to the shell is given by
\begin{equation}\label{eqn:Delta pi}
    \cos(\Delta)~=~-\frac{\dot{R}_M(\pi)}{\sqrt{h_{\phi\phi}(\pi)g(R_M(\pi))}}~=~-\frac{\dot{R}_M(\pi)}{\sqrt{-h_{\phi\phi}(\pi)h_{
    \sigma\sigma
    }(\pi)}}\,,
\end{equation}where $h_{\sigma\sigma}$ is given by~\eqref{eqn:induced metrics}.

The total angle $\delta$ around the point $\phi=\pm\pi$ on the shell is given by $\delta=\pi+2\Delta$, which may be rewritten as
\begin{subequations}\label{eqn:Defect pi}
    \begin{align}
        \delta&~=~2\pi\gamma\,,\\
        \gamma&~=~\frac{1}{2}+\frac{1}{\pi}\arccos\lr{-\frac{\dot{R}_M(\pi)}{\sqrt{-h_{\phi\phi}(\pi)h_{\sigma\sigma}(\pi)}}}\,.
    \end{align}
\end{subequations}This angular deficit ($\gamma<1$) or excess ($\gamma>1$) arises from the absence of circular symmetry in the shell. For a circularly symmetric shell, we have $R_M=const$, $\dot{R}_M=0$, in which case we have $\gamma=1$, $\delta=2\pi$.
This angular deficit or excess may be interpreted as a point mass~\cite{Deser_massdefect,Ashtekar_massdefect} lying at the end of the string, with mass
\begin{equation}\label{eqn:defect mass}
    \mu_\pi~=~\frac14(1-\gamma)\,,
\end{equation}previously referred to as the Hiscock mass~\cite{HiscockDefectMass,UnruhDefectMass} or the Deser-Jackiw-'t Hooft (DJ'tH) mass~\cite{Deser_massdefect,PelegGravCollapsePhaseTrans}. We note that the DJ'tH mass does not agree with the ADM mass~\cite{UnruhDefectMass,PelegGravCollapsePhaseTrans}. For angular deficits, this point particle has positive mass, whereas for angular excesses, the point particle has negative mass. We recall that for $z\in(-1,0)$, $\arccos z\in(\frac\pi2,\pi)$ and for $z\in(0,1)$, $\arccos z\in(0,\frac\pi2)$. Furthermore, in Class $\mrm{I}$ and $\mrm{II}$, $\dot{R}_M$ is proportional to $\mc{A}$. As Class $\mrm{I}_{\mrm{pulled}}$/$\mrm{II}_{\mrm{left}}$ is related to Class $\mrm{I}_{\mrm{pushed}}$/$\mrm{II}_{\mrm{right}}$ by mapping $\mc{A}\to-\mc{A}$, we therefore see that an angular deficit/excess in one corresponds to an angular excess/deficit in the other. Note that this consistent with the change in sign of the string tension as $\mc{A}\to-\mc{A}$~\cite{Accin3D,RuthAccBH}.

In~\Sref{sec:class III}, we construct a double-string solution, which contains an additional string at $\phi=0$. This spacetime is constructed by taking two copies of the Class $\mrm{III}$ spacetime in the interval $\phi\in(0,\pi)$, relabelling the angular coordinate in the second by $\phi\to-\phi$ and gluing the two copies at $\phi=0$ and $\phi=\pm\pi$. In this construction, we   use two copies of the spacetime and simply relabel the angular coordinate in one copy. As such, the angle between $\phi=0$ and the shell is the same in both copies of the spacetime, as is the angle between $\phi=\pm\pi$ and the shell. The tangent vectors at $\phi=\pi$ are given by~\eqref{eqn:pi tangent} and the tangent vectors at $\phi=0$ are given by
\begin{equation}
    {V^+}^\mu~=~k_1(0,1,0)\,,\quad{U^+}^\mu~=~k_2(0,\dot{R}_M(0),\dot{\Theta}_M(0))\,,
\end{equation}where
\begin{equation}
    k_1~=~\sqrt{g(R_M(0))}\,,\quad k_2~=~\frac{1}{\sqrt{h_{\phi\phi}(0)}}\,.
\end{equation}As in the Class $\mrm{I}$ and $\mrm{II}$ calculations above, the angle at $\phi=0$ and $\phi=\pm\pi$ in the interior metric is given by $\pi/2$. Hence, the angle between the tangent at $\phi=\pm\pi$ and the tangent to the shell is again given by~\eqref{eqn:Delta pi} and~\eqref{eqn:Defect pi}, whereas the angle between the tangent at $\phi=0$ and the tangent to the shell is given by
\begin{equation}
    \cos(\Delta_0)~=~\frac{\dot{R}_M(0)}{\sqrt{-h_{\phi\phi}(0)h_{\sigma\sigma}(0)}}\,.
\end{equation}The total angle around the point $\phi=0$ on the shell is given by $\delta_0=\pi+2\Delta_0$,
\begin{subequations}\label{eqn: gamma zero}
    \begin{align}
        \delta_0&~=~2\pi\gamma_0\,,\\
        \gamma_0&~=~\frac12+\frac1\pi\arccos\lr{\frac{\dot{R}_M(0)}{\sqrt{-h_{\phi\phi}(0)h_{\sigma\sigma}(0)}}}\,.
    \end{align}
\end{subequations}The associated DJ'tH mass of the point particle is given by
\begin{equation}\label{eqn:defect mass zero}
    \mu_0~=~\frac14(1-\gamma_0)\,.
\end{equation}

The angular deficit or excess is an artefact of the absence of circular symmetry. In the case of a circularly symmetric shell, one has $\dot{R}_M=0$. As such, $\gamma$~\eqref{eqn:Defect pi} and $\gamma_0$~\eqref{eqn: gamma zero} are both equal to unity and $\mu_\pi=\mu_0=0$.

\color{black}

\subsection{Angular periodicity in $\ms{V}^+$}
We comment now on the periodicity of $\theta$ in $\ms{V}^+$. \textit{A priori}, there is no reason for $\Theta_M$ to be $2\pi$ periodic and we can therefore only calculate the periodicity of $\theta$ \textit{a posteriori}. As such, $M/8$ is in general not the ADM mass parameter. One may relate the coordinates $(t,r_+,\theta)$ to the usual BTZ coordinates via
\begin{equation}\label{eqn:: ADM coordinates}
      \wt{t}~=~\frac{t}{\beta_M}\,,\qquad\wt{ r}_+~=~\beta_M r_+\,,\qquad\wt{\theta}~=~\frac{\theta}{\beta_M}\,,
\end{equation}where we have introduced
\begin{equation}\label{eqn:betaM}
    \beta_M~\coloneq~\frac{1}{2\pi}\lr{\Theta_M(\pi)-\Theta_M(-\pi)}~=~\frac{1}{2\pi}\int_{-\pi}^\pi\dd\phi'\,\frac{1}{R_M(\phi')}\sqrt{\frac{r_0^2}{\Omega^2(r_0,\phi')}-\frac{\ell^2\dot{R}_M^2(\phi')}{R_M^2(\phi')-M\ell^2}}\,,
\end{equation} such that $\wt{\theta}$ is $2\pi$ periodic. In these coordinates, the shell radial and angular coordinates are
    \begin{equation}\label{eqn:ADM shell coords}
    \wt{R}_M(\phi)~=~\beta_MR_M(\phi)\,,\quad\quad\wt{\Theta}_M(\phi)~=~\frac{\Theta_M(\phi)}{\beta_M}\,.
    \end{equation}The exterior metric is written in its usual coordinates as
\begin{subequations}\label{eqn:: BTZ ADM metric}
    \begin{align}
        \dd\wt{s}^2&~=~-\wt{g}(\wt{r}_+)\dd\wt{t}^2+\frac{\dd\wt{r}_+^2}{\wt{g}(\wt{r}_+)}+\wt{r}_+^2\dd\wt{\theta}^2\,,\\
        \wt{g}(\wt{r}_+)&~=~\frac{\wt{r}_+^2}{\ell^2}-\wt{M}\,,
    \end{align}
\end{subequations}
where  $\wt{r}_+$ is defined such that a closed loop with $\wt{r}_+=\wt{R}$ has circumference $2\pi \wt{R}$ and
\begin{equation}\label{eqn:: mass function}
        \wt{M}~=~\beta_M^2 M~=~\frac{M}{(2\pi)^2}\lr{\int_{-\pi}^{\pi}\dd\phi'\,\frac{1}{R_M(\phi')}\sqrt{\frac{r_0^2}{\Omega^2(r_0,\phi')}-\frac{\ell^2\dot{R}_M^2(\phi')}{R^2_M(\phi')-M\ell^2}}}^2\,.
    \end{equation}
    The ADM mass is then $\wt{M}/8$. If $M>0$, then \eqref{eqn:: BTZ ADM metric} is the BTZ metric.

\subsection{$\mc{A}\to0$ limit}\label{sec: A zero limit}

All three classes of metric functions listed in~\autoref{table:: Classes Polar} reduce to the   solution~\eqref{eqn:: btz metric} in the limit $\mc{A}\to0$ with the identifications $\pm m^2/\alpha^2=M$, $(\sigma,r,\phi)=(t,\alpha r_+,\theta/\alpha)$, where the plus sign corresponds to Class I and the minus sign corresponds to Class II and III. In this Section, we make some observations about the behaviour of the shell in this limit.

The metric function $f(r)$ reads
\begin{equation}
    f(r)~=~\frac{r^2}{\ell^2}\pm m^2\,,
\end{equation}where $+m^2$ denotes the Class I solution and $-m^2$ denotes the Class II and III solutions. In all classes, we have $\Omega(r,\phi)\to1$ as $\mc{A}\to0$. The shell radial and angular coordinates, {imposed by the first junction condition}, reduce to
\begin{equation}\label{eqn:BTZ limit R Theta}
    R_M(\phi)~=~R_M~\coloneq~\ell\sqrt{\frac{f(r_0)}{\alpha^2}+M}\,,\quad\quad\Theta_M(\phi)~=~\frac{r_0}{R_M}\phi\,.
\end{equation}We note that the shell is circularly symmetric and that $\Theta_M(\phi)$ is $2\pi$ periodic when $R_M=r_0$.

The minimum value of the exterior mass parameter is given by
\begin{equation}\label{eqn:BTZ Mmin}
    \Mmin~=~-\frac{f(r_0)}{\alpha^2}\,,
\end{equation}
and  as $M\to\Mmin$, we have $R_M\to0$. This is in contrast to the case when $\mc{A}\neq0$, in which case $M\to\Mmin$ does not imply $R_M(\phi)\to0$.

The ADM mass $\wt{M}/8$ is given by
\begin{equation}\label{eqn:BTZ adm}
    \wt{M}~=~\frac{r_0^2}{R_M^2}M \,,
\end{equation}
from \eqref{eqn:: mass function}. 
We see that $\wt{M}=M$ exactly when $\Theta_M(\phi)$ is $2\pi$ periodic (for which $r_0=R_M$, cf.~\eqref{eqn:BTZ limit R Theta}).

In the $\mc{A}\to0$ limit, the radial and angular coordinates of the shell $R_M$ and $\Theta_M(\phi)$ are sufficiently simple that we may find the energy density and pressure of the shell explicitly. Substituting $R_M$ and $\Theta_M(\phi)$ as given by~\eqref{eqn:BTZ limit R Theta} into the expressions between~\eqref{eqn:ext K curvature} and~\eqref{eqn:: density and pressure}, we find
\begin{subequations}
\begin{align}
    \rho&~=~\frac{1}{8\pi r_0^2}\lr{r_0\sqrt{f(r_0)}-R_M\sqrt{g(R_M)}}\,,\\
    p&~=~\frac{1}{8\pi\ell^2}\lr{\frac{R_M}{\sqrt{g(R_M)}}-\frac{r_0}{\sqrt{f(r_0)}}}\,,
\end{align}
\end{subequations}where we have used the simplified junction condition $\alpha^2=f(r_0)/g(R_M)$ {(cf.~\eqref{eqn:BTZ limit R Theta})}.

Through the junction condition, we may use $\alpha$ to specify a choice of $R_M$. A choice of interest is $R_M=r_0$ --- a choice enforcing $\Theta_M(\phi)$ to be $2\pi$ periodic and $\wt{M}=M$. The density and pressure reduce to
\begin{subequations}\label{eqn: p rho simple}
    \begin{align}
        \rho&~=~\frac{\sqrt{f(r_0)}}{8\pi r_0}\lr{1-\sqrt{\frac{g(r_0)}{f(r_0)}}}\,,\\
        p&~=~\frac{r_0}{8\pi\ell^2\sqrt{f(r_0)}}\lr{\sqrt{\frac{f(r_0)}{g(r_0)}}-1}\,.
    \end{align}
\end{subequations}This relation between $\rho$ and $p$~\eqref{eqn: p rho simple} may be restated as the equation of state
\begin{subequations}\label{eqn:EOS}
\begin{align}
    p&~=~c_s^2\rho\,,\\
    c_s^2&~=~\lr{\frac{r_0}{\ell}}^2\frac{1}{\sqrt{f(r_0)g(r_0)}}\,,
\end{align}
\end{subequations}where $c_s$ is the speed of sound. More generally, $c_s^2$ is given by $\td{p}{\rho}$ and when both the shell energy density and pressure depend on the shell angular coordinate $\phi$, we have $c_s^2=\td{p}{\phi}/\td{\rho}{\phi}$. The density and pressure may either both be positive ($\pm m^2>-M$), zero ($\pm m^2=-M$), or negative ($\pm m^2<-M$). In the case of vanishing stress energy, the interior and exterior metrics are identical. {The formulae}~\eqref{eqn: p rho simple} reproduce the results of~\cite{Mann:2006yu} in the case of a static shell and in the special case $m=0$, the formulae~\eqref{eqn: p rho simple} reproduce the density and pressure found in~\cite{LemosThinShellEntropy}, and~\cite{LemosThinShellRotating} in the case of a non-rotating shell.

We consider now which energy conditions, summarised in~\autoref{table:: energy conditions}, are respected or violated by the stress energy~\eqref{eqn: p rho simple}. As the shell energy density and pressure take the same sign, if either is negative, all energy conditions in~\autoref{table:: energy conditions} are violated. We therefore focus on the case $\rho,~p>0$, calling this condition positivity. Positive stress energy immediately satisfies the NEC, WEC, and SEC. We say stress energy is causal if $c_s^2\leq 1$. In the special case of a linear, barotropic equation of state~\eqref{eqn:EOS}, the DEC ($\rho\geq p>0$) is equivalent to causality ($c_s^2\leq1$).

\begin{table}[t!]
\centering
\caption{\justifying Energy conditions in $1+1$ dimensions.}
\begin{tabular}{|c c|c|} 
 \hline
 \textbf{Energy Condition} & & Perfect fluid\\ [0.5ex] 
 \hline
 \textit{Null energy condition} &(NEC) & $\rho+p\geq0$\\ [0.5ex]
 \textit{Weak energy condition} &(WEC) & $\rho\geq0$, $\rho+p\geq0$\\[0.5ex]
 \textit{Dominant energy condition} &(DEC) & $\rho\geq|p|$, $\rho+p\geq0$ \\ [0.5ex] 
 \textit{Strong energy condition} &(SEC) & $\rho+p\geq0$\\[0.5ex]
 \hline
\end{tabular}
\label{table:: energy conditions}
\end{table}

Consider next the Class I solution. To satisfy the energy conditions in~\autoref{table:: energy conditions}, we require positivity and causality,
\begin{subequations}\label{eqn:class I pos caus}
    \begin{align}
        -M&~<~m^2\,,&&\text{(positivity)}\\
        0&~\leq~ r_0^2(m^2-M)-m^2 M\ell^2\,,&&\text{(causality)}
    \end{align}
\end{subequations}as well as the positivity of the metric functions evaluated on the shell $f(r_0)$ and $g(r_0)$. For $M\leq0$, causality is trivially satisfied. Positivity then requires  $m^2>|M|$, yielding positive metric functions. For $M>0$, positivity is trivially satisfied, and
causality implies $m^2>M$ and $r_0^2/\ell^2>Mm^2/(m^2-M)$. In conjunction, we see that only sufficiently large shells with $m^2>|M|$ satisfy the energy conditions.

Whilst the inequality $m^2>|M|$ may seem counter-intuitive, we can explain this physically. Recall that the ADM mass interior and exterior to the shell is given by $8M^-_{\mrm{ADM}}=-m^2$ and $8M^+_{\mrm{ADM}}=M$ respectively~\cite{QuantumCorrectedBTZ,BTZNegativeSpectrum}. As such, this inequality may be rewritten as $-|M^+_{\mrm{ADM}}|>M^-_{\mrm{ADM}}$, or
$-|E^+|>E^-$, upon writing the interior and exterior energy as $E^\pm=M^\pm_{\mrm{ADM}}$. Physically, 
a shell with positive stress energy such that $\rho\geq|p|$ can source the exterior energy $E^+$ provided
the interior energy $E^-$ is sufficiently smaller than the exterior energy $E^+$. 

Turning to the Class II and III solutions (the same in the $\mc{A}\to0$ limit), to satisfy the energy conditions, we again require positivity and causality,
\begin{subequations}\label{eqn:class II pos caus}
    \begin{align}
        0&~\leq~m^2~<~M\,,&&\text{(positivity)}\\
        0&~\leq~ -r_0^2(m^2+M)+m^2 M\ell^2\,,&&\text{(causality)}
    \end{align}
\end{subequations}as well as the positivity of $f(r_0)$ and $g(r_0)$. By positivity, we have a strictly BTZ ($M>0$) solution exterior to the shell, hence $m^2+M>0$. By causality, we have
\begin{equation}\label{eqn:g ineq}
    \frac{r_0^2}{\ell^2}~\leq~\frac{m^2M}{m^2+M}\iff g(r_0)~=~\frac{r_0^2}{\ell^2}-M~\leq~-\frac{M^2}{m^2+M}\,.
\end{equation}However, as $m^2+M$ is positive, the inequality~\eqref{eqn:g ineq} implies $g(r_0)<0$. Therefore, any $r_0$ outside the event horizon will violate causality: we may either respect positivity, or causality, but not both. Hence, there are no solutions satisfying all energy conditions in~\autoref{table:: energy conditions}; in particular, positive stress energy surrounding any black-hole solution must be superluminal. This result makes sense physically, which we can see as follows. A negative cosmological constant exerts a tension on matter, forcing it to collapse~\cite{Ross:1992ba,Cruz:1994ar}. As such, for a shell to remain static, it must exert additional pressure to counteract the negative cosmological constant. This increase in pressure is sufficiently large that the DEC $\rho\geq|p|$ is violated.

A special case at the intersection of Class I, II, and III is the solution with $m=0$, corresponding to Torricelli's trumpet interior to the shell~\eqref{eqn:cylinder}. As $m=0$ belongs to Class II and III, we may appeal to the arguments above. In particular, we may respect positivity, or causality, but not both simultaneously --- the DEC. This is contrary to the claim in~\cite{LemosThinShellEntropy} that the stress energy satisfies the DEC.

Finally, we address the angular deficit or excess at $\phi=\pm\pi$, and also at $\phi=0$ in Class III. As $\mc{A}\to0$, we see $R_M(\phi)$ tends to a constant~\eqref{eqn:BTZ limit R Theta} and so $\dot{R}_M\to0$. In this case, the angular deficit or excess characterised by~\eqref{eqn:Defect pi} and~\eqref{eqn: gamma zero} vanishes: $\gamma,~\gamma_0\to1$. As such, the point particle mass~\eqref{eqn:defect mass} and~\eqref{eqn:defect mass zero} vanishes: $\mu_\pi,~\mu_0\to0$ as $\mc{A}\to0$. This is consistent with the interpretation that the angular deficit or excess is an artefact of the lack of circular symmetry of the shell.

\subsection{Large-$M$ limit}

We consider now the leading-order asymptotic behaviour of the shell for large values of the exterior mass parameter $M$.

In the limit $M\to\infty$, the shell radius and angular coordinate as viewed from outside the shell are given by
\begin{subequations}\label{eqn:large M R theta}
    \begin{align}
        R_M(\phi)&~\sim~\ell\sqrt{M}\,,\\
        \Theta_M(\phi)&~\sim~\frac{r_0}{\ell\sqrt{M}}\int_0^\phi\dd\phi'\,\frac{1}{\Omega(r_0,\phi')}\,,
    \end{align}
\end{subequations}where we have used~\eqref{eqn:: Matching Condition}. The asymptotic behaviour of $\beta_M$~\eqref{eqn:betaM} is given by
\begin{equation}
    \beta_M~\sim~\frac{r_0}{2\pi\ell\sqrt{M}}\int_{-\pi}^\pi\dd\phi'\,\frac{1}{\Omega(r_0,\phi')}\,,
\end{equation}such that  the ADM shell radius~\eqref{eqn:ADM shell coords},
\begin{equation}\label{eqn:ADM shell R}
    \beta_M R_M(\phi)~\sim~\frac{r_0}{2\pi}\int_{-\pi}^{\pi}\dd\phi'\,\frac{1}{\Omega(r_0,\phi')}\,,
\end{equation}asymptotes to a constant.

The leading-order asymptotic behaviour of the ADM mass parameter $\wt{M}/8$~\eqref{eqn:: mass function} is given by
\begin{equation}\label{eqn:mmax}
    \wt{M}_{\mrm{max}}~\sim~\lr{\frac{r_0}{2\pi\ell}}^2\lr{\int_{-\pi}^\pi\dd\phi'\,\frac{1}{\Omega(r_0,\phi')}}^2\,,
\end{equation}
which is positive, thus implying 
a black hole solution. 
Together \eqref{eqn:mmax}
and \eqref{eqn:: General min M} imply that the ADM mass of the exterior spacetime is bounded both from above and below.

In the large-$M$ limit, the energy density and pressure of the shell are given by
\begin{subequations}\label{eqn:large M SE}
    \begin{align}\label{eqn: large M rho}
        \rho(\phi)&~\sim~\frac{\sqrt{f(r_0)}}{8\pi r_0}\,,\\
        p(\phi)&~\sim~\frac{\alpha \Omega(r_0,\phi)}{8\pi\sqrt{f(r_0)}}\sqrt{M}\,.
    \end{align}
\end{subequations}As the stress energy $\rho$ in~\eqref{eqn:large M SE} is positive, we can always construct a shell with a physical stress energy for the heaviest, allowable, black-hole BTZ solution. Furthermore,  the energy density of the shell is evenly distributed since it is independent of the shell intrinsic coordinate $\phi$. However, the pressure of the constructed shell increases without bound.

In this case, we see $\rho,~p>0$; hence, the NEC, WEC, and SEC are satisfied. However, the DEC is violated: $\rho$ tends to a constant, whereas $p$ grows as $\sqrt{M}$. Furthermore, causality is also violated: in this limit, $\td{\rho}{\phi}\to0$ with $\td{p}{\phi}\neq 0$, therefore $c_s^2=\td{p}{\phi}/\td{\rho}{\phi}\to\infty$.

To consider the point mass located at the end of the string (or strings for the Class $\mrm{III}$ solution), we begin with the identity
\begin{equation}
    \dot{R}_M(\phi)~=~\frac{\ell^2 f(r_0)}{\alpha^2 R_M(\phi)}\Omega(r_0,0)\dot{\Omega}(r_0,\phi)\,.
\end{equation}
All $M$-dependence is found in $R_M\sim l\sqrt{M}$ by~\eqref{eqn:large M R theta}. As such, the angular deficit/excess
$\gamma$ in~\eqref{eqn:Defect pi}
at $\phi=\pm\pi$  is given by
\begin{equation}
    \gamma~\sim~1+\frac{\ell f(r_0)}{\alpha^2\sqrt{M}}\Omega(r_0,\pi)\dot{\Omega}(r_0,\pi)\,.
\end{equation}The leading-order behaviour of the DJ'tH mass of the point particle is then
\begin{equation}\label{eqn:mu pi large M}
    \mu_\pi~\sim~-\frac{\ell f(r_0)}{4\alpha^2\sqrt{M}}\Omega(r_0,\pi)\dot{\Omega}(r_0,\pi)\,.
\end{equation}

Similarly, in the Class $\mrm{III}$ spacetime, the leading-order behaviour of the DJ'tH mass of the point particle at $\phi=0$ is given by
\begin{equation}
    \mu_0~\sim~\frac{\ell f(r_0)}{4\alpha^2\sqrt{M}}\Omega(r_0,0)\dot{\Omega}(r_0,0)\,.
\end{equation}

The mass of the point particle due to an angular deficit or excess therefore decays as $M^{-1/2}$. That the mass tends to zero for large $M$ is consistent with the observation in~\Sref{sec:string end} that the angular deficit/excess originates from the lack of circular symmetry of the shell. In the large-$M$ limit, we see that the shell radius tends to a constant~\eqref{eqn:large M R theta} (or~\eqref{eqn:ADM shell R} in ADM coordinates). As such, the shell becomes circularly symmetric and its angular deficit or excess should vanish, along with mass of the point particle.

We now proceed to analyse the properties of the shell for each class in~\autoref{table:: Classes Polar}.

\section{Class $\text{I}_{\text{pulled}}$: Particle pulled by a string}\label{sec:: class I string}
\label{Sec3}

In this Section, we consider the Class $\mrm{I}_{\mrm{pulled}}$ C-metric solution, describing a particle pulled by a string. In polar coordinates $(\sigma,r,\phi)$, the metric reads
\begin{subequations}\label{eqn:: Class I full metric}
\begin{align}
    \dd s^2_-&~=~\frac{1}{\Omega^2(r,\phi)}\lr{-\frac{f(r)}{\alpha^2}\dd \sigma^2+\frac{\dd r^2}{f(r)}+r^2\dd\phi^2}\,,\\
    f(r)&~=~\frac{r^2}{\ell^2}+m^2(1-\mc{A}^2r^2)\,,\\
    \Omega(r,\phi)&~=~1+\mc{A}r\cos(m\phi)\,,
\end{align}    
\end{subequations}where $0<m<1$. 
The tension of the string is given by
\begin{equation}  \uptau_\pi~=~\frac1{4\pi}m\mc{A}\sin(m\pi)\,.
\end{equation}
 
We demonstrate now why this metric may be interpreted as describing the geometry around a point particle. This may be clearly seen by considering the spacetime near $r=0$. To leading order we have 
\begin{equation}
    \dd s^2~=~ -\dd \tilde{t}^2+\dd \tilde{r}^2+\tilde{r}^2\dd\tilde{\phi}^2\,,
\end{equation}where $\tilde{t}=m\sigma/\alpha$, $\tilde{r}=r/m$, and $\tilde{\phi}=m\phi$. This is $(2+1)$-dimensional Minkowski spacetime but with an angular coordinate $\tilde{\phi}$ with a periodicity of $2m\pi<2\pi$ --- the spacetime has an angular deficit. It is this angular deficit that allows us to interpret the Class I solution as a point particle: an angular deficit has an associated DJ'tH mass $\mu_{\mrm{ad}}$~\cite{Deser_massdefect,Ashtekar_massdefect},
\begin{equation}\label{eqn:mad}
    \mu_{\mrm{ad}}~=~\frac{1}{4}(1-m)\,.
\end{equation}We see that as $m\to1$, $\tilde{\phi}$ becomes $2\pi$ periodic and the DJ'tH mass vanishes.

The spacetime~\eqref{eqn:: Class I full metric} has a conformal boundary at
\begin{equation}
    \rc~=~-\frac{1}{\mc{A}\cos(m\phi)}\,.
\end{equation}Taking the limit $\mc{A}\to0$ and identifying $m^2=\alpha^2M$ with $M<0$, one recovers the geometry of a non-rotating point particle~\eqref{eqn:: btz metric} in coordinates $(\sigma,r,\phi)=(t,\alpha r_+,\theta/\alpha)$. We note that the Class I solution can be continuously deformed into Class II by letting $m^2\to-m^2$ with $\mc{A}$ fixed.

\begin{figure}[t!]
    \centering
    \begin{subfigure}[b]{0.4\textwidth}
        \centering\captionsetup{labelfont=bf}
        \begin{tikzpicture}
        \draw (0, 0) node[inner sep=0] {\includegraphics[width=\textwidth]{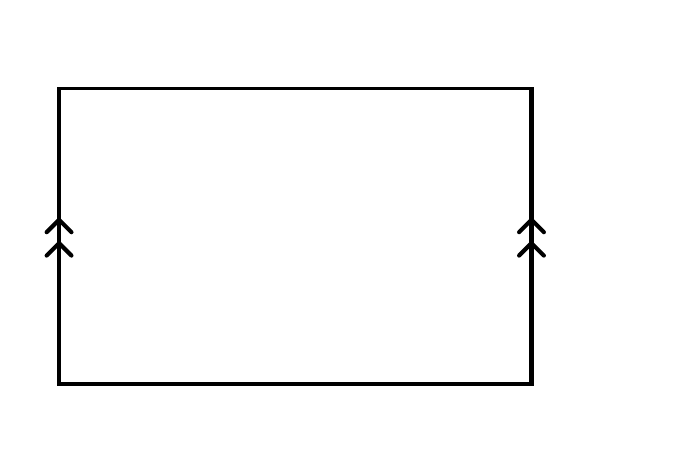}};
        \draw (2.67, 1.4) node {\large$r=\infty$};
        \draw (2.6, -1.3) node {\large$r=0$};
        \draw (-2.4, -1.85) node {\large$\phi=-\pi$};
        \draw (1.8, -1.85) node {\large$\phi=\pi$};
        \end{tikzpicture}
        \caption{\\}
    \end{subfigure}\hfill
    \begin{subfigure}[b]{0.4\textwidth}
        \centering\captionsetup{labelfont=bf}
        \begin{tikzpicture}
        \draw (0, 0) node[inner sep=0] {\includegraphics[width=\textwidth]{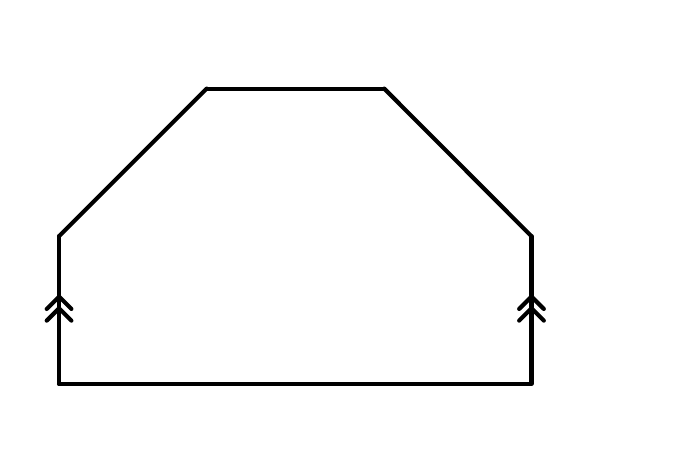}};
        \draw (1.5, 1.4) node {\large$r=\infty$};
        \node[rotate=45] at (-2.15,0.95) {\large $r=r_{\mrm{conf}}$};
        \draw (2.6, -1.3) node {\large$r=0$};
        \draw (-2.4, -1.85) node {\large$\phi=-\pi$};
        \draw (1.8, -1.85) node {\large$\phi=\pi$};
        \end{tikzpicture}
        \caption{\\}
    \end{subfigure}
    \begin{subfigure}[b]{0.4\textwidth}
        \centering\captionsetup{labelfont=bf}
        \begin{tikzpicture}
        \draw (0, 0) node[inner sep=0] {\includegraphics[width=\textwidth]{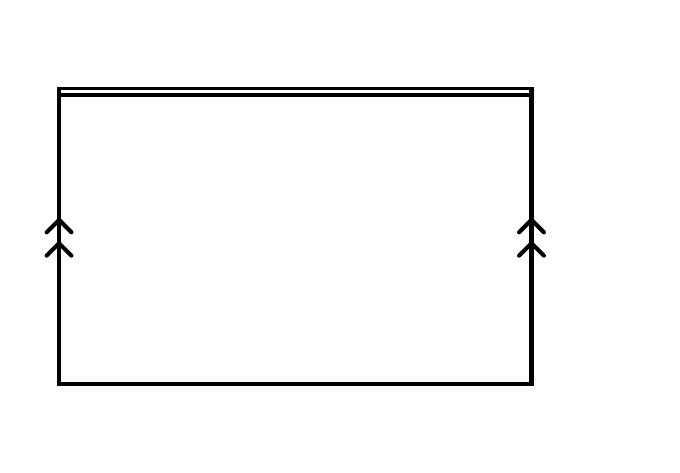}};
        \draw (2.67, 1.4) node {\large$r=\infty$};
        \draw (2.6, -1.3) node {\large$r=0$};
        \draw (-2.4, -1.85) node {\large$\phi=-\pi$};
        \draw (1.8, -1.85) node {\large$\phi=\pi$};
        \end{tikzpicture}
        \caption{\\}
    \end{subfigure}\hfill
    \begin{subfigure}[b]{0.4\textwidth}
        \centering\captionsetup{labelfont=bf}
        \begin{tikzpicture}
        \draw (0, 0) node[inner sep=0] {\includegraphics[width=\textwidth]{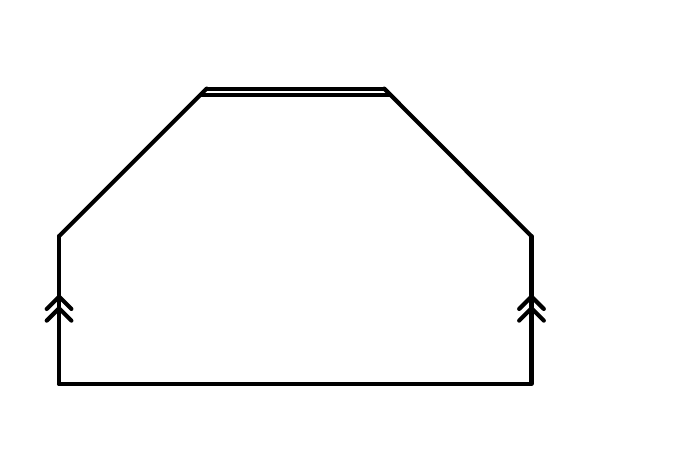}};
        \draw (1.5, 1.4) node {\large$r=\infty$};
        \node[rotate=45] at (-2.15,0.95) {\large $r=r_{\mrm{conf}}$};
        \draw (2.6, -1.3) node {\large$r=0$};
        \draw (-2.4, -1.85) node {\large$\phi=-\pi$};
        \draw (1.8, -1.85) node {\large$\phi=\pi$};
        \end{tikzpicture}
        \caption{\\}
    \end{subfigure}
    \begin{subfigure}[b]{0.4\textwidth}
        \centering\captionsetup{labelfont=bf}
        \begin{tikzpicture}
        \draw (0, 0) node[inner sep=0] {\includegraphics[width=\textwidth]{figures/SpaceDiagrams/slow_horizon.pdf}};
        \draw (2.67, 1.4) node {\large$r=r_h$};
        \draw (2.6, -1.3) node {\large$r=0$};
        \draw (-2.4, -1.85) node {\large$\phi=-\pi$};
        \draw (1.8, -1.85) node {\large$\phi=\pi$};
        \end{tikzpicture}
        \caption{\\}
    \end{subfigure}\hfill
    \begin{subfigure}[b]{0.4\textwidth}
        \centering\captionsetup{labelfont=bf}
        \begin{tikzpicture}
        \draw (0, 0) node[inner sep=0] {\includegraphics[width=\textwidth]{figures/SpaceDiagrams/conf_horizon_double.pdf}};
        \draw (1.5, 1.4) node {\large$r=r_h$};
        \node[rotate=45] at (-2.15,0.95) {\large $r=r_{\mrm{conf}}$};
        \draw (2.6, -1.3) node {\large$r=0$};
        \draw (-2.4, -1.85) node {\large$\phi=-\pi$};
        \draw (1.8, -1.85) node {\large$\phi=\pi$};
        \end{tikzpicture}
        \caption{\\}
    \end{subfigure}
    \caption{\justifying Constant time slice of the Class $\mrm{I}_{\mrm{pulled}}$ solution describing a particle being pulled by a string. \textbf{(a)} Slow acceleration with $m\leq\tfrac12$. \textbf{(b)} Slow acceleration with $m>\tfrac12$. \textbf{(c)} Saturated acceleration with $m\leq\tfrac12$. \textbf{(d)} Saturated acceleration with $m>\tfrac12$. \textbf{(e)} Rapid acceleration with $m\leq\tfrac12$. \textbf{(f)} Rapid acceleration with $m>\tfrac12$. Double lines in \textbf{(c)} to \textbf{(f)} represent a Killing horizon.
    }
    \label{fig:Class I Pulled ranges}
\end{figure}

This geometry exhibits three different phases of acceleration, dubbed \textit{slow} ($m^2\mc{A}^2\ell^2<1$), \textit{saturated} ($m^2\mc{A}^2\ell^2=1$), and \textit{rapid} ($m^2\mc{A}^2\ell^2>1$). In the slow acceleration regime, the geometry has no horizons; whereas for $m^2\mc{A}^2\ell^2\geq1$, there is a Killing horizon at 
\begin{equation}
    r_h~=~\frac{m\ell}{\sqrt{m^2\mc{A}^2\ell^2-1}}\,,
\end{equation}referred to as the acceleration horizon. The geometry of the three phases is depicted in~\autoref{fig:Class I Pulled ranges}.

For all phases of the acceleration, the minimum value of the exterior mass parameter $M$~\eqref{eqn:: General min M} is given by
\begin{subequations}\label{eqn:M min Class I}
    \begin{align}
\Mmin&~=~\frac{f(r_0)}{\alpha^2}\frac{(m^2\mc{A}^2\ell^2\sin^2(m\phi_*)-1)}{(1+\mc{A}r_0\cos(m\phi_*))^2}\,,\\
\phi_*&~=~\begin{cases}
    0&\text{for~}m\leq\frac{1}{\mc{A}\ell}\sqrt{\frac{\mc{A}r_0}{1+\mc{A}r_0}}\,,\\
    \min\lrc{\frac1m\arccos\lr{\frac{(1-m^2\mc{A}^2\ell^2)\mc{A}r_0}{m^2\mc{A}^2\ell^2}},\pi}&\text{for~}m>\frac{1}{\mc{A}\ell}\sqrt{\frac{\mc{A}r_0}{1+\mc{A}r_0}}\,.
\end{cases}
    \end{align}
\end{subequations}
We make three observations. First, for $\phi_*=\frac1m\arccos\lr{\frac{(1-m^2\mc{A}^2\ell^2)\mc{A}r_0}{m^2\mc{A}^2\ell^2}}$, we may simplify~\eqref{eqn:M min Class I}
\begin{equation}
    \Mmin~=~-\frac{m^2}{\alpha^2}\lr{1-m^2\mc{A}^2\ell^2}\,.
\end{equation}
Second, since $\mc{A}r_0/(1+\mc{A}r_0)\in (0,1)$, we may simplify~\eqref{eqn:M min Class I} for saturated accelerations,
\begin{subequations}
    \begin{align}
        \Mmin&~=~-\frac{f(r_0)}{\alpha^2}\frac{\cos^2(m\phi_*)}{\lr{1+\mc{A}r_0\cos(m\phi_*)}^2}\,,\\
        \phi_*&~=~\min\lrc{\frac{\pi}{2m},\pi}\,,
    \end{align}
\end{subequations}
in which case 
 we have $\Mmin=0$ for $1/2<m<1$. Otherwise, for $0< m\leq1/2$, we have $\Mmin<0$. Finally, 
 for rapid accelerations ($m\mc{A}\ell>1$)
 \eqref{eqn:M min Class I} simplifies to
 \begin{equation}
   \phi_*~=~
    \min\lrc{\frac1m\arccos\lr{\frac{(1-m^2\mc{A}^2\ell^2)\mc{A}r_0}{m^2\mc{A}^2\ell^2}},\pi}\,,
 \end{equation}
using again $\mc{A}r_0/(1+\mc{A}r_0)\in (0,1)$.

The shell radius $R_M(\phi)$~\eqref{eqn:: Radial matching} has the following asymptotic behaviour around $\phi=0$
\begin{equation}\label{eqn:quad behaviour}
    R_M(\phi)~\sim~\ell\lr{\sqrt{\frac{f(r_0)}{\alpha^2(1+\mc{A}r_0)^2}+M}}+\frac{\mc{A}\ell m^2 r_0 f(r_0)}{2\alpha^2(1+\mc{A}r_0)^3}\frac{1}{\sqrt{\frac{f(r_0)}{\alpha^2(1+\mc{A}r_0)^2}+M}}\phi^2+\OO(\phi^4)\,.
\end{equation}
We see that the leading  term is a strictly increasing function of $M$, whereas the coefficient of the quadratic term is a strictly decreasing function of $M$. As $M\to\Mmin$, we see the leading  term attains its minimum value, whereas the coefficient of the quadratic term attains its maximum value.

\subsection{Slow acceleration}

\begin{figure}[t!]
    \centering
        \begin{tikzpicture}
        \draw (0, 0) node[inner sep=0] {\includegraphics[width=\textwidth]{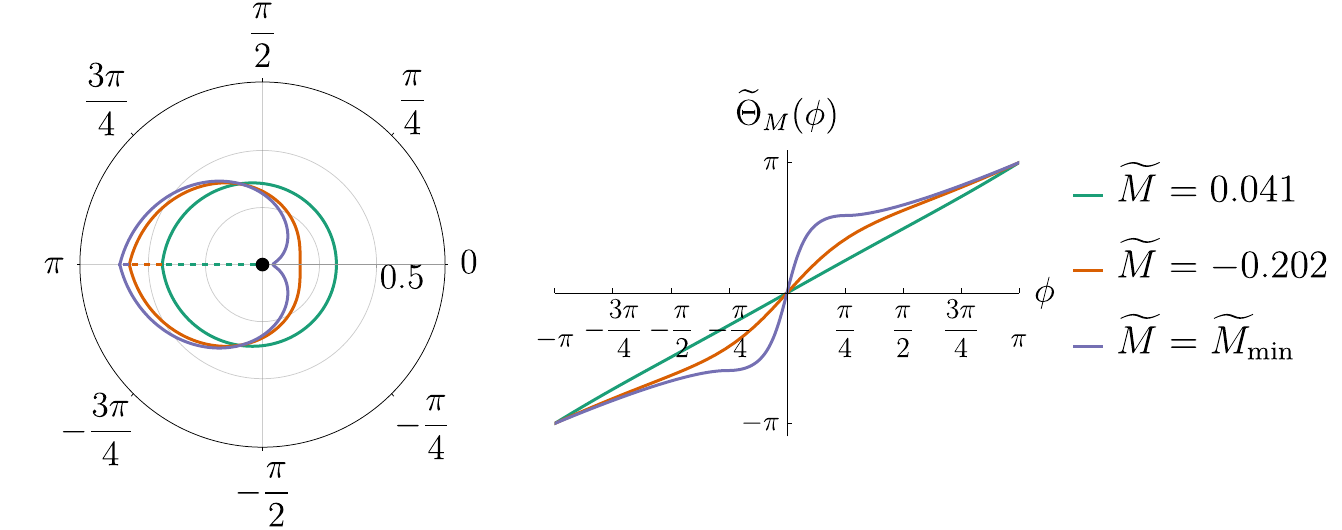}};
        \draw (-8, 2.4) node {\textbf{(a)}};
        \draw (-1.6, 2.4) node {\textbf{(b)}};
        \end{tikzpicture}
        \vspace{-1em}
        \begin{tikzpicture}
        \draw (0, 0) node[inner sep=0] {\includegraphics[width=\textwidth]{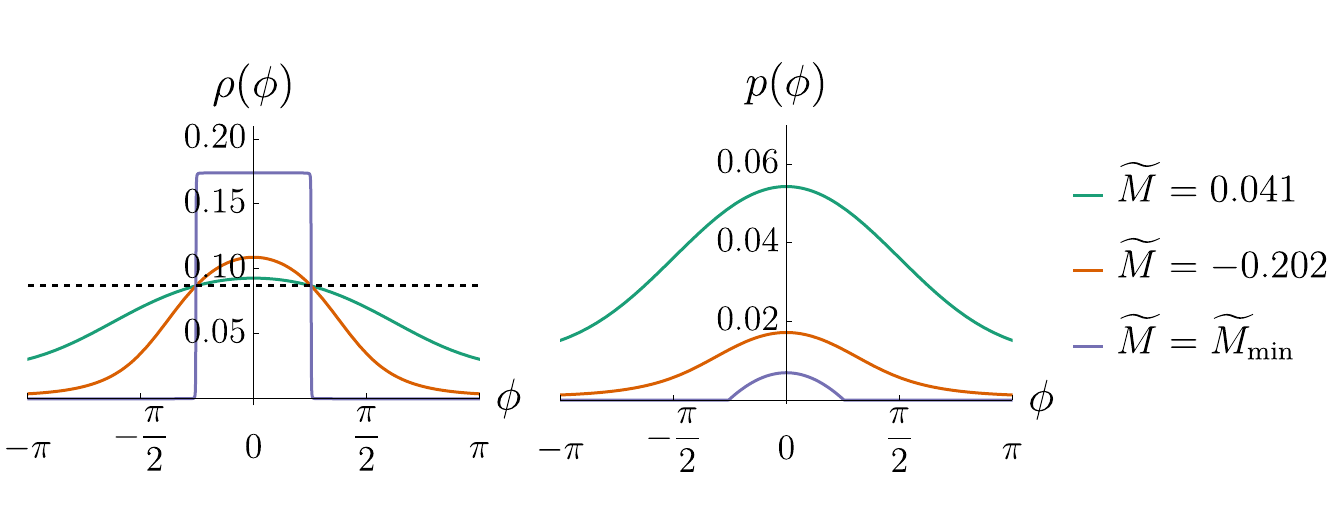}};
        \draw (-8, 2.4) node {\textbf{(c)}};
        \draw (-1.6, 2.4) node {\textbf{(d)}};
        \end{tikzpicture}
        \vspace{-1em}
        \begin{tikzpicture}
        \draw (0, 0) node[inner sep=0] {\includegraphics[width=\textwidth]{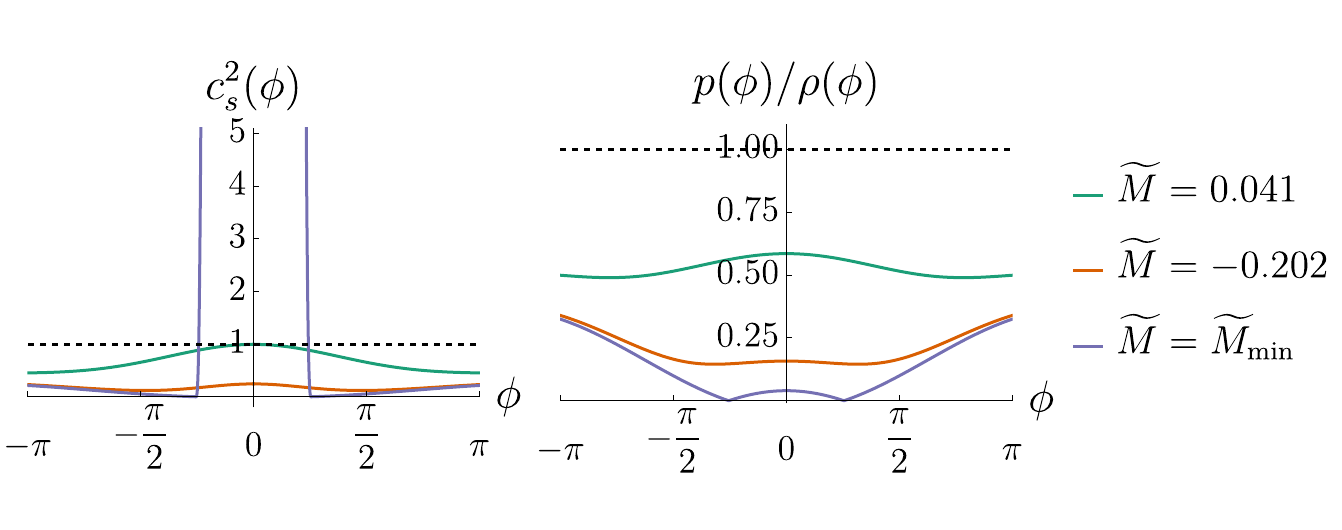}};
        \draw (-8, 2.4) node {\textbf{(e)}};
        \draw (-1.6, 2.4) node {\textbf{(f)}};
        \end{tikzpicture}
    \caption{\justifying Class $\mrm{I}_{\mrm{pulled}}$ (slow phase)
    for $r_0=0.4$, $m=0.8$, $\mc{A}=0.6$, $\alpha=0.25$, and $\ell=1$, where $\wt{M}_{\mrm{min}}\approx -0.282$ and $\wt{M}_{\mrm{max}}\approx0.149$.
    \textbf{(a)} Polar plot of shell radius $\wt{R}_M$~\eqref{eqn:ADM shell coords} over $\phi\in(-\pi,\pi)$ for varying values of $\wt{M}$; dashed line is the string and black dot represents the point particle. \textbf{(b)} Shell angular coordinate $\wt{\Theta}_{M}$~\eqref{eqn:ADM shell coords} for varying values of $\wt{M}$. \textbf{(c)} Energy density $\rho$ of the shell for varying values of $\wt{M}$. Black dashed line at asymptotic value as $\wt{M}\to\wt{M}_{\mrm{max}}$. \textbf{(d)} Pressure $p$ of the shell for varying values of $\wt{M}$. \textbf{(e)} Squared speed of sound $c_s^2$ of matter in the shell for varying values of $\wt{M}$; dashed line at $c_s=1$. \textbf{(f)} Ratio of shell pressure to shell energy density for varying values of $\wt{M}$; dashed line at $p/\rho=1$.}\label{fig:Class I slow}
\end{figure}
\begin{figure}[t!]
    \centering
    \begin{tikzpicture}
    \draw (0, 0) node[inner sep=0] {\includegraphics[width=\textwidth]{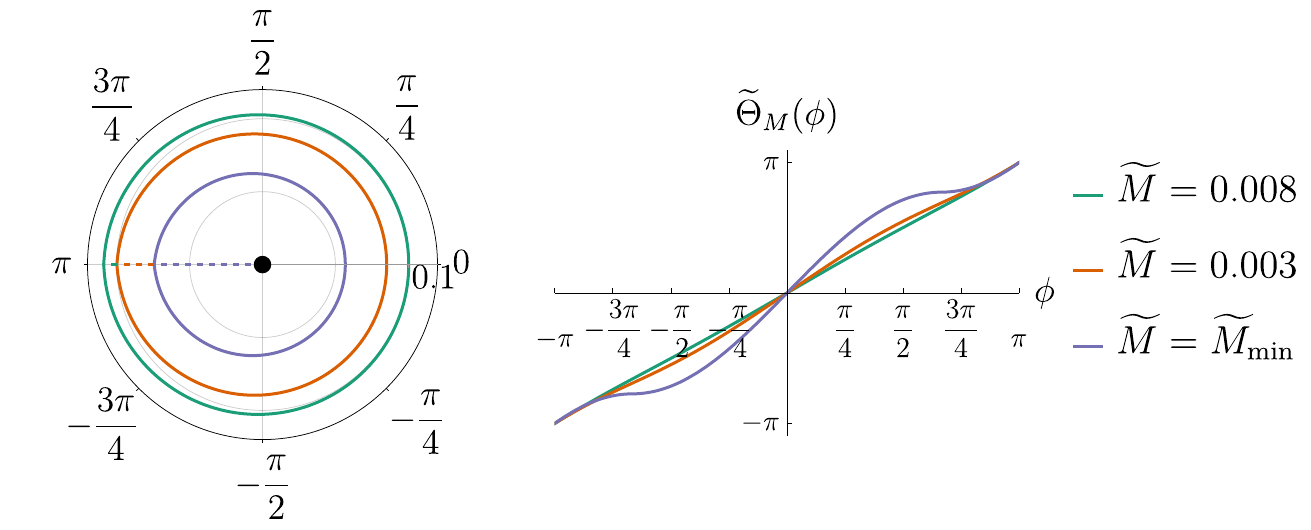}};
     \draw (-8, 2.4) node {\textbf{(a)}};
    \draw (-1.6, 2.4) node {\textbf{(b)}};
\end{tikzpicture}\vspace{-1em}
\begin{tikzpicture}
    \draw (0, 0) node[inner sep=0] {\includegraphics[width=\textwidth]{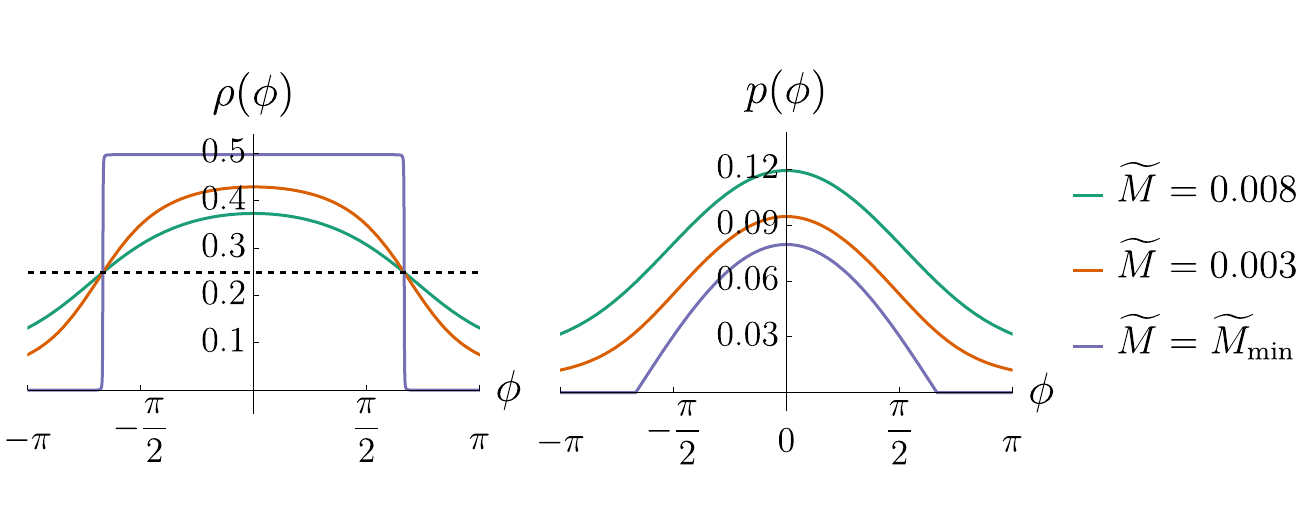}};
    \draw (-8, 2.4) node {\textbf{(c)}};
    \draw (-1.6, 2.4) node {\textbf{(d)}};
\end{tikzpicture}\vspace{-1em}
        \begin{tikzpicture}
        \draw (0, 0) node[inner sep=0] {\includegraphics[width=\textwidth]{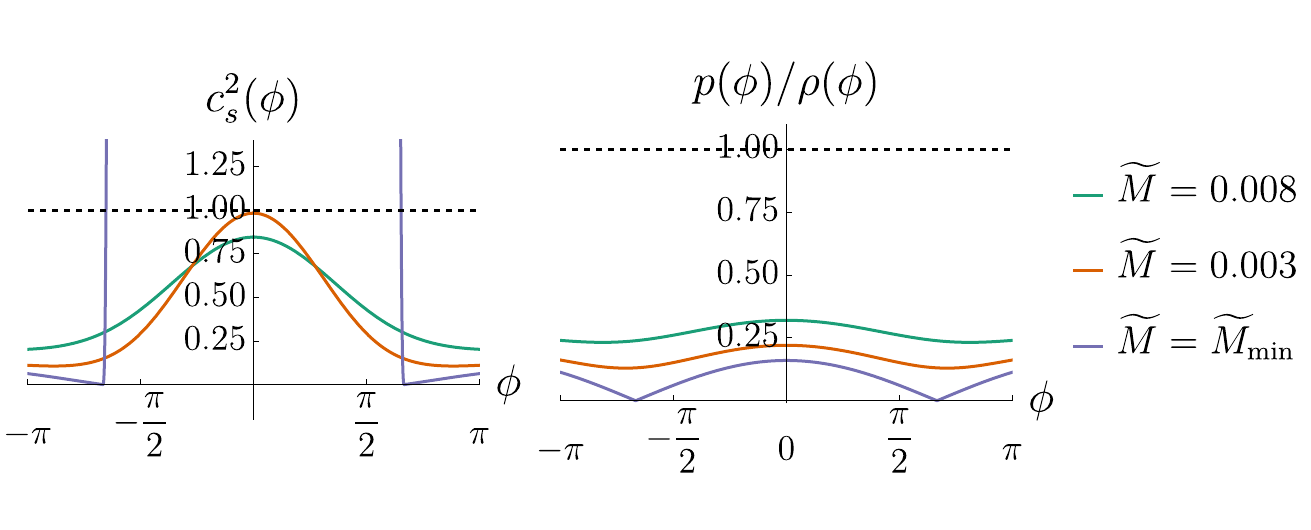}};
        \draw (-8, 2.4) node {\textbf{(e)}};
    \draw (-1.6, 2.4) node {\textbf{(f)}};
        \end{tikzpicture}
    \caption{\justifying Class $\mrm{I}_{\mrm{pulled}}$ (saturated phase), for  $r_0=0.12$, $m=0.75$, $\mc{A}=1.33$, $\alpha=0.25$, and $\ell=1$, where $\wt{M}_{\mrm{min}}=0$ and $\wt{M}_{\mrm{max}}\approx0.013$.
    \textbf{(a)} Polar plot of shell radius $\wt{R}_M$~\eqref{eqn:ADM shell coords} over $\phi\in(-\pi,\pi)$ for varying values of $\wt{M}$; dashed line indicates the string and black dot represents the point particle. \textbf{(b)} Shell angular coordinate $\wt{\Theta}_M$~\eqref{eqn:ADM shell coords} for varying values of $\wt{M}$. \textbf{(c)} Energy density $\rho$ of the shell for varying values of $\wt{M}$. Black dashed line at asymptotic value as $\wt{M}\to\wt{M}_{\mrm{max}}$. \textbf{(d)} Pressure $p$ of the shell for varying values of $\wt{M}$.\textbf{(e)} Squared speed of sound $c_s^2$ of matter in the shell for varying values of $\wt{M}$; dashed line at $c_s=1$. \textbf{(f)} Ratio of shell pressure to shell energy density for varying values of $\wt{M}$; dashed line at $p/\rho=1$.}\label{fig:Class I saturated}
\end{figure}
\begin{figure}[t!]
    \centering
    \begin{tikzpicture}
    \draw (0, 0) node[inner sep=0] {\includegraphics[width=\textwidth]{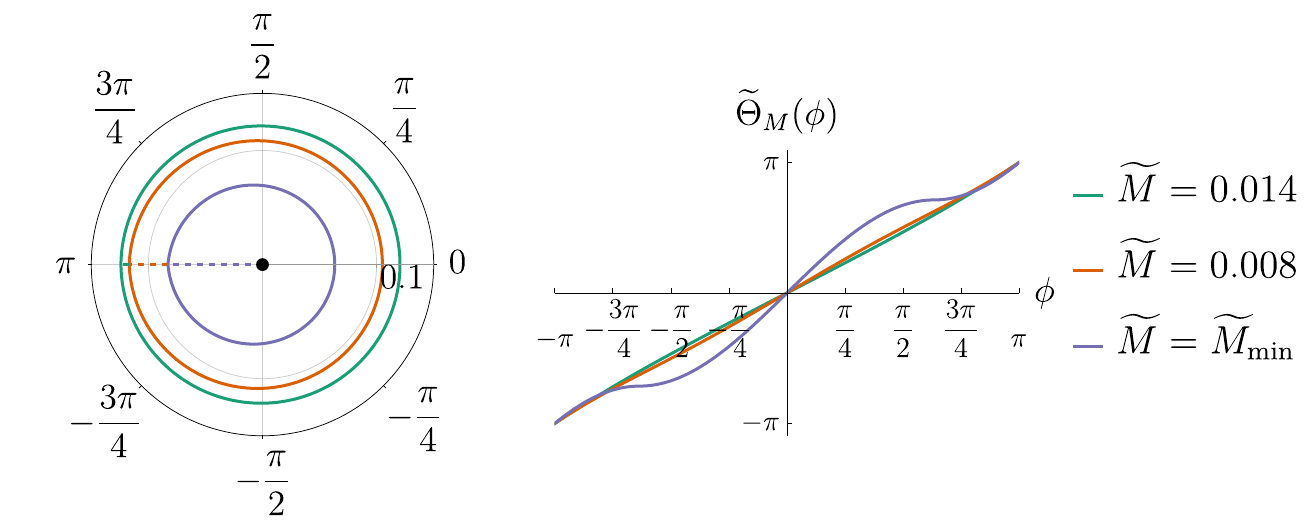}};
     \draw (-8, 2.4) node {\textbf{(a)}};
    \draw (-1.6, 2.4) node {\textbf{(b)}};
\end{tikzpicture}\vspace{-1em}
    \begin{tikzpicture}
    \draw (0, 0) node[inner sep=0] {\includegraphics[width=\textwidth]{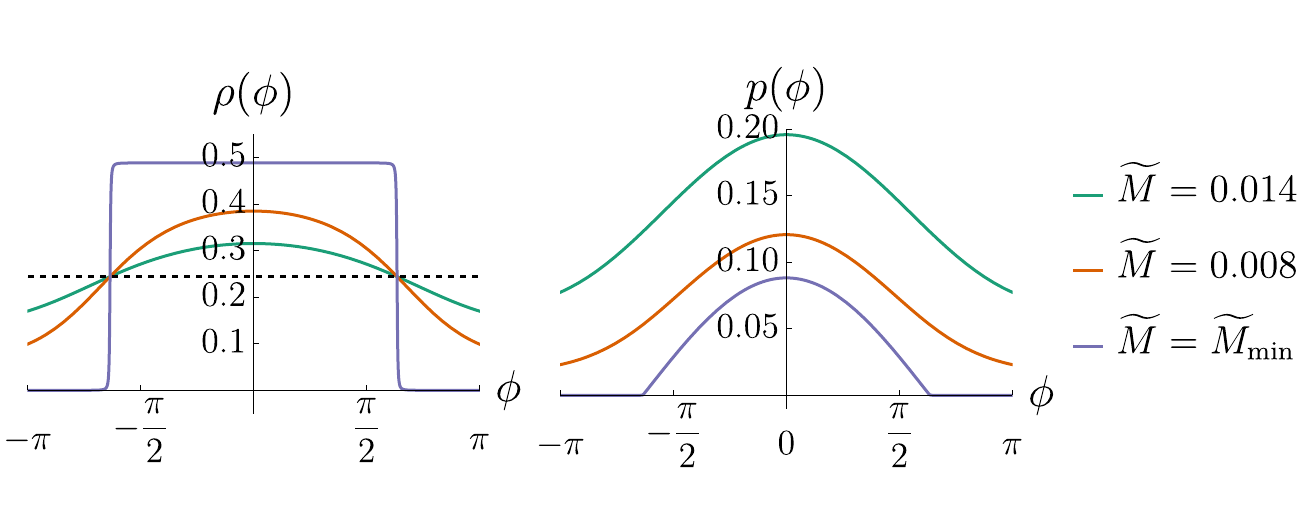}};
    \draw (-8, 2.4) node {\textbf{(c)}};
    \draw (-1.6, 2.4) node {\textbf{(d)}};
\end{tikzpicture}\vspace{-1em}
        \begin{tikzpicture}
        \draw (0, 0) node[inner sep=0] {\includegraphics[width=\textwidth]{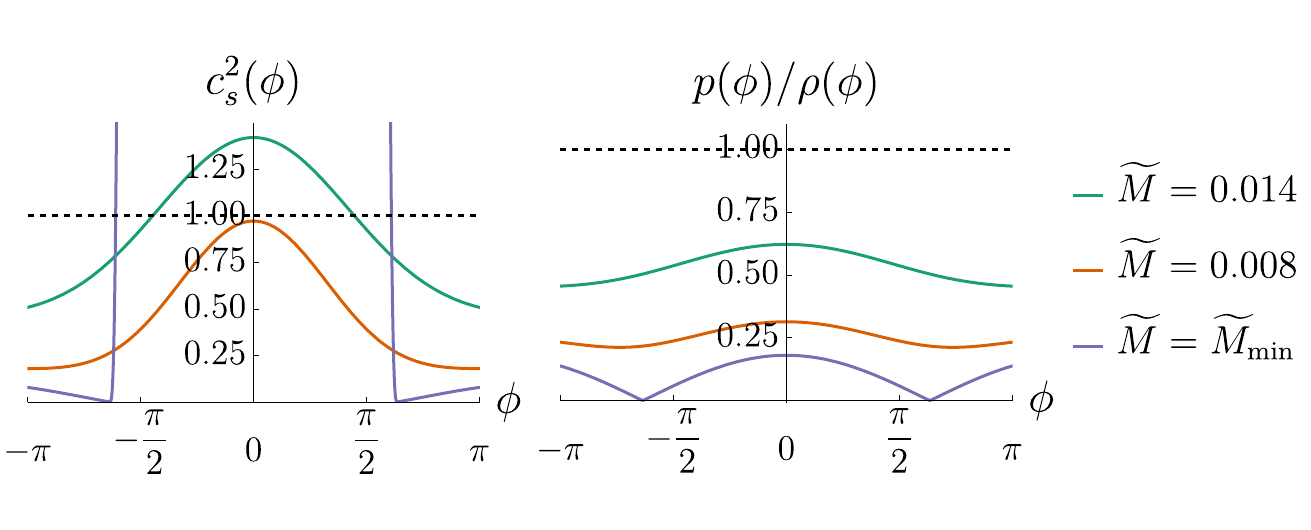}};
       \draw (-8, 2.4) node {\textbf{(e)}};
    \draw (-1.6, 2.4) node {\textbf{(f)}};
        \end{tikzpicture}
    \caption{\justifying Class $\mrm{I}_{\mrm{pulled}}$ (rapid phase)
    for $r_0=0.13$, $m=0.8$, $\mc{A}=1.35$, $\alpha=0.25$, and $\ell=1$, where $r_h\approx1.96$, $\wt{M}_{\mrm{min}}\approx0.001$, and $\wt{M}_{\mrm{max}}\approx0.016$.
    \textbf{(a)} Polar plot of shell radius $\wt{R}_M$~\eqref{eqn:ADM shell coords} over $\phi\in(-\pi,\pi)$ for varying values of $\wt{M}$; dashed line indicates the string and black dot represents the point particle. \textbf{(b)} Shell angular coordinate $\wt{\Theta}_M$~\eqref{eqn:ADM shell coords} for varying values of $\wt{M}$. \textbf{(c)} Energy density $\rho$ of the shell for varying values of $\wt{M}$. Black dashed line at asymptotic value as $\wt{M}\to\wt{M}_{\mrm{max}}$. \textbf{(d)} Pressure $p$ of the shell for varying values of $\wt{M}$. \textbf{(e)} Squared speed of sound $c_s^2$ of matter in the shell for varying values of $\wt{M}$; dashed line at $c_s=1$. \textbf{(f)} Ratio of shell pressure to shell energy density for varying values of $\wt{M}$; dashed line at $p/\rho=1$.}\label{fig:Class I rapid}    
\end{figure}

We consider first the slow acceleration regime, $m^2\mc{A}^2\ell^2<1$.

 In~\autoref{fig:Class I slow} \textbf{(a)} and \textbf{(b)}, we plot the radial coordinate $\wt{R}_M$~\eqref{eqn:ADM shell coords} and angular coordinate $\wt{\Theta}_M$~\eqref{eqn:ADM shell coords} as functions of the shell's intrinsic coordinate $\phi$. For small values of $M$, the shell resembles a cardioid, pinched outwards near the string at $\phi=\pi$ and developing a more pronounced cuspoidal shape at $\phi=0$ as $\wt{M} \to \wt{M}_{\mrm{min}}$. We see from~\eqref{eqn:quad behaviour} that the first derivative exists at $\phi=0$ and hence there is no sharp cusp. The angular coordinate $\Theta_M$ increasingly deviates from linearity as $M\to\Mmin$. As $\wt{M}$ gets larger, the shell becomes less deformed, approaching near-circular teardrop shape as $\wt{M} \to \wt{M}_{\mrm{max}}$, as anticipated by~\eqref{eqn:large M R theta}.
 The dashed line indicates the string, which 
 extends from the point particle (centred at the origin in these coordinates) out to the shell.

In~\autoref{fig:Class I slow}~\textbf{(c)} and \textbf{(d)}, we plot the shell energy density $\rho$ and pressure $p$ as function of the shell's intrinsic coordinate $\phi$. In~\autoref{fig:Class I slow}~\textbf{(e)} and~\textbf{(f)}, we plot the squared speed of sound $c_s^2$ and the ratio of pressure and energy density as functions of $\phi$. When $c_s^2\leq1$, the stress energy is causal and when $p/\rho\leq1$ with $\rho\geq0$, the stress energy satisfies the DEC. 

Both shell density and pressure, as well as speed of sound, are maximised at the cusp, being highly concentrated there as $\wt{M} \to \wt{M}_{\mrm{min}}$, seen in~\autoref{fig:Class I slow}~\textbf{(c)},~\textbf{(d)}, and~\textbf{(e)}, and. As the shell approaches the string, both density and pressure decrease super-exponentially rapidly, but do not vanish. These maxima are consistent with our modelling of the stress energy as a perfect fluid; the shell is being pulled by the string at $\phi=\pm\pi$, so one would expect the fluid to ``pool'' opposite the string, at $\phi=0$. As the ADM mass $\wt{M}/8$ gets larger, the pressure generally increases, but becomes greater on the part of the shell farthest from the string. The stress energy is positive, thereby satisfying the NEC, WEC, and SEC (cf.~\autoref{table:: energy conditions}). In~\autoref{fig:Class I slow}~\textbf{(e)} and~\textbf{(f)}, we see there exists a parameter range in which the stress energy satisfies the DEC and causality ($c_s\leq1$) simultaneously. The stress energy may also satisfy the DEC, whilst violating causality, as in the case $\wt{M}=\wt{M}_{\mrm{min}}$. As $\wt{M}\to\wt{M}_{\mrm{min}}$, the stress energy of the shell is positive but decays faster than exponentially. The behaviour of the stress energy as $\wt{M}\to\wt{M}_{\mrm{max}}$ is given by~\eqref{eqn:large M SE}; the energy density $\rho$ tends to a constant, whereas the pressure still depends on $\phi$. As $\wt{M}$ varies between $\wt{M}_{\mrm{min}}$ and $\wt{M}_{\mrm{max}}$, the speed of sound $c_s^2=\td{p}{\phi}/\td{\rho}{\phi}$ responds non-linearly, diverging both as $\wt{M}\to\wt{M}_{\mrm{min}}$ and $\wt{M}\to\wt{M}_{\mrm{max}}$.

\subsection{Saturated acceleration}

In the saturated acceleration phase, $m^2\mc{A}^2\ell^2=1$, we find that the shell has features similar to the slow acceleration case, the most notable distinction being the absence of a cusp 
as $\wt{M} \to \wt{M}_{\mrm{min}}$. We remark that $\wt{M}_{\mrm{min}}=0$; hence the $\wt{M}=\wt{M}_{\mrm{min}}$ plot depicts a shell with an exterior geometry of Torricelli's trumpet~\eqref{eqn:cylinder}.

We plot the various results in~\autoref{fig:Class I saturated} \textbf{(a)} to \textbf{(f)}.  We see from~\autoref{fig:Class I saturated}
\textbf{(a)} that, in contrast to the slow acceleration phase, the shell remains regular near $\phi=0$ as $\wt{M}\to\wt{M}_{\mrm{min}}$, maintaining a slight teardrop shape, pinched at the string $\phi=\pi$, for all values of $\wt{M}$. We note that the teardrop shape becomes more prominent for larger values of $r_0$ with the consequence $c_s>1$ for all values of $\wt{M}$.
The angular coordinate~\eqref{eqn:ADM shell coords} grows monotonically with $\phi$
(\autoref{fig:Class I saturated} \textbf{(b)}), but with notably reduced non-linear behaviour for small $\wt{M}$. The density is notably larger on the part of the shell that is away from its pinch (\autoref{fig:Class I saturated}~\textbf{(c)}). The pressure,
shown in~\autoref{fig:Class I saturated}~\textbf{(d)}, is always maximal opposite the pinch.  This is again consistent with our modelling of the stress energy as a perfect fluid. Again, the stress energy of the shell is positive and satisfies the NEC, WEC, and SEC for all $\wt{M}$. As seen in~\autoref{fig:Class I saturated}~\textbf{(e)} and~\textbf{(f)}, there exists a parameter range for which causality and the DEC are respected simultaneously. We recall from~\eqref{eqn:large M SE} that $p$ diverges as $\wt{M}\to\wt{M}_{\mrm{max}}$, hence for sufficiently large $\wt{M}$ both causality and the DEC will be violated. We see there exist also parameters for which the DEC is respected, whilst causality is violated. As $\wt{M}\to\wt{M}_{\mrm{min}}$, the stress energy of the shell is positive but decays faster than exponentially. The behaviour of the stress energy as $\wt{M}\to\wt{M}_{\mrm{max}}$ is given by~\eqref{eqn:large M SE}; the energy density $\rho$ tends to a constant, whereas the pressure still depends on $\phi$, and $c_s^2=\frac{\dd p}{\dd\phi}/\frac{\dd\rho}{\dd\phi}$ diverges.

\subsection{Rapid acceleration}
\noindent The rapid acceleration phase,  $m^2\mc{A}^2\ell^2>1$, has features quite similar to the saturated phase. Results are shown in~\autoref{fig:Class I rapid} \textbf{(a)} to \textbf{(f)}.

The teardrop shape of the shell is still present, and the density and pressure are still maximised opposite the pinch, with a high-density region of the shell extending well past half its circumference, shown
in~\autoref{fig:Class I rapid} \textbf{(c)}. Again, we note that the teardrop shape becomes more prominent for larger values of $r_0$ with the consequence $c_s>1$ for all values of $\wt{M}$. By positivity of the stress energy, the NEC, WEC, and SEC are still respected, as well as causality and the DEC for certain choices of the parameters. As $\wt{M}\to\wt{M}_{\mrm{min}}$, the stress energy of the shell is positive but decays faster than exponentially. The behaviour of the stress energy as $\wt{M}\to\wt{M}_{\mrm{max}}$ is given by~\eqref{eqn:large M SE}; the energy density $\rho$ tends to a constant, whereas the pressure still depends on $\phi$, and $c_s^2=\td{p}{\phi}/\td{\rho}{\phi}$ diverges.

\begin{figure}[t]
    \centering
    \begin{subfigure}[b]{0.4\textwidth}
        \centering\captionsetup{labelfont=bf}
        \begin{tikzpicture}
        \draw (0, 0) node[inner sep=0] {\includegraphics[width=\textwidth]{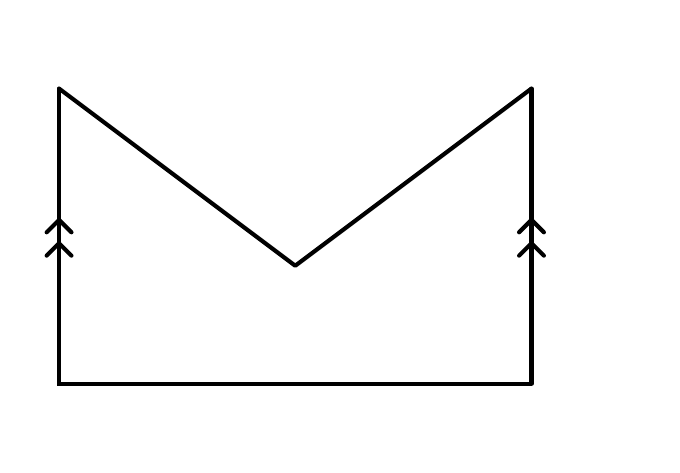}};
        \node[rotate=38] at (0.3,0.5) {\large $r=r_{\mrm{conf}}$};
        \draw (2.6, -1.3) node {\large$r=0$};
        \draw (-2.4, -1.85) node {\large$\phi=-\pi$};
        \draw (1.8, -1.85) node {\large$\phi=\pi$};
        \end{tikzpicture}
        \caption{\\}
    \end{subfigure}\hfill
    \begin{subfigure}[b]{0.4\textwidth}
        \centering\captionsetup{labelfont=bf}
        \begin{tikzpicture}
        \draw (0, 0) node[inner sep=0] {\includegraphics[width=\textwidth]{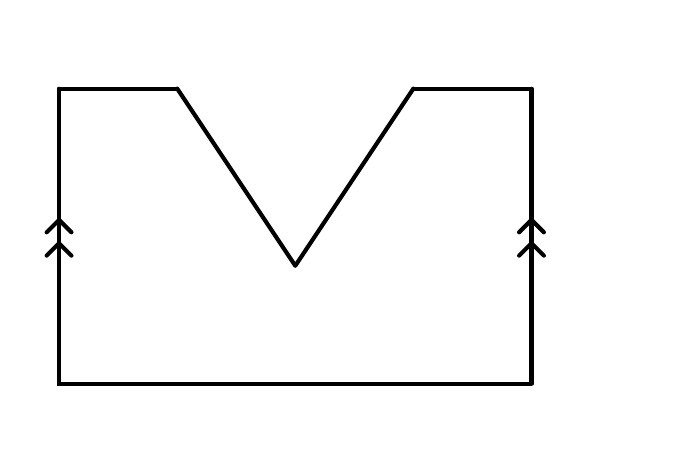}};
        \draw (2.67, 1.4) node {\large$r=\infty$};
        \node[rotate=58] at (0.05,0.8) {\large $r=r_{\mrm{conf}}$};
        \draw (2.6, -1.3) node {\large$r=0$};
        \draw (-2.4, -1.85) node {\large$\phi=-\pi$};
        \draw (1.8, -1.85) node {\large$\phi=\pi$};
        \end{tikzpicture}
        \caption{\\}
    \end{subfigure}\vspace{1em}
    \begin{subfigure}[b]{0.4\textwidth}
        \centering\captionsetup{labelfont=bf}
        \begin{tikzpicture}
        \draw (0, 0) node[inner sep=0] {\includegraphics[width=\textwidth]{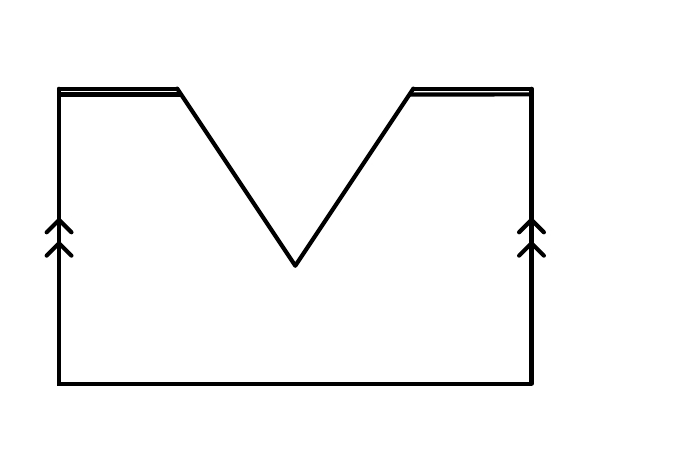}};
        \draw (2.67, 1.4) node {\large$r=\infty$};
        \node[rotate=58] at (0.05,0.8) {\large $r=r_{\mrm{conf}}$};
        \draw (2.6, -1.3) node {\large$r=0$};
        \draw (-2.4, -1.85) node {\large$\phi=-\pi$};
        \draw (1.8, -1.85) node {\large$\phi=\pi$};
        \end{tikzpicture}
        \caption{\\}
    \end{subfigure}\hfill
    \begin{subfigure}[b]{0.4\textwidth}
        \centering\captionsetup{labelfont=bf}
        \begin{tikzpicture}
        \draw (0, 0) node[inner sep=0] {\includegraphics[width=\textwidth]{figures/SpaceDiagrams/IpushDouble.pdf}};
        \draw (2.67, 1.4) node {\large$r=r_h$};
        \node[rotate=58] at (0.05,0.8) {\large $r=r_{\mrm{conf}}$};
        \draw (2.6, -1.3) node {\large$r=0$};
        \draw (-2.4, -1.85) node {\large$\phi=-\pi$};
        \draw (1.8, -1.85) node {\large$\phi=\pi$};
        \end{tikzpicture}
        \caption{\\}
    \end{subfigure}
    \caption{\justifying Constant time slice of the Class $\mrm{I}_{\mrm{pushed}}$ solution describing a particle being pushed by a strut. \textbf{(a)} Slow and saturated acceleration regimes with $m\leq\tfrac12$ and rapid acceleration regime with $m\mc{A}\ell\sin(m\pi)\leq1$. \textbf{(b)} Slow acceleration with $m>\tfrac12$. \textbf{(c)} Saturated acceleration with $m>\tfrac12$. \textbf{(d)} Rapid acceleration with $m\mc{A}\ell\sin(m\pi)>1$. Double lines in \textbf{(c)} and \textbf{(d)} represent a Killing horizon.}
    \label{fig:Class I Pushed ranges}
\end{figure}
\section{Class $\text{I}_{\text{pushed}}$: Particle pushed by a strut}\label{sec:: class I strut}
\label{Sec4}
We consider now the Class $\mrm{I}_{\mrm{pushed}}$ metric solution describing a particle pushed by a strut. The metric of this spacetime is found by mapping $\mc{A}\to-\mc{A}$ in~\eqref{eqn:: Class I full metric},
\begin{subequations}\label{eqn:I push metric}
    \begin{align}
        \dd s^2&~=~\frac{1}{\Omega^2(r,\phi)}\lr{-\frac{f(r)}{\alpha^2}\dd \sigma^2+\frac{\dd r^2}{f(r)}+r^2\dd\phi}\,,\\
        f(r)&~=~\frac{r^2}{\ell^2}+m^2\lr{1-\mc{A}^2r^2}\,,\\
        \Omega(r,\phi)&~=~1-\mc{A}r\cos(m\phi)\,.
    \end{align}
\end{subequations}

The tension of the strut is given by
\begin{equation}
    \uptau_\pi~=~-\frac1{4\pi}m\mc{A}\sin(m\pi)\,.
\end{equation}

As in the Class $\text{I}_{\mrm{pulled}}$ solution in~\Sref{sec:: class I string}, the Class $\text{I}_{\mrm{pushed}}$ spacetime also admits slow, saturated, and rapid acceleration regimes depending on whether $m^2\mc{A}^2\ell^2$ is greater than, less than, or equal to unity. For $m^2\mc{A}^2\ell^2\geq 1$, there is a Killing horizon at
\begin{equation}
    r_h~=~\frac{m\ell}{\sqrt{m^2\mc{A}^2\ell^2-1}}\,.
\end{equation}
The geometry of the three phases of acceleration is depicted in~\autoref{fig:Class I Pushed ranges}.

We interpret the Class $\text{I}_{\mrm{pushed}}$ geometry as a point particle owing to the angular deficit in the metric~\eqref{eqn:I push metric}. This angular deficit has the same associated mass as the Class $\text{I}_{\mrm{pulled}}$ solution, given by~\eqref{eqn:mad}.

For all phases of acceleration, the minimum value of the exterior mass parameter is given by
\begin{subequations}\label{eqn:Mmin I push}
    \begin{align}
        \Mmin&~=~\frac{f(r_0)}{\alpha^2}\frac{\lr{m^2\mc{A}^2\ell^2\sin^2(m\phi_*)-1}}{\lr{1-\mc{A}r_0\cos(m\phi_*)}^2}\,,\\
        \phi_*&~=~\min\lrc{\frac{1}{m}\arccos\lr{\frac{(m^2\mc{A}^2\ell^2-1)\mc{A}r_0}{m^2\mc{A}^2\ell^2}},\pi}\,.
    \end{align}
\end{subequations}
In the case $\phi_*<\pi$, this reduces to
\begin{equation}
    \Mmin~=~-\frac{m^2}{\alpha^2}\lr{1-m^2\mc{A}^2\ell^2}\,.
\end{equation}


\subsection{Slow acceleration}
We consider first the slow acceleration regime, $m^2\mc{A}^2\ell^2<1$. In this regime,~\eqref{eqn:Mmin I push} implies that $\Mmin<0$.

\begin{figure}[t!]
    \centering
    \begin{tikzpicture}
    \draw (0, 0) node[inner sep=0] {\includegraphics[width=\textwidth]{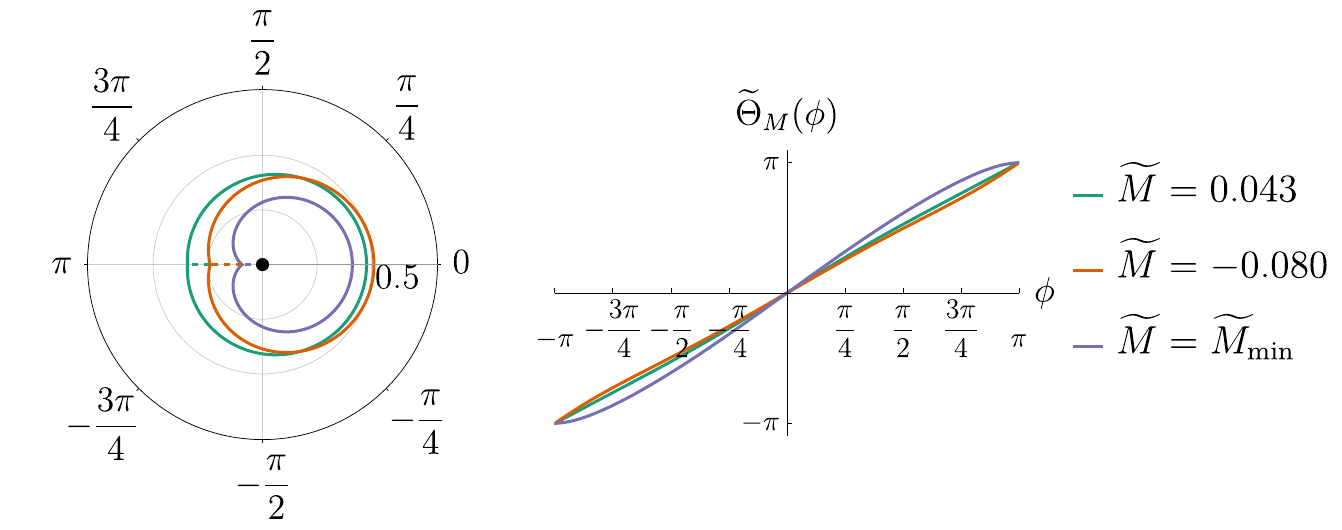}};
     \draw (-8, 2.4) node {\textbf{(a)}};
    \draw (-1.6, 2.4) node {\textbf{(b)}};
\end{tikzpicture}\vspace{-1em}
    \begin{tikzpicture}
    \draw (0, 0) node[inner sep=0] {\includegraphics[width=\textwidth]{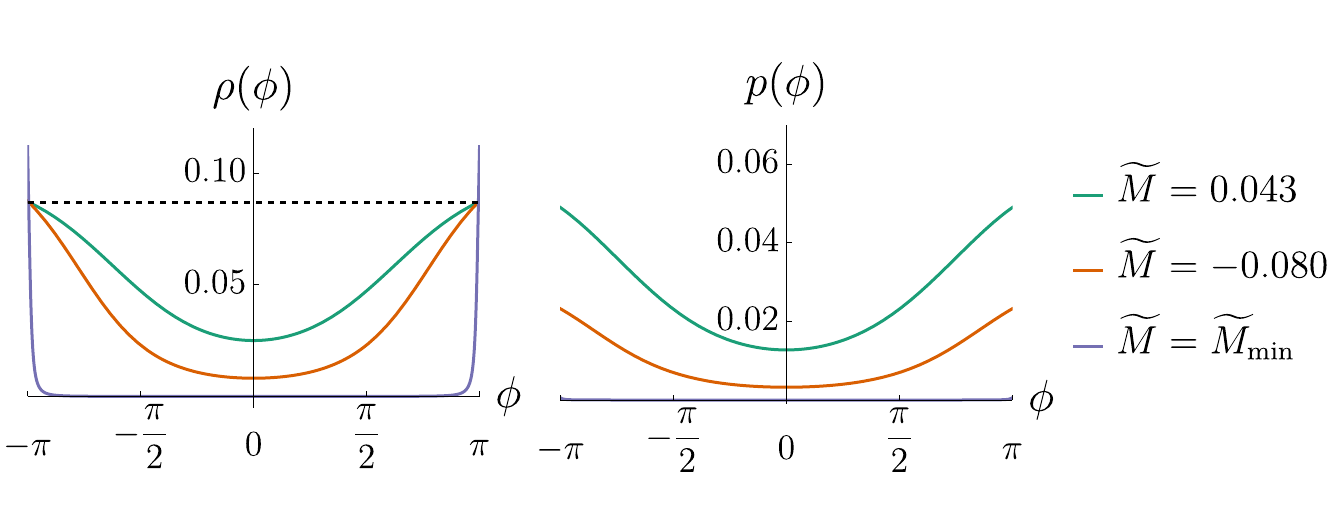}};
    \draw (-8, 2.4) node {\textbf{(c)}};
    \draw (-1.6, 2.4) node {\textbf{(d)}};
\end{tikzpicture}\vspace{-1em}
        \begin{tikzpicture}
        \draw (0, 0) node[inner sep=0] {\includegraphics[width=\textwidth]{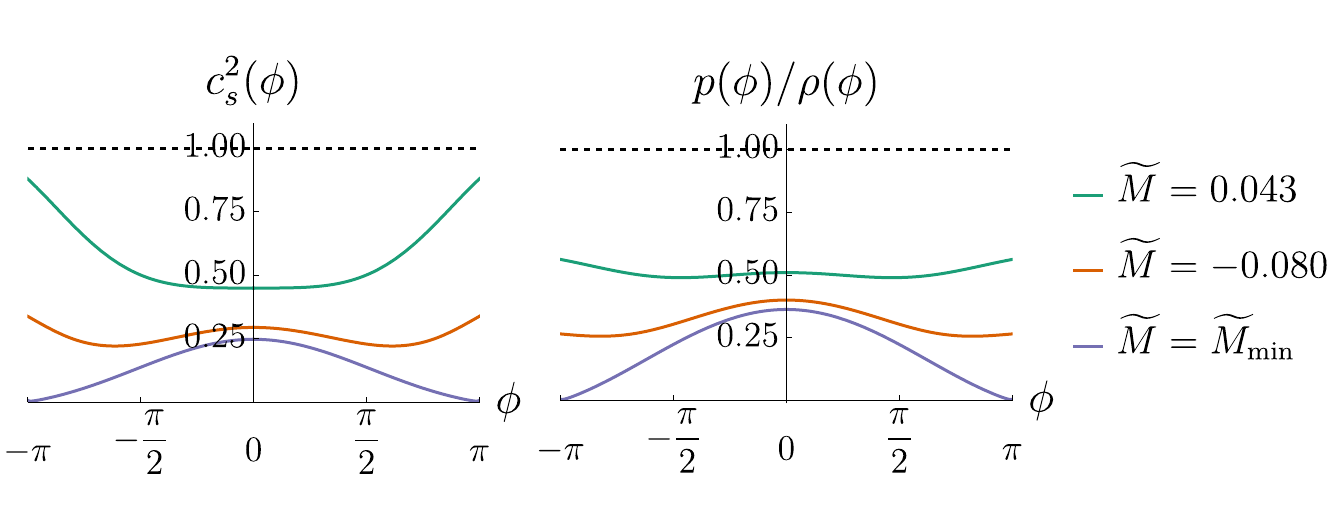}};
        \draw (-8, 2.4) node {\textbf{(e)}};
    \draw (-1.6, 2.4) node {\textbf{(f)}};
        \end{tikzpicture}
    \caption{\justifying Class $\mrm{I}_{\mrm{pushed}}$ (slow phase) for $r_0=0.4$, $m=0.8$, $\mc{A}=0.6$, $\alpha=0.25$, and $\ell=1$, where $\wt{M}_{\mrm{min}}\approx -0.100$ and $\wt{M}_{\mrm{max}}\approx0.188$. \textbf{(a)} Polar plot of shell radius $\wt{R}_M$~\eqref{eqn:ADM shell coords} over $\phi\in(-\pi,\pi)$ for varying values of $\wt{M}$; dashed line indicates the strut and black dot represents the point particle. \textbf{(b)} Shell angular coordinate $\wt{\Theta}_M$~\eqref{eqn:ADM shell coords} for varying values of $\wt{M}$. \textbf{(c)} Energy density $\rho$ of the shell for varying values of $\wt{M}$. Black dashed line at asymptotic value as $\wt{M}\to\wt{M}_{\mrm{max}}$. \textbf{(d)} Pressure $p$ of the shell for varying values of $\wt{M}$. \textbf{(e)} Squared speed of sound $c_s^2$ of matter in the shell for varying values of $\wt{M}$; dashed line at $c_s=1$. \textbf{(f)} Ratio of shell pressure to shell energy density for varying values of $\wt{M}$; dashed line at $p/\rho=1$.}\label{fig:Class I slow strut}
\end{figure}

In~\autoref{fig:Class I slow strut} \textbf{(a)} and \textbf{(b)}, we plot the radial coordinate $\wt{R}_M$~\eqref{eqn:: Radial matching} and angular coordinate $\wt{\Theta}_M$~\eqref{eqn:: Theta matching} as functions of the shell's intrinsic coordinate $\phi$. For small values of $M$, the shell again resembles a cardioid, but with a cusp now at the strut  $\phi=\pi$. In contrast to the cuspoidal shape in the Class $\text{I}_{\mrm{pulled}}$ solution, this is a true cusp in the sense that the first derivatives at $\phi=\pm\pi$ do not agree. The cusp becomes more pronounced as $M\to\Mmin$. The angular coordinate $\wt{\Theta}_M$ has strongest deviations from linearity as $M\to\Mmin$. As the ADM mass $\wt{M}/8$ gets larger, the shell becomes less deformed and approaches a circular shape as $\wt{M}\to\wt{M}_{\mrm{max}}$. The dashed line indicates the strut extending from the point particle out to the shell.

In~\autoref{fig:Class I slow strut}~\textbf{(c)} and \textbf{(d)}, we plot the shell energy density $\rho$ and pressure $p$ as functions of the shell's intrinsic coordinate $\phi$ and in~\autoref{fig:Class I slow strut}~\textbf{(e)} and \textbf{(f)}, we plot the squared speed of sound $c_s^2$ and the ratio of pressure and energy density as functions of the shell's intrinsic coordinate $\phi$. When $c_s^2\leq1$, the stress energy is causal and when $p/\rho\leq1$ with $\rho\geq0$, the stress energy satisfies the DEC.

We find that both the shell density and pressure, as well as the speed of sound, are maximised at the strut at $\phi=\pi$ pushing the point particle, being extremely concentrated there as $\wt{M}\to\wt{M}_{\mrm{min}}$, shown in~\autoref{fig:Class I slow strut} \textbf{(c)},~\textbf{(d)}, and~\textbf{(e)}. 
As before,
these maxima are consistent with our modelling of the stress energy as a perfect fluid; since the shell is being pushed by the strut at $\phi=\pm\pi$, we expect the perfect fluid to ``pool'' at the location of strut at $\phi=\pm\pi$ As $\wt{M}\to\wt{M}_{\mrm{min}}$, the energy density of the shell is almost entirely localised where the strut meets the shell. The pressure of the shell increases as the ADM mass $\wt{M}/8$ get larger. This shell also satisfies the NEC, WEC, and SEC. The stress energy may either satisfy or violate causality and the DEC, depending on the choice of parameters (recalling that both $c_s^2$ and $p/\rho$ diverge as $\wt{M}\to\wt{M}_{\mrm{max}}$). As $\wt{M}\to\wt{M}_{\mrm{min}}$, the stress energy of the shell is positive but decays faster than exponentially. The behaviour of the stress energy as $\wt{M}\to\wt{M}_{\mrm{max}}$ is given by~\eqref{eqn:large M SE}; the energy density $\rho$ tends to a constant, whereas the pressure still depends on $\phi$.

\begin{figure}[t!]
    \centering
    \begin{tikzpicture}
    \draw (0, 0) node[inner sep=0] {\includegraphics[width=\textwidth]{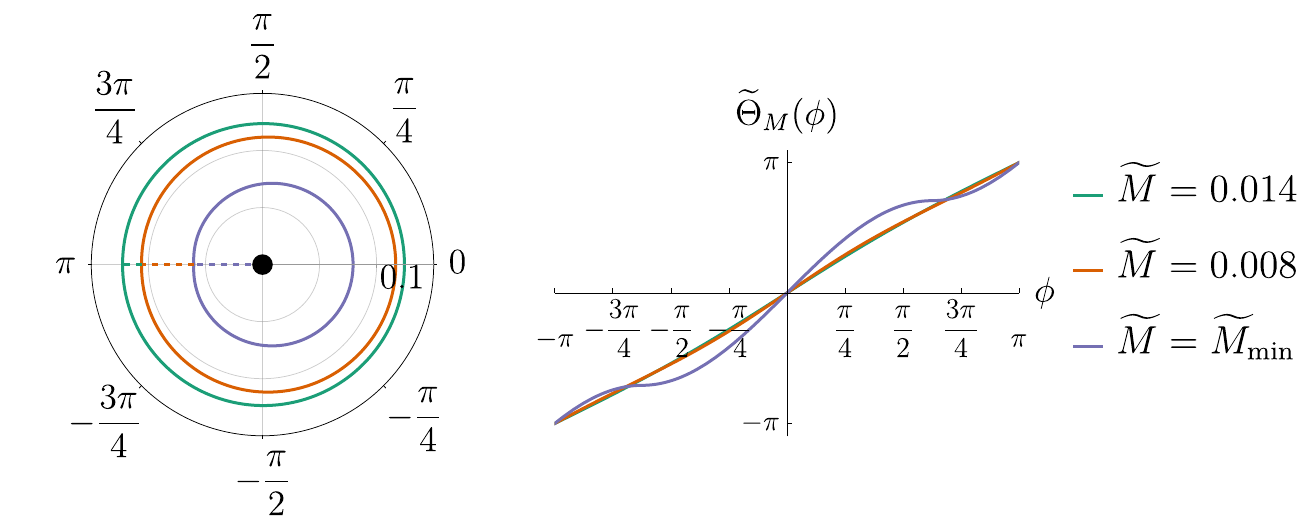}};
     \draw (-8, 2.4) node {\textbf{(a)}};
    \draw (-1.6, 2.4) node {\textbf{(b)}};
\end{tikzpicture}\vspace{-1em}
\begin{tikzpicture}
    \draw (0, 0) node[inner sep=0] {\includegraphics[width=\textwidth]{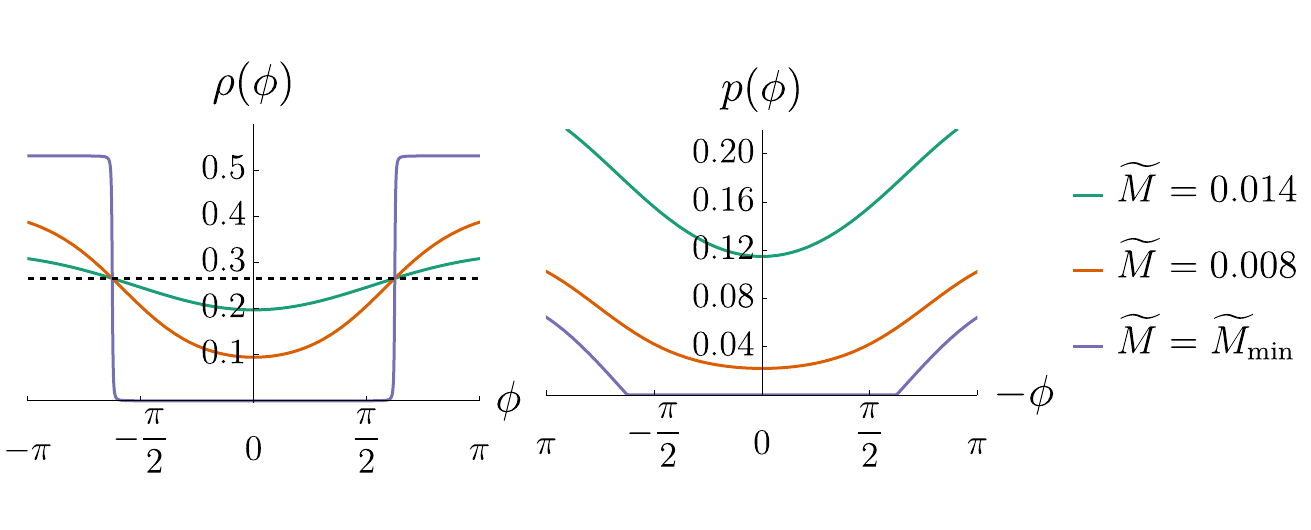}};
    \draw (-8, 2.4) node {\textbf{(c)}};
    \draw (-1.6, 2.4) node {\textbf{(d)}};
\end{tikzpicture}\vspace{-1em}
        \begin{tikzpicture}
        \draw (0, 0) node[inner sep=0] {\includegraphics[width=\textwidth]{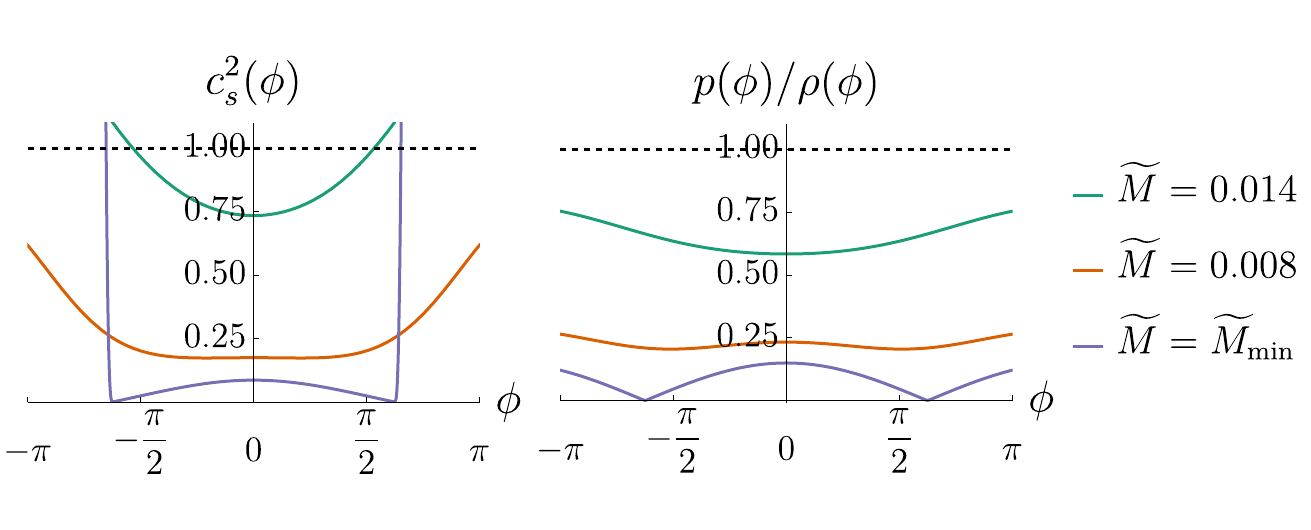}};
        \draw (-8, 2.4) node {\textbf{(e)}};
    \draw (-1.6, 2.4) node {\textbf{(f)}};
        \end{tikzpicture}
    \caption{\justifying Class $\mrm{I}_{\mrm{pushed}}$ (saturated phase) for $r_0=0.12$, $m=0.8$, $\mc{A}=1.25$, $\alpha=0.25$, and $\ell=1$, where $\wt{M}_{\mrm{min}}=0$ and $\wt{M}_{\mrm{max}}\approx0.016$. \textbf{(a)} Polar plot of shell radius $\wt{R}_M$~\eqref{eqn:ADM shell coords} over $\phi\in(-\pi,\pi)$ for varying values of $\wt{M}$; dashed line indicates the strut and black dot represents the point particle. \textbf{(b)} Shell angular coordinate $\wt{\Theta}_M$~\eqref{eqn:ADM shell coords} for varying values of $\wt{M}$. \textbf{(c)} Energy density $\rho$ of the shell for varying values of $\wt{M}$. Black dashed line at asymptotic value as $\wt{M}\to\wt{M}_{\mrm{max}}$. \textbf{(d)} Pressure $p$ of the shell for varying values of $\wt{M}$. \textbf{(e)} Squared speed of sound $c_s^2$ of matter in the shell for varying values of $\wt{M}$; dashed line at $c_s=1$. \textbf{(f)} Ratio of shell pressure to shell energy density for varying values of $\wt{M}$; dashed line at $p/\rho=1$. }\label{fig:Class I saturated strut}
\end{figure}
\begin{figure}[t!]
    \centering
    \begin{tikzpicture}
    \draw (0, 0) node[inner sep=0] {\includegraphics[width=\textwidth]{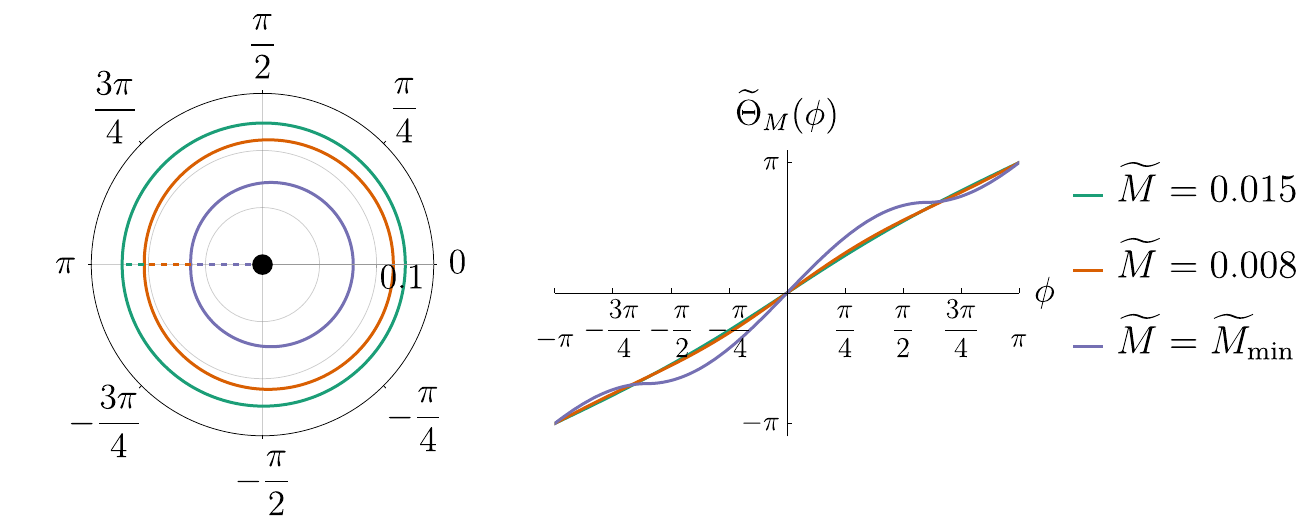}};
     \draw (-8, 2.4) node {\textbf{(a)}};
    \draw (-1.6, 2.4) node {\textbf{(b)}};
\end{tikzpicture}\vspace{-1em}
    \begin{tikzpicture}
    \draw (0, 0) node[inner sep=0] {\includegraphics[width=\textwidth]{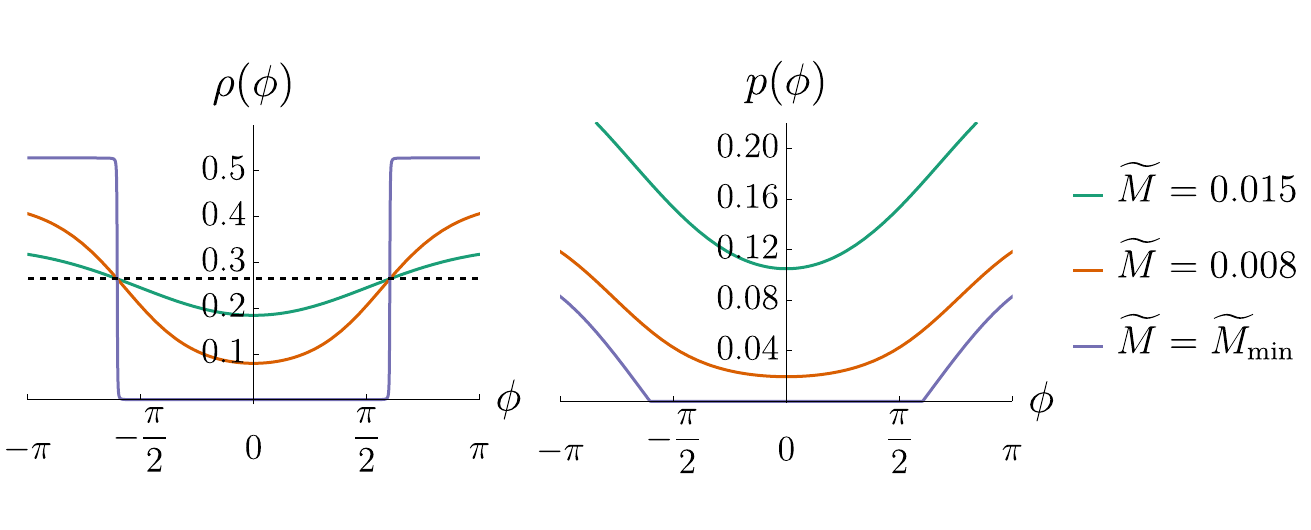}};
    \draw (-8, 2.4) node {\textbf{(c)}};
    \draw (-1.6, 2.4) node {\textbf{(d)}};
\end{tikzpicture}\vspace{-1em}
        \begin{tikzpicture}
        \draw (0, 0) node[inner sep=0] {\includegraphics[width=\textwidth]{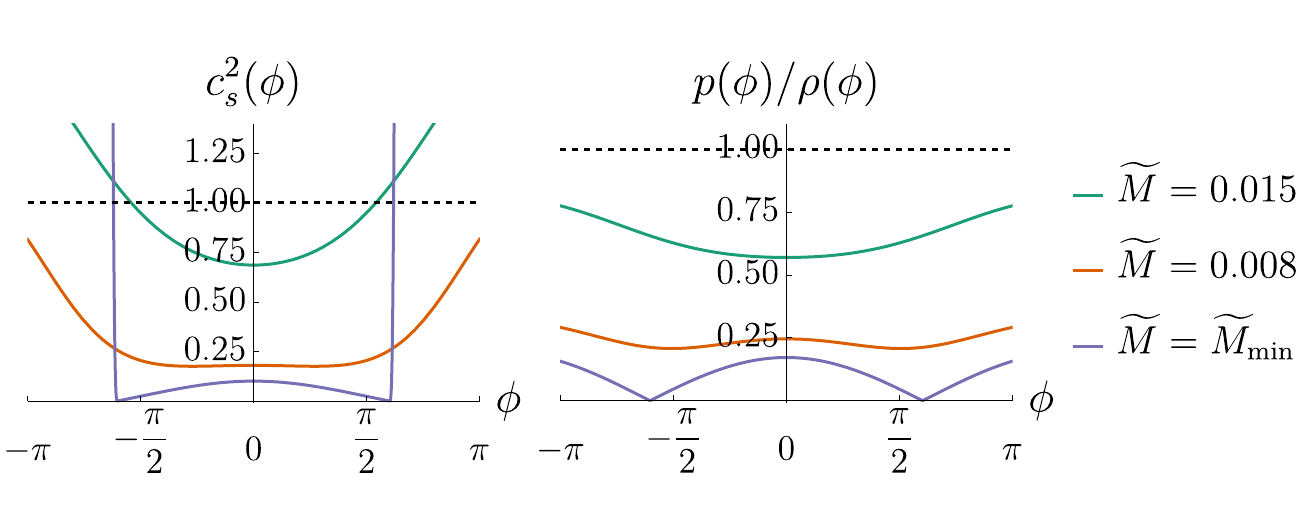}};
        \draw (-8, 2.4) node {\textbf{(e)}};
    \draw (-1.6, 2.4) node {\textbf{(f)}};
        \end{tikzpicture}
    \caption{\justifying Class $\mrm{I}_{\mrm{pushed}}$ (rapid phase) for $r_0=0.12$, $m=0.8$, $\mc{A}=1.5$, $\alpha=0.25$, and $\ell=1$, where $r_h\approx0.019$, $\wt{M}_{\mrm{min}}\approx0.001$, and $\wt{M}_{\mrm{max}}\approx0.016$. \textbf{(a)} Polar plot of shell radius $\wt{R}_M$~\eqref{eqn:ADM shell coords} over $\phi\in(-\pi,\pi)$ for varying values of $\wt{M}$; dashed line indicates the strut and black dot represents the point particle. \textbf{(b)} Shell angular coordinate $\wt{\Theta}_M$~\eqref{eqn:ADM shell coords} for varying values of $\wt{M}$. \textbf{(c)} Energy density $\rho$ of the shell for varying values of $\wt{M}$. Black dashed line at asymptotic value as $\wt{M}\to\wt{M}_{\mrm{max}}$. \textbf{(d)} Pressure $p$ of the shell for varying values of $\wt{M}$. \textbf{(e)} Squared speed of sound $c_s^2$ of matter in the shell for varying values of $\wt{M}$; dashed line at $c_s=1$.  \textbf{(f)} Ratio of shell pressure to shell energy density for varying values of $\wt{M}$; dashed line at $p/\rho=1$.}\label{fig:Class I rapid strut}
\end{figure}

\subsection{Saturated acceleration}
In the saturated acceleration phase, $m^2\mc{A}^2\ell^2=1$, we find that the shell has features similar to the slow acceleration case. A noticeable difference is the decreased prominence of the cusp as the shell meets the strut. We remark that $\wt{M}_{\mrm{min}}=0$; hence the $\wt{M}=\wt{M}_{\mrm{min}}$ plot depicts a shell with an exterior geometry of Torricelli's trumpet~\eqref{eqn:cylinder}.

We plot the shell coordinates, stress energy, speed of sound, and ratio of pressure to energy density in~\autoref{fig:Class I saturated strut} \textbf{(a)} to \textbf{(f)}. We see from~\autoref{fig:Class I saturated strut} that all shells are regular near $\phi=0$. The angular coordinate $\wt{\Theta}_M$~\eqref{eqn:ADM shell coords} grows monotonically with $\phi$ (\autoref{fig:Class I saturated strut} \textbf{(b)}), but with noticeable non-linear behaviour in comparison to the slow acceleration regime. This non-linearity is emphasised for small $\wt{M}$. As shown in~\autoref{fig:Class I saturated strut} \textbf{(c)}, \textbf{(d)}, and \textbf{(e)}, the energy density and pressure are concentrated on the part of the shell closest to the strut, again consistent with our modelling of the stress energy as a perfect fluid. As $\wt{M}\to\wt{M}_{\mrm{min}}$, the energy density behaves like a bump function, with negligible energy density opposite the cusp. Again, the stress energy of the shell satisfies the NEC, WEC, and SEC, and may either respect or violate causality and the DEC. As $\wt{M}\to\wt{M}_{\mrm{min}}$, the stress energy is positive but decays faster than exponentially. The behaviour of the stress energy as $\wt{M}\to\wt{M}_{\mrm{max}}$ is given by~\eqref{eqn:large M SE}; the energy density tends to a constant, whereas the pressure still depends on $\phi$. As $\wt{M}$ varies, the speed of sound responds non-linearly, diverging everywhere as $\wt{M}\to\wt{M}_{\mrm{max}}$ and in the region around the strut at $\phi=\pm\pi$, where the energy density is approximately constant, as $\wt{M}\to\wt{M}_{\mrm{min}}$.

\subsection{Rapid acceleration}

As in Class $\text{I}_{\mrm{pulled}}$, the rapid acceleration phase, $m^2\mc{A}^2\ell^2>1$, has features quite similar to the saturated phase. We show results in~\autoref{fig:Class I rapid strut} \textbf{(a)} to \textbf{(f)}.

In this phase of acceleration, the shells have a more pronounced elliptical shape for larger values of $r_0$, becoming more prominent as $\wt{M}\to\wt{M}_{\mrm{min}}$, shown in~\autoref{fig:Class I rapid strut} \textbf{(a)}. The angular coordinate is highly nonlinear for all considered masses, owing to the elliptical shape (\autoref{fig:Class I rapid strut} \textbf{(b)}). The high-energy density region of the shell extends almost to the arc from $\phi=-\pi/2$ to $\phi=\pi/2$ (\autoref{fig:Class I rapid strut} \textbf{(c)}). The NEC, WEC, and SEC are also still respected, and causality and the DEC may be either violated or respected. 
As $\wt{M}\to\wt{M}_{\mrm{min}}$, the stress energy of the shell is positive but decays faster than exponentially. The behaviour of the stress energy as $\wt{M}\to\wt{M}_{\mrm{max}}$ is given by~\eqref{eqn:large M SE}; the energy density tends to a constant, whereas the pressure still depends on $\phi$, and the speed of sound $c_s^2$ diverges.

\subsection{Class $\text{I}_{\text{C}}$}
We consider now the Class $\text{I}_{\text{C}}$ solution, describing a black hole being pulled by a string, first described in~\cite{Accin3D}. This solution is a subregion within the parameter space of the rapid acceleration phase of the Class $\text{I}_{\mrm{pushed}}$ solution.

A necessary requirement for this solution is $0<m<1/2$. Combined with the rapid acceleration phase, we require also $\mc{A}\ell>2$. Because of this, the Class $\text{I}_{\mrm{C}}$ black hole solution is disconnected from the non-accelerating BTZ solution given by the limit $\mc{A}\to0$.

\noindent The geometry of the Class $\text{I}_{\text{C}}$ spacetime is described by the metric
\begin{subequations}\label{eqn:: Class Ic metric}
\begin{align}
    \dd s^2_-&~=~\frac{1}{\Omega^2(r,\phi)}\lr{-\frac{f(r)}{\alpha^2}\dd \sigma^2+\frac{\dd r^2}{f(r)}+r^2\dd\phi^2}\,,\\
    f(r)&~=~\frac{r^2}{\ell^2}+m^2(1-\mc{A}^2r^2)\,,\\
    \Omega(r,\phi)&~=~\mc{A}r\cos(m\phi)-1\,.
\end{align}    
\end{subequations}Though the metric~\eqref{eqn:: Class Ic metric} resembles the Class $\text{I}_{\text{pushed}}$ metric~\eqref{eqn:I push metric}, it describes the geometry of a black hole being pulled by a string, whose tension  
\begin{equation}  \uptau_\pi~=~\frac1{4\pi}m\mc{A}\sin(m\pi)
\end{equation}
is positive~\cite{Accin3D}.

In the rapid acceleration regime, there is a Killing horizon at 
\begin{equation}\label{eqn:KHorizon}
    r_h~=~\frac{m\ell}{\sqrt{m^2\mc{A}^2\ell^2-1}}\,,
\end{equation}and the conformal boundary is found at
\begin{equation}\label{eqn:Ic rc}
    \rc~=~\frac{1}{\mc{A}\cos(m\phi)}\,.
\end{equation}

\begin{figure}[t!]
    \centering
    \begin{tikzpicture}
        \draw (0, 0) node[inner sep=0] {\includegraphics[width=0.4\textwidth]{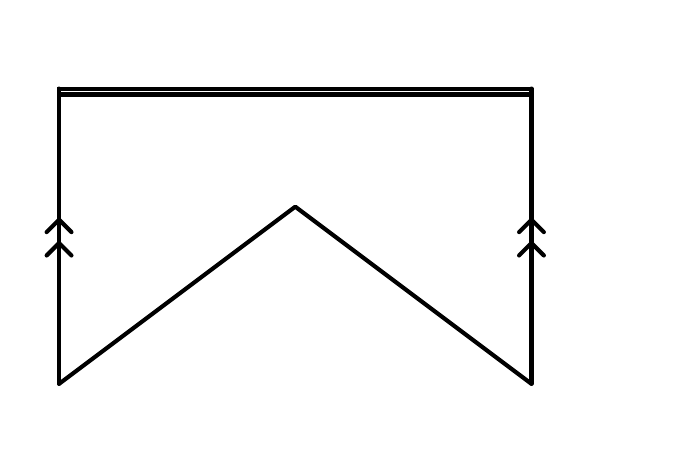}};
        \node[rotate=38] at (-1.25,-0.65) {\large $r=r_{\mrm{conf}}$};
        \draw (2.7, 1.3) node {\large$r=r_h$};
        \draw (-2.4, -1.85) node {\large$\phi=-\pi$};
        \draw (1.8, -1.85) node {\large$\phi=\pi$};
        \end{tikzpicture}
    \caption{\justifying Constant time slice of the Class $\mrm{I}_{\mrm{C}}$ solution. Double line represents a Killing horizon.}
    \label{fig:Class Ic ranges}
\end{figure}

We consider values of the radial coordinate restricted between the conformal boundary~\eqref{eqn:KHorizon} and the horizon~\eqref{eqn:Ic rc}, $\rc<r<r_h$. We remark that $r<r_h$   ensures the positivity of the metric function $f(r)$, whereas $r>\rc$ ensures the positivity of the conformal factor $\Omega(r,\phi)$. The geometry of this solution is depicted in~\autoref{fig:Class Ic ranges}.

A quantity of interest is the circumference of a closed loop of constant radial coordinate, given by
\begin{equation}\label{eqn:circum}
    \int_{-\pi}^\pi\dd\phi\, \sqrt{g_{\phi\phi}}~=~\int_{-\pi}^\pi\dd\phi\,\frac{r}{\mc{A}r\cos(m\phi)-1}~=~\frac{4 r}{m\sqrt{\mc{A}^2r^2-1}}\arctanh\lr{\sqrt{\frac{\mc{A}r+1}{\mc{A}r-1}}\tan\lr{\frac{m\pi}{2}}}\,.
\end{equation}
It was shown in~\cite{Accin3D} that~\eqref{eqn:circum} is a decreasing function of $r$. As such, closed loops near to the horizon are smaller than those close to the conformal boundary. Now we can identify the horizon $r=r_h$ as a black hole horizon and we consider only $r$ that satisfy $0<r_{\mrm{conf}}<r<r_h$. We therefore interpret larger values of the coordinate $r$ as being closer to the black hole horizon. The radial coordinate $r$ is therefore physically unintuitive but a convenient coordinate for the purpose of calculation and comparison with the other C-metric classes.

We summarise the constants on the parameters $\mc{A}$, $m$, and $\ell$ as
\begin{subequations}
    \begin{align}\label{eqn:: class I c constrain a}
        m\mc{A}\ell&~>~1\,,\\
        \label{eqn:: class I c constrain b}0~<~m&~<~\frac{1}{2}\,,\\
        m\mc{A}\ell\sin(m\pi)&~<~1\,.\label{eqn:: class I c constrain c}
    \end{align}
\end{subequations}Constraints~\eqref{eqn:: class I c constrain a} and~\eqref{eqn:: class I c constrain b} are necessary for the existence of the Class $\text{I}_{\text{C}}$ solution~\cite{Accin3D}, whereas constraint~\eqref{eqn:: class I c constrain c} ensures that $r_h>r_{\mrm{conf}}$. The physical properties of this black-hole solution are described in~\cite{Accin3D}.

\begin{figure}[t!]
    \centering
    \begin{tikzpicture}
    \draw (0, 0) node[inner sep=0] {\includegraphics[width=\textwidth]{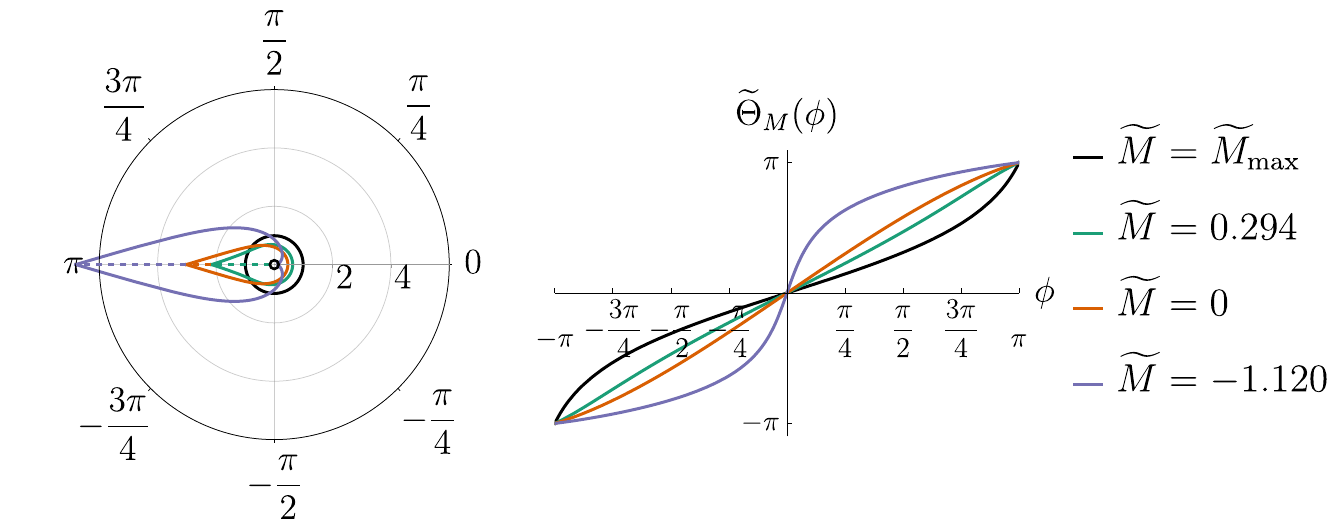}};
     \draw (-8, 2.4) node {\textbf{(a)}};
    \draw (-1.6, 2.4) node {\textbf{(b)}};
\end{tikzpicture}\vspace{-1em}
    \begin{tikzpicture}
    \draw (0, 0) node[inner sep=0] {\includegraphics[width=\textwidth]{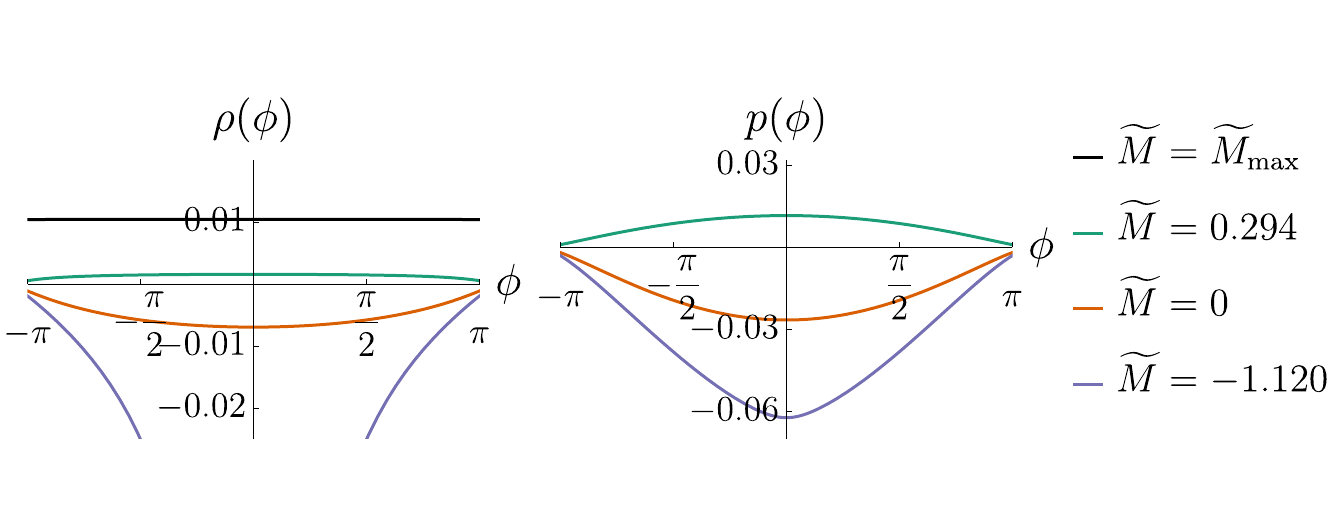}};
    \draw (-8, 2.4) node {\textbf{(c)}};
    \draw (-1.6, 2.4) node {\textbf{(d)}};
\end{tikzpicture}\vspace{-1em}
        \begin{tikzpicture}
        \draw (0, 0) node[inner sep=0] {\includegraphics[width=\textwidth]{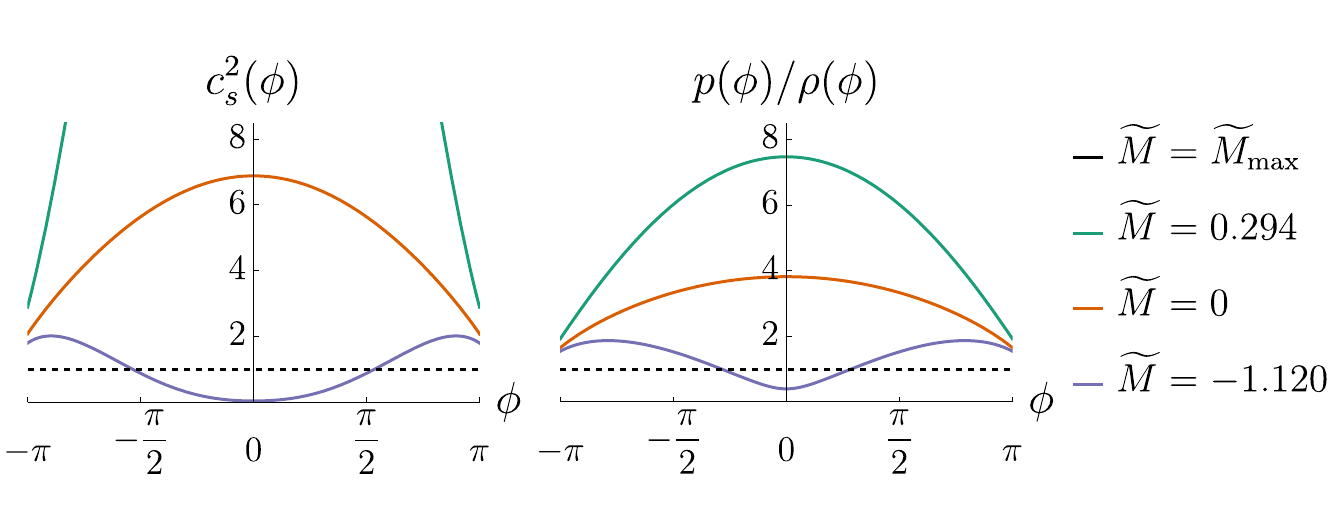}};
        \draw (-8, 2.4) node {\textbf{(e)}};
    \draw (-1.6, 2.4) node {\textbf{(f)}};
        \end{tikzpicture}
    \caption{\justifying Class $\text{I}_{\text{C}}$ for $r_0=0.3$, $m=0.25$, $\mc{A}=5.1$, $\alpha=0.25$, and $\ell=1$, where $r_h\approx0.316$, $\wt{M}_{\mrm{min}}=-\infty$, and $\wt{M}_{\mrm{max}}\approx0.978$. \textbf{(a)} Polar plot of shell radius $\wt{R}_M$~\eqref{eqn:ADM shell coords} over $\phi\in(-\pi,\pi)$ for varying values of $\wt{M}$; dashed line indicates the string and the black ring represents the black hole. \textbf{(b)} Shell angular coordinate $\wt{\Theta}_M$~\eqref{eqn:ADM shell coords} for varying values of $\wt{M}$. \textbf{(c)} Energy density $\rho$ of the shell for varying values of $\wt{M}$; the curve for $\wt{M}_{\mrm{min}}$ is divergent as $1/\phi$ as $\phi\to0$. \textbf{(d)} Pressure $p$ of the shell for varying values of $\wt{M}$. \textbf{(e)} Squared speed of sound $c_s^2$ of matter in the shell for varying values of $\wt{M}$; dashed line at $c_s=1$. \textbf{(f)} Ratio of shell pressure to shell energy density for varying values of $\wt{M}$; dashed line at $p/\rho=1$.}\label{fig:Class Ic}
\end{figure}

The minimum value of the exterior mass parameter is given by
\begin{subequations}\label{eqn:: min M class Ic}
    \begin{align}
        \Mmin&~=~\frac{f(r_0)}{\alpha^2}\frac{\lr{m^2\mc{A}^2\ell^2\sin^2(m\phi_*)-1}}{\lr{1-
        \mc{A}r_0\cos(m\phi_*)}^2}\,,\\
          \phi_*&~=~\begin{cases}
          0&\text{for~}m\leq\frac{1}{\mc{A}\ell}\sqrt{\frac{\mc{A}r_0}{\mc{A}r_0-1}}\,,\\
          \min\lrc{\frac 1m\arccos\lr{\frac{(m^2\mc{A}^2\ell^2-1)\mc{A}r_0}{m^2\mc{A}^2\ell^2}},\pi}&\text{for~}m>\frac{1}{\mc{A}\ell}\sqrt{\frac{\mc{A}r_0}{\mc{A}r_0-1}}\,.
      \end{cases}
    \end{align}
\end{subequations}
 We note two special cases. First, for $\phi_*=\frac{1}{m}\arccos\lr{\frac{(m^2\mc{A}^2\ell^2-1)\mc{A}r_0}{m^2\mc{A}^2\ell^2}}$, we have
\begin{equation}
    \Mmin~=~-\frac{m^2}{\alpha^2}\lr{1-m^2\mc{A}^2\ell^2}\,,
\end{equation} whereas for $\phi_*=0$, we have
\begin{equation}\label{eqn:Mmin zero Ic}
    \Mmin~=~-\frac{f(r_0)}{\alpha^2 (1-\mc{A}r_0)^2}\,.
\end{equation}
In this case, with $M=\Mmin$,   $\rho(\phi)\propto-1/\phi$ around $\phi=0$ and hence $\rho(\phi)\to-\infty$ as $\phi\to0$. Furthermore, we have $\wt{M}\to-\infty$ as $M\to\Mmin$. This may be seen by noting that the integrand in the definition of $\wt{M}$~\eqref{eqn:: mass function} is even and so may be written over the interval $\phi'\in(0,\pi)$. For arbitrary values of $M$, the integrand is order unity near zero. However, as $M\to\Mmin$~\eqref{eqn:Mmin zero Ic}, the integrand is order $1/\phi'$ near $\phi'=0$ and the integral diverges.

The shell constructed around the Class $\mrm{I}_{\mrm{C}}$ solution is markedly different from that of the rapid acceleration phase of the Class $\mrm{I}_{\mrm{pushed}}$ shell. We remark that $\wt{M}_{\mrm{min}}=-\infty$; hence, between the curves depicting the shell with $\wt{M}=\wt{M}_{\mrm{min}}$ and $\wt{M}=0$ lies the behaviour of the shell with global AdS exterior to the shell, $\wt{M}=-1$. 

We plot the radial and angular coordinates of the shell in~\autoref{fig:Class Ic} \textbf{(a)} and \textbf{(b)}. For small values of $M$, the shell resembles a teardrop with a pinch near the string at $\phi=\pi$. As $M\to\Mmin$, the shell becomes cuspoidal at $\phi=0$, similar to the slow acceleration regime in the Class $\mrm{I}_{\mrm{pulled}}$ solution depicted in~\autoref{fig:Class I slow} \textbf{(a)}. However, as $\beta_M$ diverges in this limit, the shell ADM radius $\wt{R}_M$ also increases and the ADM angular coordinate becomes increasingly nonlinear. As $\wt{M}\to\wt{M}_{\mrm{max}}$, the teardrop rounds off, but the angular coordinate remains nonlinear. 

In~\autoref{fig:Class Ic} \textbf{(c)} and \textbf{(d)}, we plot the shell energy density $\rho$ and pressure $p$. The energy density and pressure may take any sign and increase with $M$. When the energy density and pressure are negative, they are maximised where the string meets the shell at $\phi=\pi$, whereas when the energy density and pressure are positive, they are maximised around $\phi=0$. For $M=\Mmin$, we have cut off the graph of the energy density so as to depict the other curves clearly; the energy density decreases, reaching a minimum at $\phi=0$ before increasing. As predicted in~\eqref{eqn:large M SE}, the energy density and pressure are
positive for large values of $M$, as $\wt{M}\to\wt{M}_{\mrm{max}}$. We recall that $p(\phi)\to\infty$ as $\wt{M}\to\wt{M}_{\mrm{max}}$, hence this curve is not shown.

As the exterior spacetime mass $M$ varies, we find both physical ($\rho>0$) and unphysical ($\rho<0$) solutions, depicted in~\autoref{fig:Class Ic}~\textbf{(c)}. The physical and unphysical shells are separated by a critical mass parameter $M_{\mrm{crit}}$, for which the stress energy of the shell --- both the energy density and pressure --- vanishes,
\begin{equation}\label{eqn:IC crit mass}
    M_{\mrm{crit}}~\coloneq~\frac{m^2}{\alpha^2}(m^2\mc{A}^2\ell^2-1)~>~0\,.
\end{equation}
In this case, physically there is no shell. However, the shell radial and angular functions are still regular for this critical exterior mass. We may therefore interpret this solution with $M=M_{\mrm{crit}}$ as a spacetime consisting of an accelerated black hole, pulled by a finite-length string at the end of which lies a point particle with mass $\mu_\pi$~\eqref{eqn:defect mass}. The NEC, WEC, and SEC are respected for $M>M_{\mrm{crit}}$.

In~\autoref{fig:Class Ic}~\textbf{(e)} and~\textbf{(f)}, we plot the squared speed of sound $c_s^2$ and the ratio of pressure and energy density as functions of the shell's intrinsic coordinate $\phi$. When $c_s^2\leq1$, the stress energy is causal and when $p/\rho\leq1$ with $\rho\geq0$, the stress energy satisfies the DEC. We find that the squared speed of sound and ratio $p/\rho$ behave similarly. Furthermore, we find that positive stress energy ($M>M_{\mrm{crit}}$) violates causality and the DEC. This aligns with the intuition of Section~\ref{sec: A zero limit} that any black-hole solution surrounded by a shell with positive stress energy must have superluminal stress energy.

\begin{figure}[t!]
    \centering
    \begin{tikzpicture}
    \draw (0, 0) node[inner sep=0] {\includegraphics[width=0.9\textwidth]{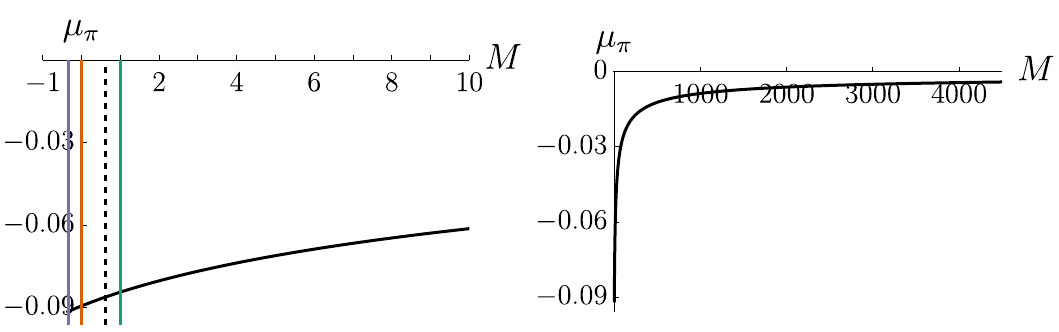}};
     \draw (-7, 2.4) node {\textbf{(a)}};
    \draw (0.6, 2.4) node {\textbf{(b)}};
\end{tikzpicture}
    \caption{\justifying Class $\text{I}_{\text{C}}$ for $r_0=0.3$, $m=0.25$, $\mc{A}=5.1$, $\alpha=0.25$, and $\ell=1$, where $r_h\approx0.316$, $\wt{M}_{\mrm{min}}= -\infty$, and $\wt{M}_{\mrm{max}}\approx0.978$. \textbf{(a)} Point particle mass $\mu_\pi$ at $\phi=\pm\pi$~\eqref{eqn:defect mass} for varying values of $M$. Black vertical dashed line corresponds to $M=M_{\mrm{crit}}$~\eqref{eqn:IC crit mass}. Coloured vertical lines correspond to the shells plotted in~\autoref{fig:Class Ic}. \textbf{(b)} Point particle mass for larger range of $M$ values, demonstrating the large-$M$ behaviour of $\mu_\pi$ predicted in~\eqref{eqn:mu pi large M}.}\label{fig:Class Ic mu pi}
\end{figure}
 In~\autoref{fig:Class Ic mu pi}, we plot the DJ'tH mass of the point particle at the end of the string
 as a function of the exterior mass parameter $M$. We find that the mass is negative for all values of $M$, indicating an angular excess at $\phi=\pm\pi$. As $M\to\infty$, we see $\mu_\pi\to0$. The point particle has negative mass, in contrast to the positive tension of the string.

\section{Class $\text{II}_{\text{right}}$}
\label{Sec5}

We consider now the Class $\text{II}_{\mrm{right}}$ spacetime, describing a BTZ black hole pushed by a strut. In polar coordinates $(\sigma,r,\phi)$, the metric reads
\begin{subequations}\label{eqn:Class II coords}
    \begin{align}
        \dd s_-^2&~=~\frac{1}{\Omega^2(r,\phi)}\lr{-\frac{f(r)}{\alpha^2}\dd\sigma^2+\frac{\dd r^2}{f(r)}+r^2\dd\phi^2}\,,\\
        f(r)&~=~\frac{r^2}{\ell^2}-m^2\lr{1-\mc{A}^2r^2}\,,\\
        \Omega(r,\phi)&~=~1+\mc{A}r\cosh(m\phi)\,,
    \end{align}
\end{subequations}
where $m>0$.  We now obtain
\begin{equation}\label{eqn: tau II right}
    \uptau_\pi~=~-\frac1{4\pi}m\mc{A}\sinh(m\pi) 
\end{equation}
for the tension of the strut.
 Taking the limit $\mc{A}\to0$ and identifying $m^2=\alpha^2M$ with $M>0$, one recovers the geometry of a non-rotating BTZ black hole~\eqref{eqn:: btz metric} in coordinates $(\sigma,r,\phi)=(t,\alpha r_+,\theta/\alpha)$.

\begin{figure}[t]
    \centering
    \begin{subfigure}[b]{0.4\textwidth}
        \centering\captionsetup{labelfont=bf}
        \begin{tikzpicture}
        \draw (0, 0) node[inner sep=0] {\includegraphics[width=\textwidth]{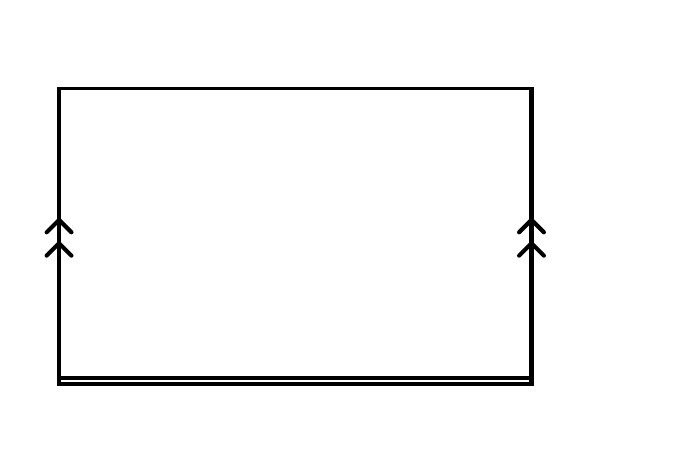}};
        \draw (2.7, 1.4) node {\large$r=\infty$};
        \draw (2.7, -1.3) node {\large$r=r_h$};
        \draw (-2.4, -1.85) node {\large$\phi=-\pi$};
        \draw (1.8, -1.85) node {\large$\phi=\pi$};
        \end{tikzpicture}
        \caption{\\}
    \end{subfigure}\hfill
    \begin{subfigure}[b]{0.4\textwidth}
        \centering\captionsetup{labelfont=bf}
        \begin{tikzpicture}
        \draw (0, 0) node[inner sep=0] {\includegraphics[width=\textwidth]{figures/SpaceDiagrams/Ipushslowless.pdf}};
        \node[rotate=38] at (0.3,0.5) {\large $r=r_{\mrm{conf}}$};
        \draw (2.9, -1.3) node {\large$r=-\infty$};
        \draw (-2.4, -1.85) node {\large$\phi=-\pi$};
        \draw (1.8, -1.85) node {\large$\phi=\pi$};
        \end{tikzpicture}
        \caption{\\}
    \end{subfigure}
    \begin{subfigure}[b]{0.4\textwidth}
        \centering\captionsetup{labelfont=bf}
        \begin{tikzpicture}
        \draw (0, 0) node[inner sep=0] {\includegraphics[width=\textwidth]{figures/SpaceDiagrams/IpushDouble.pdf}};
        \draw (2.67, 1.4) node {\large$r=r_{\mrm{D}}$};
        \node[rotate=58] at (0.05,0.8) {\large $r=r_{\mrm{conf}}$};
        \draw (2.9, -1.3) node {\large$r=-\infty$};
        \draw (-2.4, -1.85) node {\large$\phi=-\pi$};
        \draw (1.8, -1.85) node {\large$\phi=\pi$};
        \end{tikzpicture}
        \caption{}
    \end{subfigure}
    \caption{\justifying Constant time slice of the Class $\mrm{II}_{\mrm{right}}$ solution. \textbf{(a)} $r>0$ patch. \textbf{(b)} $r<0$ patch, slow acceleration phase. \textbf{(c)} $r<0$ patch, rapid acceleration phase. Double lines in \textbf{(a)} and \textbf{(c)} represent a Killing horizon.}
    \label{fig:Class II right ranges}
\end{figure}
The Class $\mrm{II}_{\mrm{right}}$ spacetime is covered by two charts, one for $r>0$ and a second for $r<0$, that are glued together across $r=\pm\infty$. We note that the proper distance to $r=\pm\infty$ is finite,
\begin{equation}
    \int_{\pm r_0}^{\pm\infty}\dd r\,\sqrt{g_{rr}}~=~\int_{\pm r_0}^{\pm\infty}\dd r\,\frac{1}{\sqrt{\Omega^2(r,\phi)f(r)}}~<~\infty\,,
\end{equation}for any fixed $\phi$. This follows from the asymptotic behaviour $1/\sqrt{\Omega^2(r,\phi)f(r)}=\OO(r^{-2})$ as $r\to\infty$. The stress energy of a shell formed in the $r<0$ chart is always unphysical and comprised of exotic matter. One may notice that for $r<0$, the energy density in the large-$M$ limit is negative~\eqref{eqn: large M rho}.

The conformal boundary is located at
\begin{equation}\label{eqn:Class IIrc}
    \rc~=~-\frac{1}{\mc{A}\cosh(m\phi)}\,,
\end{equation}and the metric~\eqref{eqn:Class II coords} has two Killing horizons
\begin{subequations}\label{eqn:Class II horizons}
    \begin{align}
         r_h&~=~\frac{m\ell}{\sqrt{1+m^2\mc{A}^2\ell^2}}\,,\\
        r_{\mrm{D}}&~=~-\frac{m\ell}{\sqrt{1+m^2\mc{A}^2\ell^2}}\,.
    \end{align}
\end{subequations}The chart $r>0$ describes the exterior of a black hole with horizon at $r=r_h$. The Class $\mrm{II}_{\mrm{right}}$ spacetime exhibits two phases of acceleration, slow and rapid, depending on the relative positions of $r_{\mrm{D}}$ and $\rc$. The rapid phase of acceleration is indicated by the presence of a Killing horizon in the $r<0$ chart.

\begin{figure}[t!]
    \centering
    \begin{tikzpicture}
    \draw (0, 0) node[inner sep=0] {\includegraphics[width=\textwidth]{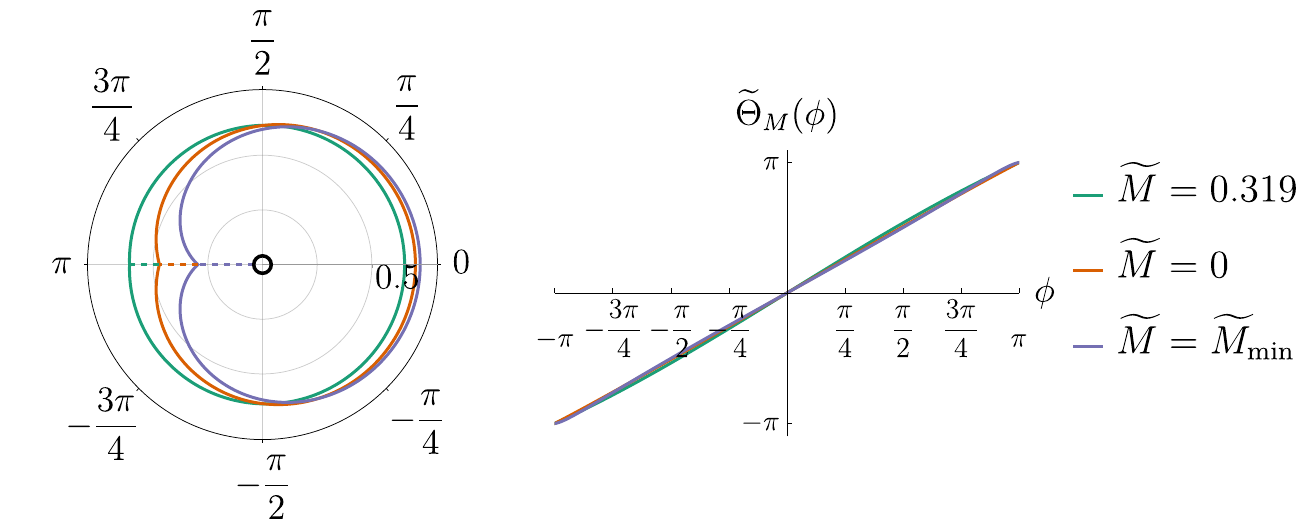}};
     \draw (-8, 2.4) node {\textbf{(a)}};
    \draw (-1.6, 2.4) node {\textbf{(b)}};
\end{tikzpicture}\vspace{-1em} 
    \begin{tikzpicture}
    \draw (0, 0) node[inner sep=0] {\includegraphics[width=\textwidth]{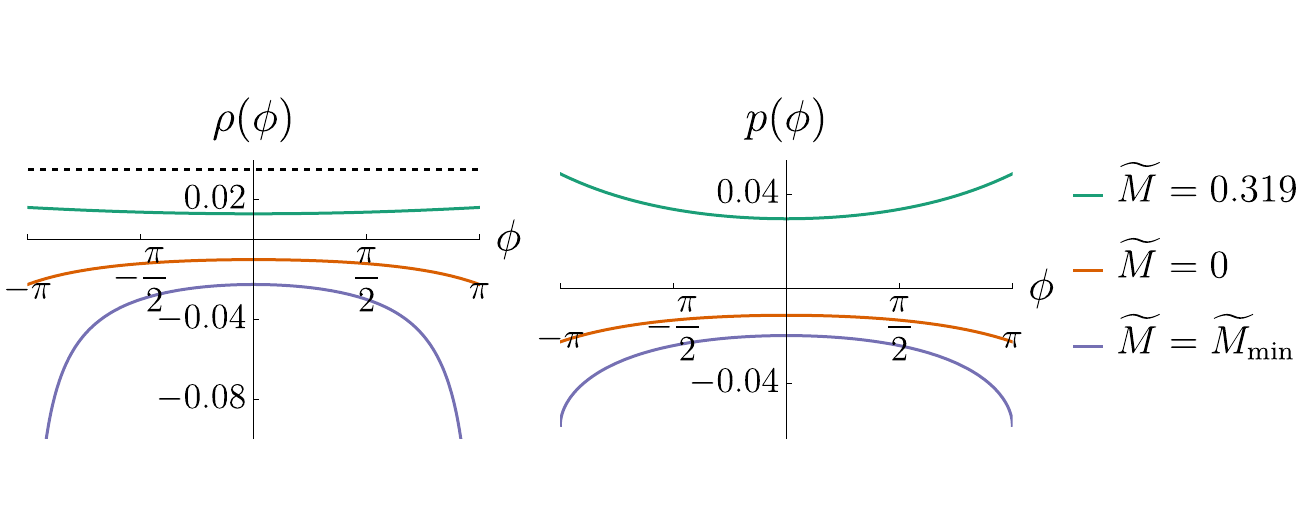}};
    \draw (-8, 2.4) node {\textbf{(c)}};
    \draw (-1.6, 2.4) node {\textbf{(d)}};
\end{tikzpicture}\vspace{-1em} 
\begin{tikzpicture}
    \draw (0, 0) node[inner sep=0] {\includegraphics[width=\textwidth]{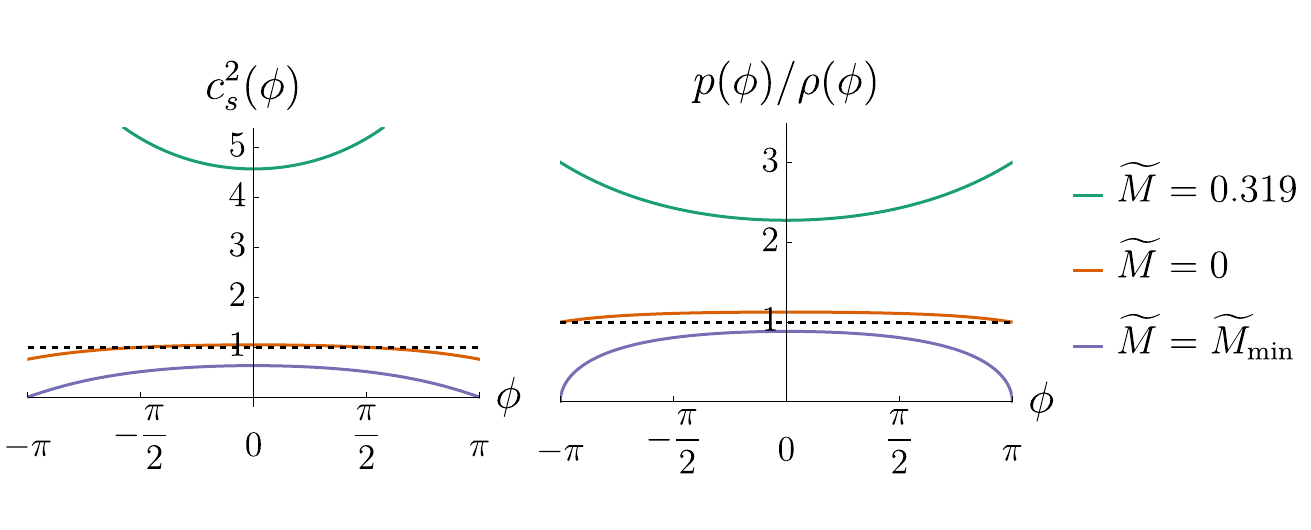}};
    \draw (-8, 2.4) node {\textbf{(e)}};
    \draw (-1.6, 2.4) node {\textbf{(f)}};
\end{tikzpicture}
    \caption{\justifying Class $\mrm{II}_{\mrm{right}}$ ($r>0$, slow phase) for $r_0=1$, $m=0.5$, $\mc{A}=0.4$, $\alpha=0.25$, and $\ell=1$, where $r_h\approx0.490$, $\wt{M}_{\mrm{min}}\approx-0.325$, and $\wt{M}_{\mrm{max}}\approx0.407$ \textbf{(a)} Polar plot of shell radius $\wt{R}_M$~\eqref{eqn:ADM shell coords} over $\phi\in(-\pi,\pi)$ for varying values of $\wt{M}$; dashed line indicates the strut and black ring represents the black hole. \textbf{(b)} Shell angular coordinate $\wt{\Theta}_M$~\eqref{eqn:ADM shell coords} for varying values of $\wt{M}$. \textbf{(c)} Energy density $\rho$ of the shell for varying values of $\wt{M}$. Black dashed line at asymptotic value as $\wt{M}\to\wt{M}_{\mrm{max}}$. \textbf{(d)} Pressure $p$ of the shell for varying values of $\wt{M}$. \textbf{(e)} Squared speed of sound $c_s^2$ of matter in the shell for varying values of $\wt{M}$; dashed line at $c_s=1$. \textbf{(f)} Ratio of shell pressure to shell energy density for varying values of $\wt{M}$; dashed line at $p/\rho=1$. }\label{fig:Class II slow pos}
\end{figure}
The slow acceleration phase, defined by
\begin{equation}\label{eqn: II slow cond}
    m\mc{A}\ell\sinh(m\pi)~<~1\,,
\end{equation}occurs when $r_{\mrm{D}}$ is hidden behind the conformal boundary $\rc$, $r<\rc<r_{\mrm{D}}<0$. However, in the rapid acceleration phase, a ``droplet'' horizon -- a noncompact horizon intersecting conformal infinity (also known as a black droplet)~\cite{BlackDroplet,BlackDroplet1,Accin3D} -- forms. The geometry of the Class $\mrm{II}_{\mrm{right}}$ solution is depicted in~\autoref{fig:Class II right ranges}.

The minimum value of the exterior mass parameter $M$ depends on the chart. In the $r>0$ chart, the
minimum value of the exterior mass parameter is given by
\begin{equation}\label{eqn:Mmin II pos}
        \Mmin~=~\frac{f(r_0)}{\alpha^2}\frac{\lr{m^2\mc{A}^2\ell^2\sinh^2(m\pi)-1}}{\lr{1+\mc{A}r_0\cosh(m\pi)}^2}\,,
\end{equation}which is negative in the slow acceleration regime and positive in the rapid acceleration regime.

\subsection{Slow acceleration}
We consider first the slow acceleration regime, $m\mc{A}\ell\sinh(m\pi)<1$. We remark that in the $r>0$ chart, we depict a shell with $\wt{M}=0$, which has an exterior geometry of Torricelli's trumpet~\eqref{eqn:cylinder}.

We plot in~\autoref{fig:Class II slow pos} \textbf{(a)} and \textbf{(b)} the shell radial and angular coordinates~\eqref{eqn:ADM shell coords} as functions of the shell's intrinsic coordinate $\phi$. For small values of $M$, the shell resembles a cardioid with a true cusp where the strut (indicated by the dashed line) meets the shell at $\phi=\pi$. As $\wt{M}\to\wt{M}_{\mrm{max}}$, the shell becomes less deformed and tends to a circular shape, as expected from~\eqref{eqn:large M R theta}.   The angular coordinate exhibits very little deviation from linearity.

\begin{figure}[t!]
    \centering
    \begin{tikzpicture}
    \draw (0, 0) node[inner sep=0] {\includegraphics[width=0.9\textwidth]{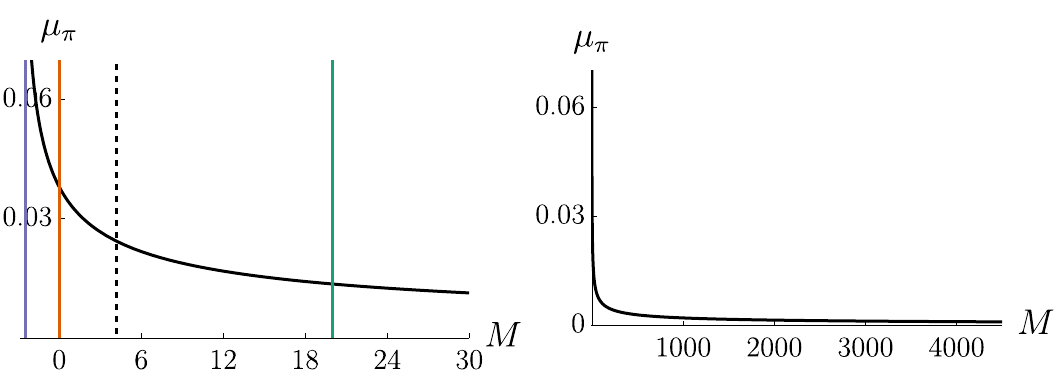}};
     \draw (-7.5, 2.4) node {\textbf{(a)}};
    \draw (0.1, 2.4) node {\textbf{(b)}};
\end{tikzpicture}
    \caption{\justifying Class $\mrm{II}_{\mrm{right}}$ ($r>0$, slow phase) for $r_0=1$, $m=0.5$, $\mc{A}=0.4$, $\alpha=0.25$, and $\ell=1$, where $r_h\approx0.490$, $\wt{M}_{\mrm{min}}\approx-0.325$, and $\wt{M}_{\mrm{max}}\approx0.407$ \textbf{(a)} Point particle mass $\mu_\pi$ at $\phi=\pm\pi$~\eqref{eqn:defect mass} for varying values of $M$. Black vertical dashed line corresponds to $M=M_{\mrm{crit}}$~\eqref{eqn:II crit mass}. Coloured vertical lines correspond to the shells plotted in~\autoref{fig:Class II slow pos}. \textbf{(b)} Point particle mass for larger range of $M$ values, demonstrating the large-$M$ behaviour of $\mu_\pi$ predicted in~\eqref{eqn:mu pi large M}.}\label{fig:Class II mu pi}
\end{figure}
\begin{figure}[t!]
    \centering
    \begin{tikzpicture}
    \draw (0, 0) node[inner sep=0] {\includegraphics[width=\textwidth]{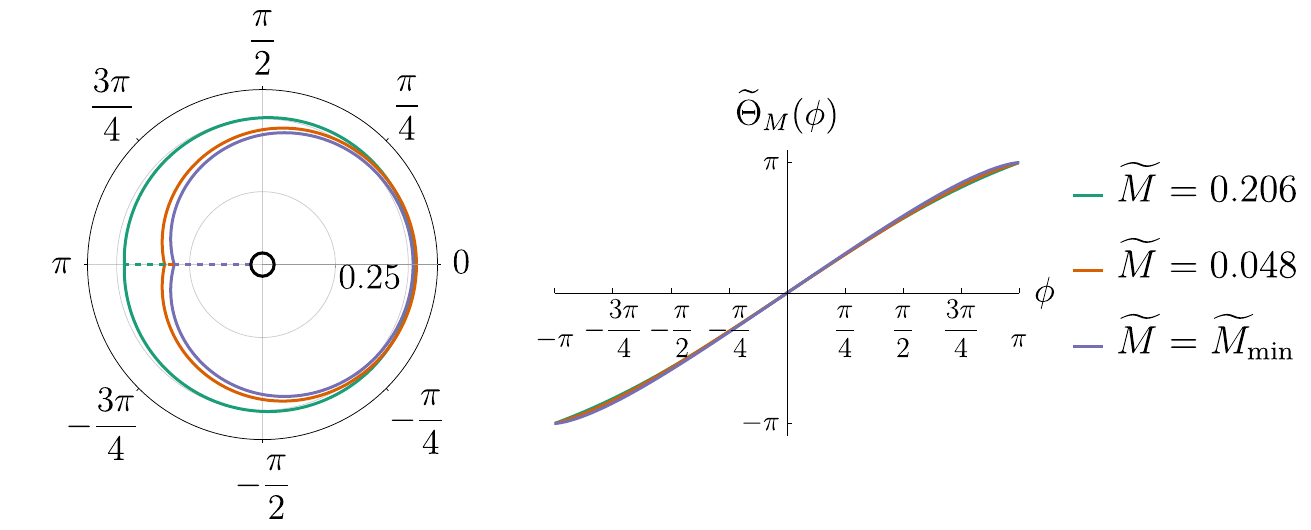}};
     \draw (-8, 2.4) node {\textbf{(a)}};
    \draw (-1.6, 2.4) node {\textbf{(b)}};
\end{tikzpicture}\vspace{-1em} 
    \begin{tikzpicture}
    \draw (0, 0) node[inner sep=0] {\includegraphics[width=\textwidth]{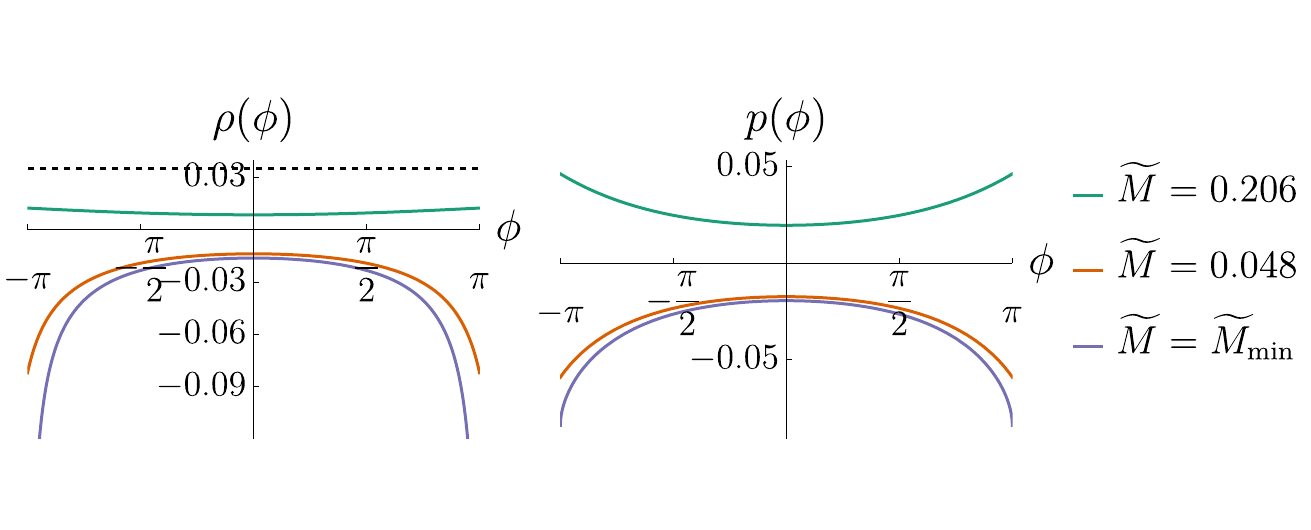}};
    \draw (-8, 2.4) node {\textbf{(c)}};
    \draw (-1.6, 2.4) node {\textbf{(d)}};
\end{tikzpicture}\vspace{-1em} 
\begin{tikzpicture}
    \draw (0, 0) node[inner sep=0] {\includegraphics[width=\textwidth]{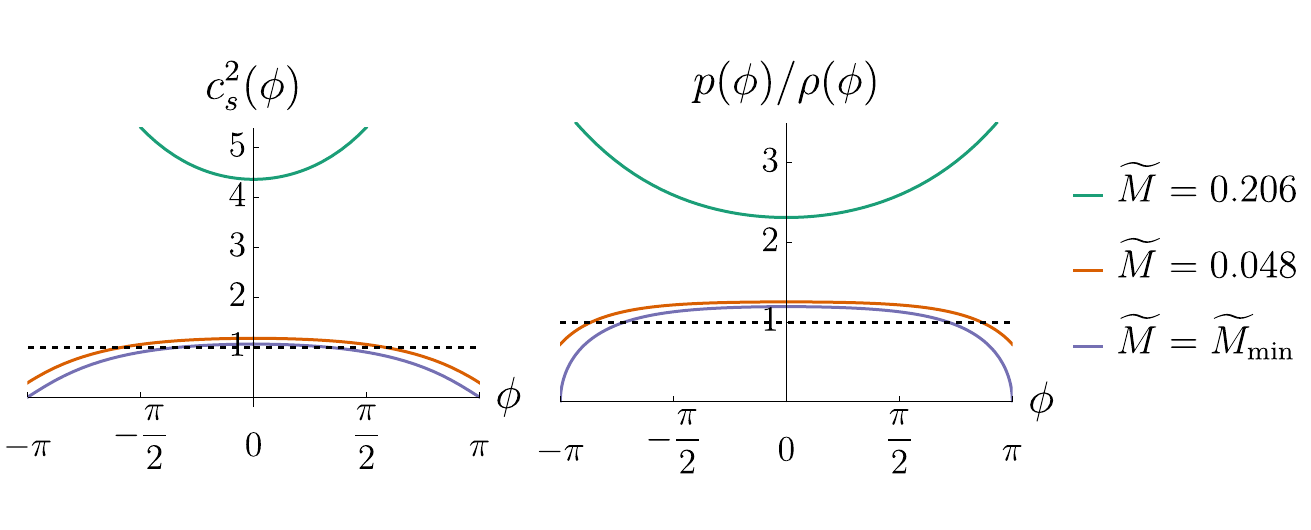}};
    \draw (-8, 2.4) node {\textbf{(e)}};
    \draw (-1.6, 2.4) node {\textbf{(f)}};
\end{tikzpicture}
    \caption{\justifying Class $\mrm{II}_{\mrm{right}}$ ($r>0$, rapid phase) for $r_0=1$, $m=0.6$, $\mc{A}=0.6$, $\alpha=0.25$, and $\ell=1$, where $r_h\approx0.565$, $\wt{M}_{\mrm{min}}\approx0.023$, and $\wt{M}_{\mrm{max}}\approx0.263$. \textbf{(a)} Polar plot of shell radius $\wt{R}_M$~\eqref{eqn:ADM shell coords} over $\phi\in(-\pi,\pi)$ for varying values of $\wt{M}$; dashed line indicates the strut and black ring represents the black hole. \textbf{(b)} Shell angular coordinate $\wt{\Theta}_M$~\eqref{eqn:ADM shell coords} for varying values of $\wt{M}$. \textbf{(c)} Energy density $\rho$ of the shell for varying values of $\wt{M}$. Black dashed line at asymptotic value as $\wt{M}\to\wt{M}_{\mrm{max}}$. \textbf{(d)} Pressure $p$ of the shell for varying values of $\wt{M}$. \textbf{(e)} Squared speed of sound $c_s^2$ of matter in the shell for varying values of $\wt{M}$; dashed line at $c_s=1$. \textbf{(f)} Ratio of shell pressure to shell energy density for varying values of $\wt{M}$; dashed line at $p/\rho=1$. }\label{fig:Class II rapid pos}
\end{figure} 

As shown in~\autoref{fig:Class II slow pos} \textbf{(c)} and \textbf{(d)}, the energy density $\rho$ and pressure $p$ of the shell may take either sign. For $\rho,~p>0$, the energy density and pressure are maximised where the strut meets the shell at $\phi=\pi$, whereas for $\rho,~p<0$, the energy density and pressure are maximised opposite the cusp, at $\phi=0$. Both the energy density and pressure increase with $M$. Over the range of $M$  we find both physical ($\rho>0$) and unphysical ($\rho<0$) shells. Between these, we find a critical exterior mass parameter $M_{\mrm{crit}}$, for which the stress energy of the shell vanishes,
\begin{equation}\label{eqn:II crit mass}
    M_{\mrm{crit}}~\coloneq~\frac{m^2}{\alpha^2}(m^2\mc{A}^2\ell^2+1)~>~0\,.
\end{equation} Only for $M>M_{\mrm{crit}}$ does the stress energy satisfy the NEC, WEC, and SEC. As $\wt{M}$ increases, the energy density is more evenly distributed across the shell.

The similarity between~\eqref{eqn:II crit mass} and the Class $\mrm{I}_{\mrm{C}}$ critical mass~\eqref{eqn:IC crit mass}
is noteworthy.
As in the Class $\mrm{I}_{\mrm{C}}$ case, there is no shell. We interpret this solution with $M=M_{\mrm{crit}}$ as a spacetime consisting of an accelerated black hole, pushed by a finite-length strut with a point particle of mass $\mu_\pi$~\eqref{eqn:defect mass} at its end.

In~\autoref{fig:Class II slow pos}~\textbf{(e)} and \textbf{(f)}, we plot the squared speed of sound $c_s^2$ and the ratio of pressure and energy density as functions of the shell's intrinsic coordinate $\phi$. When $c_s^2\leq1$, the stress energy is causal and when $p/\rho\leq1$ with $\rho\geq0$, the stress energy satisfies the DEC. We find that the two quantities behave similarly: for $\rho\leq0$, both are maximised opposite the strut and for $\rho\geq0$, both are maximised where the strut meets the shell. As with the Class $\mrm{I}_{\mrm{C}}$ black-hole solution, we find that positive stress energy ($M>M_{\mrm{crit}}$) violates causality and the DEC, aligning with the intuition of Section~\ref{sec: A zero limit} that a static shell must exert additional pressure to counteract the negative cosmological constant, in turn yielding
superluminal positive stress energy.

In~\autoref{fig:Class II mu pi}, we plot the DJ'tH mass of the point particle as a function of the exterior mass parameter $M$. We find that $\mu_\pi>0$ for all values of $M$, indicating an angular deficit at $\phi=\pm\pi$. As $\wt{M}\to\wt{M}_{\mrm{max}}$ ($M\to\infty$), we see $\mu_\pi\to0$. In contrast to the Class $\mrm{I}_{\mrm{C}}$ solution, the point particle has positive mass, however the strut at $\phi=\pm\pi$ in Class $\mrm{II}_{\mrm{right}}$ spacetimes has negative tension.

In the limit of zero acceleration, the string tension and deficit mass vanish and the critical mass~\eqref{eqn:II crit mass} tends to $M_{\mrm{crit}}\to m^2/\alpha^2$. As analysed in~\Sref{sec: A zero limit}, this corresponds to the interior geometry matching the outer geometry in coordinates $(\sigma,r,\phi)=(t,r_+\alpha,\theta/\alpha)$ and one would expect no shell. A similar correspondence is not available, however, for the Class $\mrm{I}_{\mrm{C}}$ critical mass because the Class $\mrm{I}_{\mrm{C}}$ solution is disconnected from the BTZ black hole.

\subsection{Rapid acceleration}
The rapid acceleration phase has features quite similar to the slow phase. Results for the $r>0$ chart are shown in~\autoref{fig:Class II rapid pos} \textbf{(a)} to \textbf{(f)}.

The cusp at $\phi=\pi$ is noticeably less pronounced for small values of $M$ than in the slow acceleration phase and the angular coordinate is more non-linear. The energy density, pressure, speed of sound, and the ratio $p/\rho$, however, are all characteristically similar to the slow acceleration phase. In particular, for sufficiently large values of the exterior mass parameter ($M>M_{\mrm{crit}}$), the NEC, WEC, and SEC are satisfied, whilst violating causality and the DEC. Again, the shell vanishes for $M=M_{\mrm{crit}}$~\eqref{eqn:II crit mass}. The mass of the point particle is again positive and tends to zero as $M\to\infty$.

\section{Class $\text{II}_{\text{left}}$}
\label{Sec6}

\begin{figure}[t]
        \centering
        \begin{tikzpicture}
        \draw (0, 0) node[inner sep=0] {\includegraphics[width=0.4\textwidth]{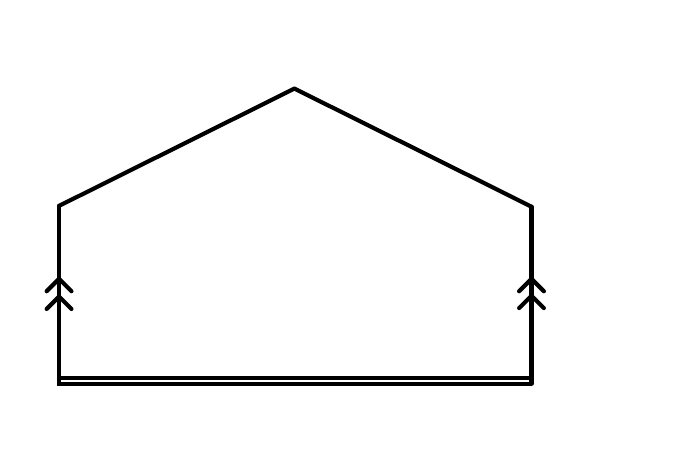}};
        \draw (2.7, -1.3) node {\large$r=r_h$};
        \draw (-2.4, -1.85) node {\large$\phi=-\pi$};
        \draw (1.8, -1.85) node {\large$\phi=\pi$};
        \node[rotate=28] at (-1.6,1.1) {\large $r=r_{\mrm{conf}}$};
        \end{tikzpicture}
    \caption{\justifying Constant time slice of the Class $\mrm{II}_{\mrm{left}}$ solution. Double lines represent a Killing horizon.}
    \label{fig:Class II left ranges}
\end{figure}

The Class $\mrm{II}_{\mrm{left}}$ solution describes an accelerating BTZ black hole pulled by a string. The Class $\mrm{II}_{\mrm{left}}$ metric is found by the mapping $\mc{A}\to-\mc{A}$ in~\eqref{eqn:Class II coords},
\begin{subequations}
    \begin{align}
        \dd s^2~&=~\frac{1}{\Omega^2(r,\phi)}\lr{-\frac{f(r)}{\alpha^2}\dd\sigma^2+\frac{\dd r^2}{f(r)}+r^2\dd\phi^2}\,,\\
        f(r)~&=~\frac{r^2}{\ell^2}-m^2\lr{1-\mc{A}^2r^2}\,,\\
        \Omega(r,\phi)~&=~1-\mc{A}r\cosh(m\phi) 
    \end{align}
\end{subequations}
and
\begin{equation}
\uptau_\pi~=~\frac1{4\pi}m\mc{A}\sinh(m\pi)\,.
\end{equation}
is the tension of the string. The conformal boundary is found at
\begin{equation}
    \rc~=~\frac{1}{\mc{A}\cosh(m\phi)}\,.
\end{equation}The spacetime is covered by a single chart with $r>0$. There is a single Killing horizon at
\begin{equation}
    r_h~=~\frac{m\ell}{\sqrt{1+m^2\mc{A}^2\ell^2}}\,.
\end{equation}To ensure $r_h<\rc$, we require
\begin{equation}
    m\mc{A}\ell\sinh(m\pi)~<~1\,.
\end{equation}Therefore, there is no equivalent rapid phase of acceleration in the $\mrm{II}_{\mrm{left}}$ solution. The geometry of this spacetime is depicted in~\autoref{fig:Class II left ranges}.

The minimum value of the exterior mass parameter is given by
\begin{subequations}
    \begin{align}
        \Mmin&~=~\frac{f(r_0)}{\alpha^2}\frac{\lr{m^2\mc{A}^2\ell^2\sinh^2(m\phi_*)-1}}{\lr{1-\mc{A}r_0\cosh(m\phi_*)}^2}\,,\\
        \phi_*&~=~\begin{cases}
            0&\text{for~}m\leq\frac{1}{\mc{A}\ell}\sqrt{\frac{\mc{A}r_0}{1-\mc{A}r_0}}\,,\\
            \min\lrc{\frac{1}{m}\arccosh\lr{\frac{(m^2\mc{A}^2\ell^2+1)\mc{A}r_0}{m^2\mc{A}^2\ell^2}},\pi}\,&\text{for~}m>\frac{1}{\mc{A}\ell}\sqrt{\frac{\mc{A}r_0}{1-\mc{A}r_0}}\,.
        \end{cases}
    \end{align}
\end{subequations}We note two special cases. First, for $\phi_*=\frac{1}{m}\arccosh\lr{\frac{(1+m^2\mc{A}^2\ell^2)\mc{A}r_0}{m^2\mc{A}^2\ell^2}}$, we have
\begin{equation}
    \Mmin~=~\frac{m^2}{\alpha^2}\lr{m^2\mc{A}^2\ell^2+1}\,,
\end{equation}which we may identify as the critical mass $M_{\mrm{crit}}$~\eqref{eqn:II crit mass} from Section~\ref{Sec5}. For masses $M>M_{\mrm{crit}}$, we recall that the stress energy of the shell is positive.

Second, for $\phi_*=0$, we have
\begin{equation}\label{eqn:II left zero diverge}
    \Mmin~=~-\frac{f(r_0)}{\alpha^2(1-\mc{A}r_0)^2}\,.
\end{equation}In the limit $M\to\Mmin$~\eqref{eqn:II left zero diverge}, we have $\rho(\phi)\propto-1/\phi$ around $\phi=0$ and hence $\rho(\phi)\to-\infty$ as $\phi\to0$. Furthermore, we have $\wt{M}_{\mrm{min}}\to-\infty$ as $M\to\Mmin$~\eqref{eqn:II left zero diverge}. This may be seen by noting that the integrand in the definition of $\wt{M}$~\eqref{eqn:: mass function} is even and may be written over the interval $\phi'\in(0,\pi)$. For arbitrary values of $M$, the integrand is order unity near zero. However, for $M=\Mmin$~\eqref{eqn:II left zero diverge}, the integrand is order $1/\phi'$ near $\phi'=0$ and the integral diverges.

\begin{figure}[t!]
    \centering
    \begin{tikzpicture}
    \draw (0, 0) node[inner sep=0] {\includegraphics[width=\textwidth]{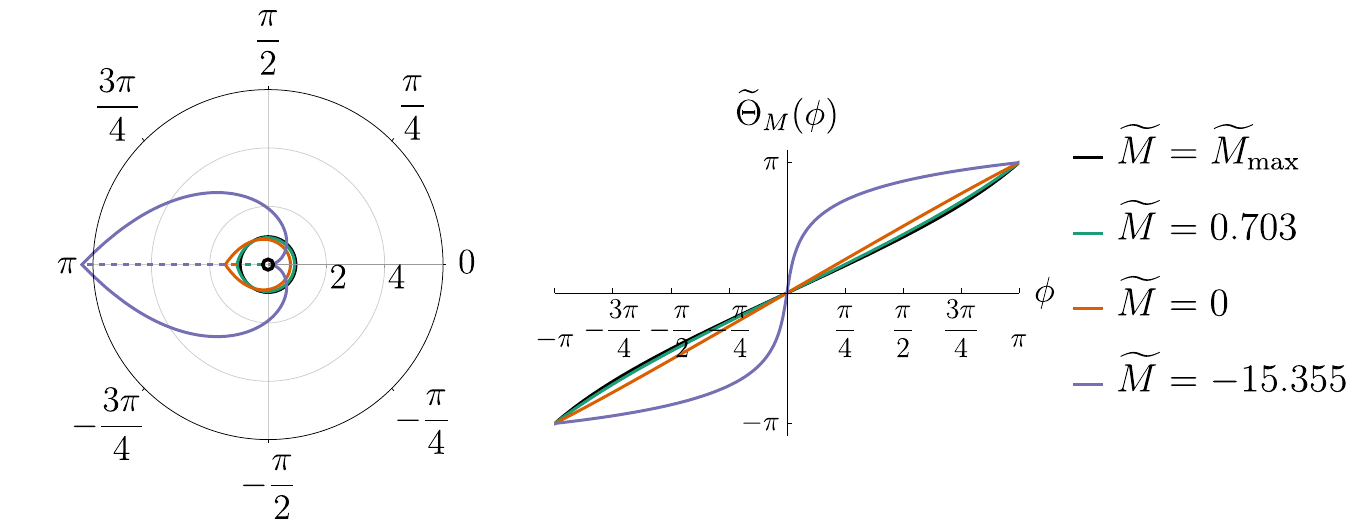}};
     \draw (-8, 2.4) node {\textbf{(a)}};
    \draw (-1.6, 2.4) node {\textbf{(b)}};
\end{tikzpicture}\vspace{-0em} 
    \begin{tikzpicture}
    \draw (0, 0) node[inner sep=0] {\includegraphics[width=\textwidth]{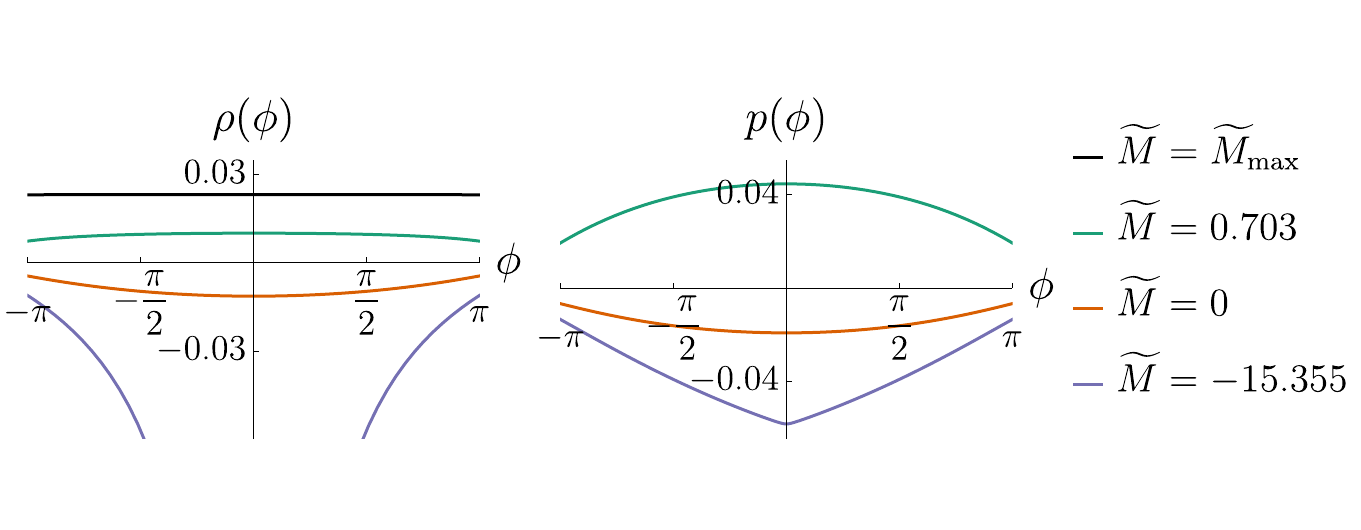}};
    \draw (-8, 2.4) node {\textbf{(c)}};
    \draw (-1.6, 2.4) node {\textbf{(d)}};
\end{tikzpicture}
\begin{tikzpicture}
    \draw (0, 0) node[inner sep=0] {\includegraphics[width=\textwidth]{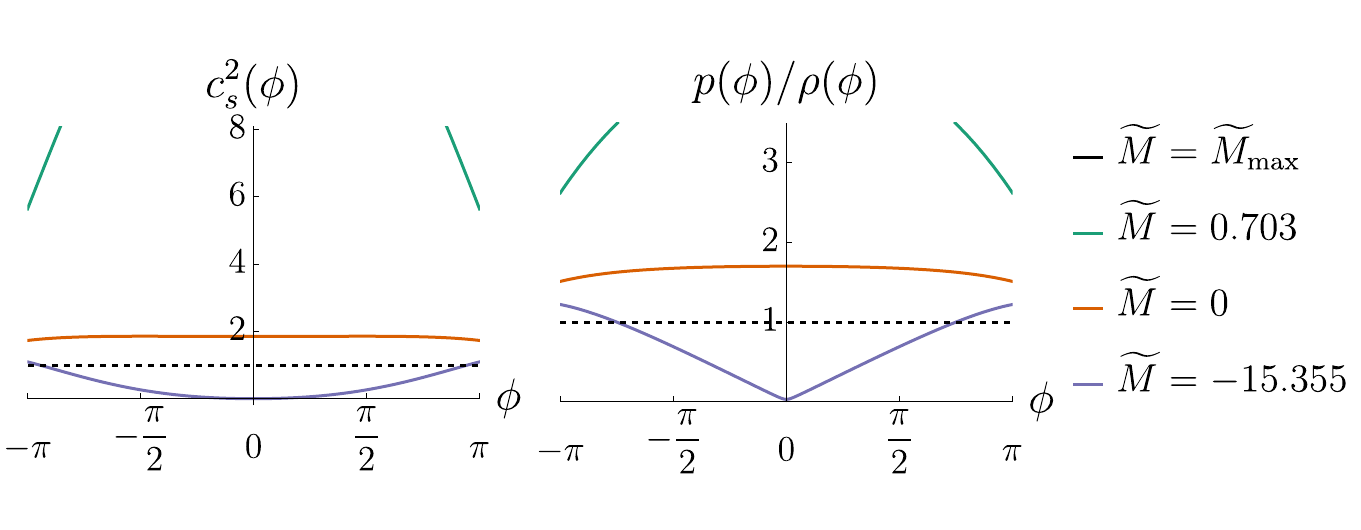}};
    \draw (-8, 2.4) node {\textbf{(e)}};
    \draw (-1.6, 2.4) node {\textbf{(f)}};
\end{tikzpicture}\vspace{-0em} 
    \caption{\justifying Class $\mrm{II}_{\mrm{left}}$ for $r_0=0.6$, $m=0.5$, $\mc{A}=0.4$, $\alpha=0.25$, and $\ell=1$, where $r_h\approx0.490$, $\wt{M}_{\mrm{min}}=-\infty$, and $\wt{M}_{\mrm{max}}\approx0.912$. \textbf{(a)} Polar plot of shell radius $\wt{R}_M$~\eqref{eqn:ADM shell coords} over $\phi\in(-\pi,\pi)$ for varying values of $\wt{M}$; dashed line indicates the string and black ring represents the black hole. \textbf{(b)} Shell angular coordinate $\wt{\Theta}_M$~\eqref{eqn:ADM shell coords} for varying values of $\wt{M}$. \textbf{(c)} Energy density $\rho$ of the shell for varying values of $\wt{M}$. \textbf{(d)} Pressure $p$ of the shell for varying values of $\wt{M}$. \textbf{(e)} Squared speed of sound $c_s^2$ of matter in the shell for varying values of $\wt{M}$; dashed line at $c_s=1$. \textbf{(f)} Ratio of shell pressure to shell energy density for varying values of $\wt{M}$; dashed line at $p/\rho=1$. }\label{fig:Class II left}
\end{figure}

In~\autoref{fig:Class II left} \textbf{(a)} to \textbf{(f)}, we plot the shell coordinates, stress energy, speed of sound, and ratio of pressure to energy density. The results are qualitatively similar to the Class I$_{\mrm{C}}$ solution, describing an accelerated black hole pulled by a string. We recall, however, that these two black-hole solutions are fundamentally different: the Class II$_{\mrm{left}}$ solution is a one-parameter extension of the BTZ black hole, which may be recovered through the limit $\mc{A}\to0$, whilst the Class I$_{\mrm{C}}$ is disconnected from the BTZ solution as we may not take the $\mc{A}\to0$ limit.

In~\autoref{fig:Class II left} \textbf{(a)} and \textbf{(b)}, we plot the shell radial and angular coordinates as functions of the shell's intrinsic coordinate $\phi$. As $M\to\Mmin$, the shell resembles a cardioid, pinched where the string meets the shell at $\phi=\pi$ and with a cuspoidal shape at $\phi=0$. We note that in~\autoref{fig:Class II left} \textbf{(a)} though it appears graphically that the shell meets the horizon, this does not happen; in the interior spacetime, the shell is placed at $r=r_0>r_h$, which is smoothly joined to the exterior spacetime coordinates.

As $\wt{M}\to\wt{M}_{\mrm{max}}$, the shell becomes less deformed and tends to a teardrop shape, pinched at $\phi=\pi$. The angular coordinate $\wt{\Theta}_M$ is strongly nonlinear for negative values of $\wt{M}$ and becomes increasingly linear as $\wt{M}\to0$; however, the angular coordinate is nonlinear as $\wt{M}\to\wt{M}_{\mrm{max}}$. We remark that $\wt{M}_{\mrm{min}}=-\infty$; hence, between the curves depicting the shell with $\wt{M}=\wt{M}_{\mrm{min}}$ and $\wt{M}=0$ there exists a shell whose exterior is global AdS, with $\wt{M}=-1$. As $M\to\Mmin$, the ADM factor $\beta_M$ diverges and the ADM shell radius $\beta_MR_M(\phi)$ increases without bound.

We show plots of the shell energy density and pressure in~\autoref{fig:Class II left} \textbf{(c)} and \textbf{(d)}. The energy density and pressure may both be positive or negative. For $\rho,~p<0$, the energy density and pressure are maximised where the string meets the shell at $\phi=\pi$. For $\rho,~p > 0$, the energy density and pressure are maximised at $\phi=0$. Both the energy density and pressure increase with $M$ and are positive for sufficiently large $M$, as expected by~\eqref{eqn:large M SE}. In~\autoref{fig:Class II left} \textbf{(c)}, the graph of the energy density for $\wt{M}=\wt{M}_{\mrm{min}}$ is cut off to see the other curves clearly; as $\Mmin$ is given by~\eqref{eqn:II left zero diverge} for the plotted parameters, the energy density diverges as $\phi\to0$. The NEC, WEC, and SEC are satisfied for sufficiently large values of $M$.
\begin{figure}[t!]
    \centering
    \begin{tikzpicture}
    \draw (0, 0) node[inner sep=0] {\includegraphics[width=0.9\textwidth]{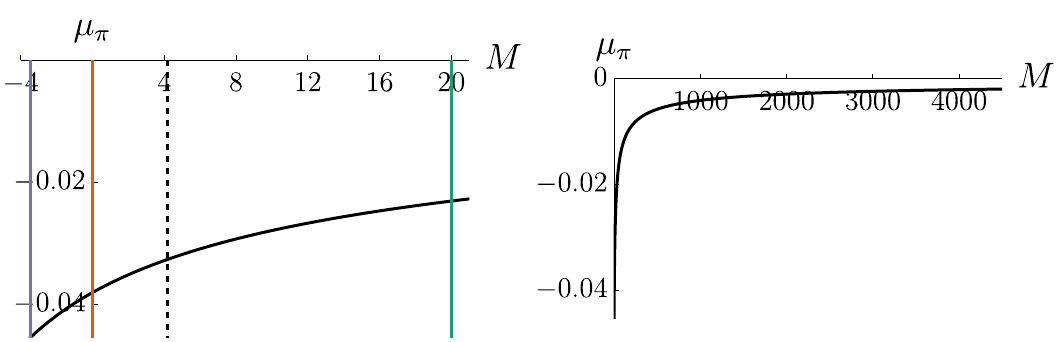}};
     \draw (-7, 2.4) node {\textbf{(a)}};
    \draw (0.6, 2.4) node {\textbf{(b)}};
\end{tikzpicture}
    \caption{\justifying Class $\mrm{II}_{\mrm{left}}$ for $r_0=0.6$, $m=0.5$, $\mc{A}=0.4$, $\alpha=0.25$, and $\ell=1$, where $r_h\approx0.490$, $\wt{M}_{\mrm{min}}=-\infty$, and $\wt{M}_{\mrm{max}}\approx0.912$. \textbf{(a)} Point particle mass $\mu_\pi$ at $\phi=\pm\pi$ for varying values of $M$. Black vertical dashed line corresponds to $M=M_{\mrm{crit}}$~\eqref{eqn:II crit mass}. Coloured vertical lines correspond to the shells plotted in~\autoref{fig:Class II left}. \textbf{(b)} Point particle mass for larger range of $M$ values, demonstrating the large-$M$ behaviour of $\mu_\pi$.}\label{fig:Class II left mu pi}
\end{figure}

In~\autoref{fig:Class II left}~\textbf{(e)} and \textbf{(f)}, we plot the shell speed of sound and ratio of pressure to energy density. We find that positive stress energy ($M>M_{\mrm{crit}}$) violates causality the the DEC. Again, this aligns with the intuition of Section~\ref{sec: A zero limit} that any black-hole solution surrounded by a shell with positive stress energy must have superluminal stress energy.

 As in Class $\mrm{II}_{\mrm{right}}$ and Class $\mrm{I}_{\mrm{C}}$, we find a critical mass parameter between the physical ($\rho>0$) and unphysical ($\rho<0$) solutions for which the stress energy of the shell vanishes. The critical mass is~\eqref{eqn:II crit mass} in both Class $\mrm{II}_{\mrm{right}}$ and
Class $\mrm{II}_{\mrm{left}}$, since each is obtained from the other by mapping $\mc{A}\to-\mc{A}$ and the critical mass is quadratic in $\mc{A}$. For this choice of exterior mass parameter, the solution describes a spacetime consisting of an accelerated black hole, pulled by a finite-length string at whose end  lies a point particle with mass $\mu_\pi$~\eqref{eqn:defect mass}.

In~\autoref{fig:Class II left mu pi}, we plot the DJ'tH mass of the point particle as a function of the exterior mass parameter $M$. We find that the mass in negative for all values of $M$, indicating an angular excess at $\phi=\pm\pi$. As $\wt{M}\to\wt{M}_{\mrm{max}}$, we see $\mu_\pi\to0$. In contrast to the Class $\mrm{II}_{\mrm{right}}$ solution, the point particle has negative mass, however the domain wall at $\phi=\pm\pi$ in Class $\mrm{II}_{\mrm{left}}$ spacetimes has positive tension. We note again a similarity to the Class I$_{\mrm{C}}$ solution, which has positive tension and a point particle of negative mass.


\section{Class III}\label{sec:class III}
\label{Sec7}

We consider now the Class III spacetime, which has received the least attention in the literature. By making cuts in the spacetime and identifying two edges, one may form a double-strut solution~\cite{Accin3D}.  
This procedure, however, does not result in a periodic angular coordinate.

Instead, we consider two copies of the patch in the Class III spacetime from $\phi=0$ to $\phi=\pi$. In the second copy, we relabel the angular coordinate $\phi\to-\phi$ and glue along the $\phi=0$ edges and identify the $\phi=\pi$ and $\phi=-\pi$ edges. This is now a periodic solution described by the metric
\begin{subequations}
    \begin{align}
        \dd s^2&~=~\frac{1}{\Omega^2(r,\phi)}\lr{-\frac{f(r)}{\alpha^2}\dd\sigma^2+\frac{\dd r^2}{f(r)}+r^2\dd\phi^2}\,,\\
        f(r)&~=~\frac{r^2}{\ell^2}-m^2\lr{1+\mc{A}^2r^2}\,,\\
        \Omega(r,\phi)&~=~1+\mc{A}r|\sinh(m\phi)|\,.
    \end{align}
\end{subequations}
  We call this solution the Class $\mrm{III}_{\mrm{right}}$ spacetime.

\subsection{Class $\text{III}_{\text{right}}$}

\begin{figure}[t]
    \centering
    \begin{subfigure}[b]{0.4\textwidth}
        \centering\captionsetup{labelfont=bf}
        \begin{tikzpicture}
        \draw (0, 0) node[inner sep=0] {\includegraphics[width=\textwidth]{figures/SpaceDiagrams/box_low_horizon.pdf}};
        \draw (2.7, 1.4) node {\large$r=\infty$};
        \draw (2.7, -1.3) node {\large$r=r_h$};
        \draw (-2.4, -1.85) node {\large$\phi=-\pi$};
        \draw (1.8, -1.85) node {\large$\phi=\pi$};
        \end{tikzpicture}
        \caption{\\}
    \end{subfigure}\hfill
    \begin{subfigure}[b]{0.4\textwidth}
        \centering\captionsetup{labelfont=bf}
        \begin{tikzpicture}
        \draw (0, 0) node[inner sep=0] {\includegraphics[width=\textwidth]{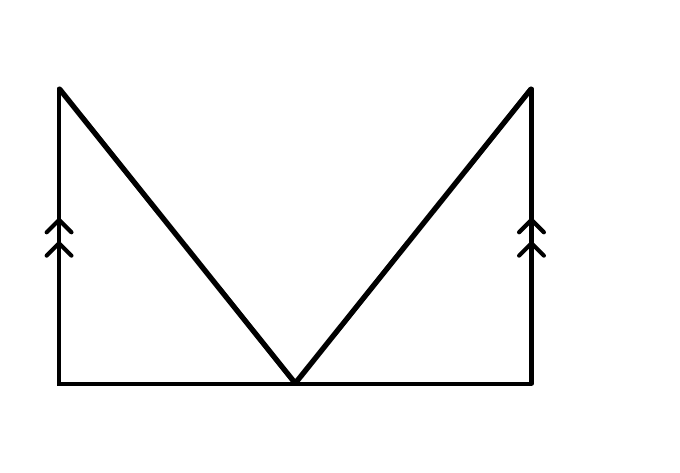}};
        \node[rotate=51] at (0.3,-0.2) {\large $r=r_{\mrm{conf}}$};
        \draw (2.9, -1.3) node {\large$r=-\infty$};
        \draw (-2.4, -1.85) node {\large$\phi=-\pi$};
        \draw (1.8, -1.85) node {\large$\phi=\pi$};
        \end{tikzpicture}
        \caption{\\}
    \end{subfigure}
    \begin{subfigure}[b]{0.4\textwidth}
        \centering\captionsetup{labelfont=bf}
        \begin{tikzpicture}
        \draw (0, 0) node[inner sep=0] {\includegraphics[width=\textwidth]{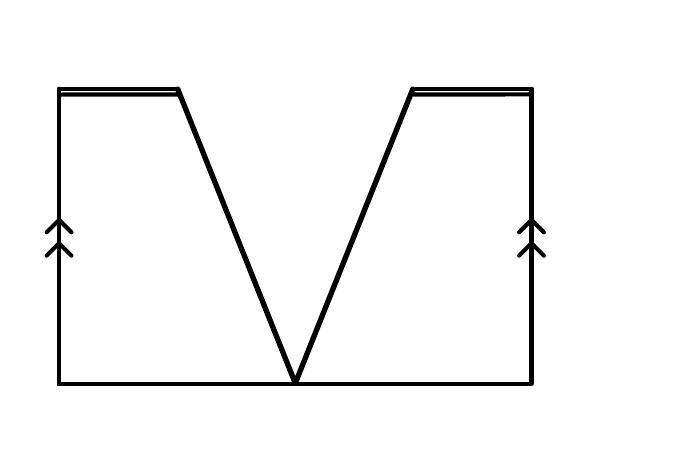}};
        \draw (2.67, 1.4) node {\large$r=r_{\mrm{D}}$};
        \node[rotate=69] at (0.05,0.3) {\large $r=r_{\mrm{conf}}$};
        \draw (2.9, -1.3) node {\large$r=-\infty$};
        \draw (-2.4, -1.85) node {\large$\phi=-\pi$};
        \draw (1.8, -1.85) node {\large$\phi=\pi$};
        \end{tikzpicture}
        \caption{}
    \end{subfigure}
    \caption{\justifying Constant time slice of the Class $\mrm{III}_{\mrm{right}}$ solution. \textbf{(a)} $r>0$ patch. \textbf{(b)} $r<0$ patch, slow acceleration phase. \textbf{(c)} $r<0$ patch, rapid acceleration phase. Double lines in \textbf{(a)} and \textbf{(c)} represent a Killing horizon.}
    \label{fig:Class III right ranges}
\end{figure}
This spacetime describes an accelerating black-hole being pushed by a strut at $\phi=\pi$ and pulled by a string at $\phi=0$. The corresponding strut/string tensions are
\begin{subequations}
    \begin{align}
        \uptau_\pi&~=~-\frac{m\mc{A}}{4\pi}\cosh(m\pi)\,,\\
        \uptau_0&~=~\frac{m\mc{A}}{4\pi}\,.
    \end{align}
\end{subequations}\color{black}
Positivity of the metric function $f(r)$ requires $m\mc{A}\ell<1$. The conformal boundary is found at
\begin{equation}
    \rc~=~-\frac{1}{\mc{A}|\sinh(m\phi)|}\,.
\end{equation}

The geometry of the Class $\mrm{III}_{\mrm{right}}$, depicted in~\autoref{fig:Class III right ranges}, is remarkably similar to that of the Class $\mrm{II}_{\mrm{right}}$ spacetime. As in the Class $\mrm{II}_{\mrm{right}}$ solution, this spacetime is covered two charts, one each for $r>0$ and $r<0$, that are glued together across $r=\pm\infty$. 
This is motivated by the
observation  that the proper distance to $r=\pm\infty$ is finite,
\begin{equation}
    \int_{\pm r_0}^{\pm\infty}\dd r\,\sqrt{g_{rr}}~=~\int_{\pm r_0}^{\pm\infty}\dd r\,\frac{1}{\sqrt{\Omega^2(r,\phi)f(r)}}~<~\infty\,,
\end{equation}for any fixed $\phi\neq0$. This follows from the asymptotic behaviour $1/\sqrt{\Omega^2(r,\phi)f(r)}=\OO(r^{-2})$ as $r\to\infty$. For $\phi=0$, we have $1/\sqrt{\Omega^2(r,\phi)f(r)}=\OO(r^{-1})$ as $r\to\infty$ and the proper length is infinite. This may be seen in~\autoref{fig:Class III right ranges} as the point at which the conformal boundary meets $r=-\infty$. The stress energy of a shell formed in the $r<0$ chart is always unphysical and comprised of exotic matter. One may notice that for $r<0$, the energy density in the large-$M$ limit is negative~\eqref{eqn: large M rho}.

There are two horizons,
\begin{subequations}
    \begin{align}
        r_h&~=~\frac{m\ell}{\sqrt{1-m^2\mc{A}^2\ell^2}}\,,\\
        r_{\mrm{D}}&~=~-\frac{m\ell}{\sqrt{1-m^2\mc{A}^2\ell^2}}\,,
    \end{align}
\end{subequations}
and the minimum value of the exterior mass parameter is given by
\begin{subequations}
    \begin{align}
        \Mmin&~=~\frac{f(r_0)}{\alpha^2}\frac{\lr{m^2\mc{A}^2\ell^2\cosh^2(m\pi)-1}}{\lr{1+\mc{A}r_0|\sinh(m\pi)|}^2}\,.
    \end{align}
\end{subequations}

The $r>0$ chart describes the exterior of a black hole whose horizon is at $r=r_h$. The $r<0$ chart exhibits two phases of acceleration, slow and rapid, depending on the relative positions of $r_{\mrm{D}}$ and $\rc$. 

In the slow phase of acceleration, defined by
\begin{equation}
    m\mc{A}\ell\cosh(m\pi)~<~1\,,
\end{equation}the radial coordinate $r$ is bounded between spatial infinity at $r=-\infty$ and the conformal boundary, $-\infty<r<\rc$. In the rapid acceleration phase, a droplet horizon forms --- as in the Class $\mrm{II}_{\mrm{right}}$ rapid acceleration regime.

\begin{figure}[t!]
    \centering
    \begin{tikzpicture}
    \draw (0, 0) node[inner sep=0] {\includegraphics[width=\textwidth]{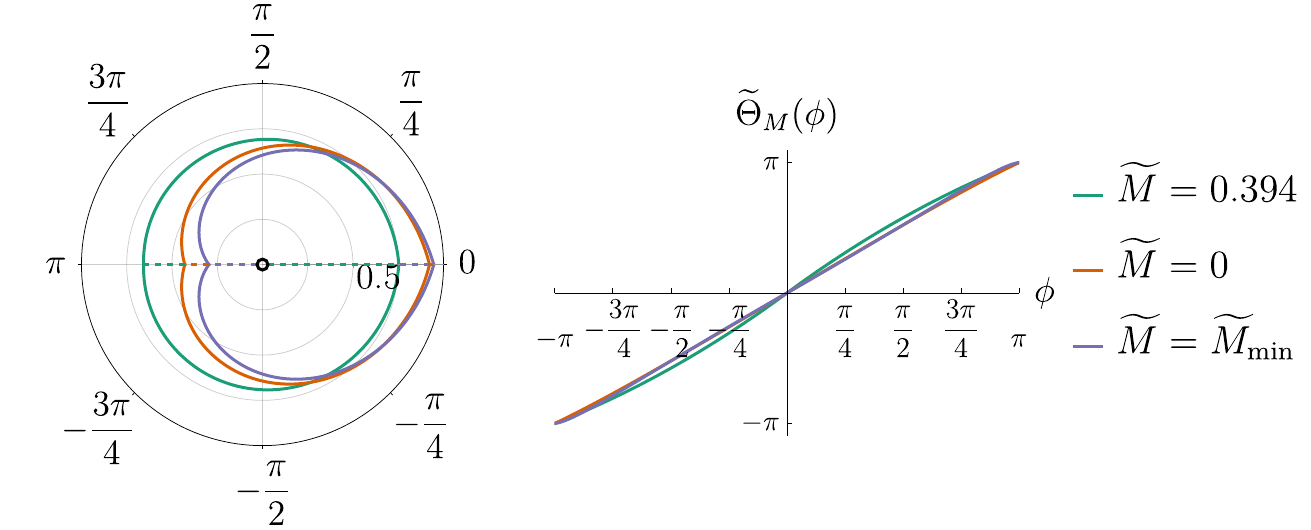}};
     \draw (-8, 2.4) node {\textbf{(a)}};
    \draw (-1.6, 2.4) node {\textbf{(b)}};
\end{tikzpicture}\vspace{-1em} 
    \begin{tikzpicture}
    \draw (0, 0) node[inner sep=0] {\includegraphics[width=\textwidth]{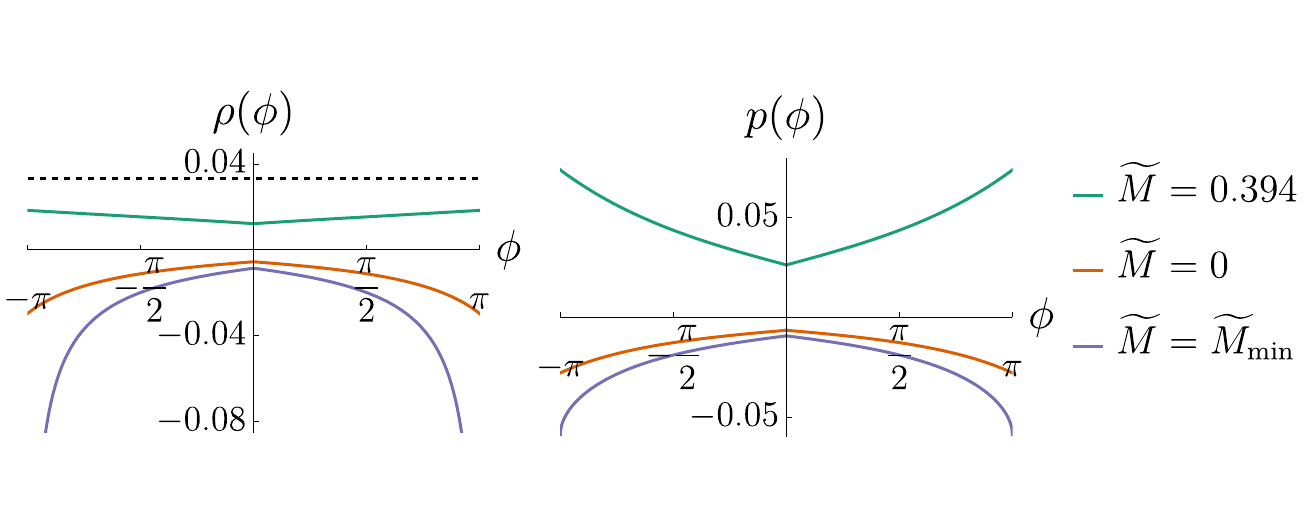}};
    \draw (-8, 2.4) node {\textbf{(c)}};
    \draw (-1.6, 2.4) node {\textbf{(d)}};
\end{tikzpicture}\vspace{-1em} 
\begin{tikzpicture}
    \draw (0, 0) node[inner sep=0] {\includegraphics[width=\textwidth]{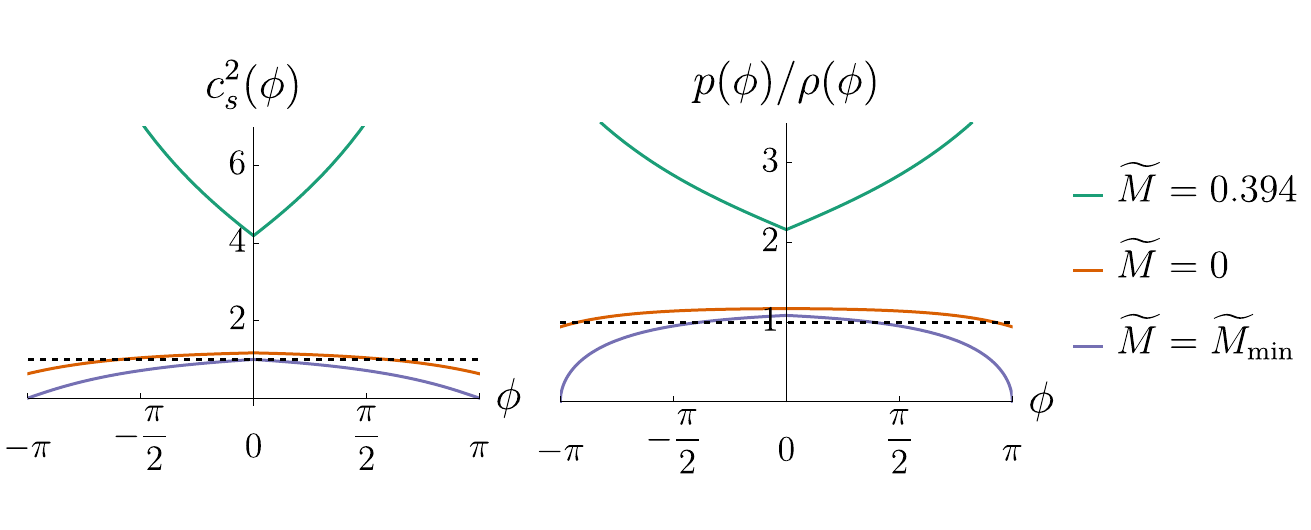}};
    \draw (-8, 2.4) node {\textbf{(e)}};
    \draw (-1.6, 2.4) node {\textbf{(f)}};
\end{tikzpicture}
    \caption{\justifying Class $\mrm{III}_{\mrm{right}}$ for $r_0=1$, $m=0.5$, $\mc{A}=0.5$, $\alpha=0.25$, and $\ell=1$, where $r_h\approx0.516$, $\wt{M}_{\mrm{min}}\approx-0.135$, and $\wt{M}_{\mrm{max}}\approx0.501$. \textbf{(a)} Polar plot of shell radius $\wt{R}_M$~\eqref{eqn:ADM shell coords} over $\phi\in(-\pi,\pi)$ for varying values of $\wt{M}$; dashed lines at $\phi=0,\pm\pi$ indicate the string and strut respectively and black ring represents the black hole. \textbf{(b)} Shell angular coordinate $\wt{\Theta}_M$~\eqref{eqn:ADM shell coords} for varying values of $\wt{M}$. \textbf{(c)} Energy density $\rho$ of the shell for varying values of $\wt{M}$. Black dashed line at asymptotic value as $\wt{M}\to\wt{M}_{\mrm{max}}$. \textbf{(d)} Pressure $p$ of the shell for varying values of $\wt{M}$. \textbf{(e)} Squared speed of sound $c_s^2$ of matter in the shell for varying values of $\wt{M}$; dashed line at $c_s=1$. \textbf{(f)} Ratio of shell pressure to shell energy density for varying values of $\wt{M}$; dashed line at $p/\rho=1$.
    }\label{fig:Class III right}
\end{figure}
\begin{figure}[t!]
    \centering
    \begin{tikzpicture}
    \draw (0, 0) node[inner sep=0] {\includegraphics[width=0.9\textwidth]{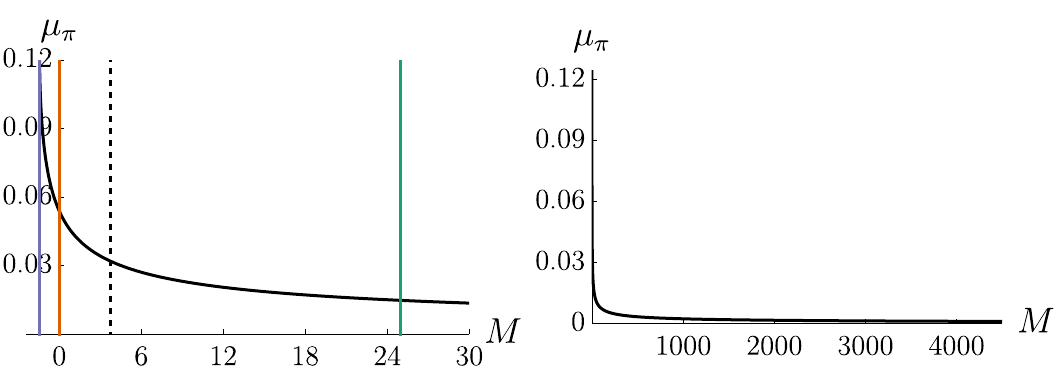}};
     \draw (-7.5, 2.4) node {\textbf{(a)}};
    \draw (0.2, 2.4) node {\textbf{(b)}};
\end{tikzpicture}
    \begin{tikzpicture}
    \draw (0, 0) node[inner sep=0] {\includegraphics[width=0.9\textwidth]{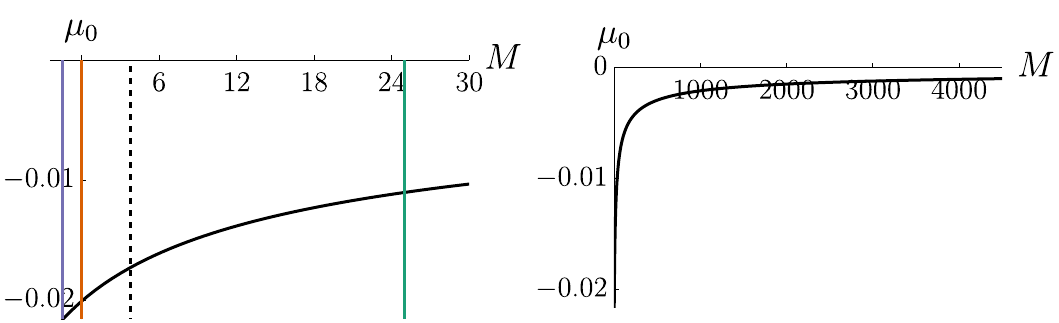}};
    \draw (-7.5, 1.8) node {\textbf{(c)}};
    \draw (0.2, 1.8) node {\textbf{(d)}};
\end{tikzpicture}
    \caption{\justifying Class $\mrm{III}_{\mrm{right}}$ for $r_0=1$, $m=0.5$, $\mc{A}=0.5$, $\alpha=0.25$, and $\ell=1$, where $r_h\approx0.516$, $\wt{M}_{\mrm{min}}\approx-0.135$, and $\wt{M}_{\mrm{max}}\approx0.501$. \textbf{(a)} Point particle mass $\mu_\pi$ at $\phi=\pm\pi$ for varying values of $M$. Black vertical dashed line corresponds to $M=M_{\mrm{crit}}$~\eqref{eqn:III crit mass}. Coloured vertical lines correspond to the shells plotted in~\autoref{fig:Class III right}. \textbf{(b)} Point particle mass $\mu_\pi$ for larger range of $M$ values, demonstrating the large-$M$ behaviour of $\mu_\pi$. \textbf{(c)} Point particle mass $\mu_0$ at $\phi=0$ for varying values of $M$. Black vertical dashed line corresponds to $M=M_{\mrm{crit}}$~\eqref{eqn:III crit mass}. Coloured vertical lines correspond to the shells plotted in~\autoref{fig:Class III right}. \textbf{(d)} Point particle mass $\mu_0$ for larger range of $M$ values, demonstrating the large-$M$ behaviour of $\mu_0$.
    }\label{fig:Class III right pi mu}
\end{figure}

There is, however, a key distinction between the $r<0$ patches of the Class $\mrm{II}_{\mrm{right}}$ and the periodic Class III solutions; in the Class $\mrm{III}_{\mrm{right}}$ solution, the conformal boundary reaches spatial infinity at $\phi=0$. As such, the $r<0$ patch has the spatial topology of (an open subset of) $\RR^2$, rather than $\RR\times\SSS^1$, and no closed loops of constant $r$ may form. Because of this, we may not form a shell in the $r<0$ patch and we must restrict our attention to the $r>0$ black hole exterior.

In~\autoref{fig:Class III right} \textbf{(a)} and \textbf{(b)}, we plot the shell radial and angular coordinates as functions of the shell's intrinsic coordinate $\phi$. For small values of $M$, the shell resembles a cardioid and is pinched at $\phi=0$ with a cusp at $\phi=\pi$, at which points the string and strut (indicated by the dashed lines) meet the shell respectively. The shell becomes less deformed as $\wt{M}\to\wt{M}_{\mrm{max}}$. The angular coordinate exhibits slight nonlinearities. We remark that $\wt{M}_{\mrm{min}}=0$; hence the $\wt{M}=\wt{M}_{\mrm{min}}$ plot depicts a shell with an exterior geometry of Torricelli's trumpet~\eqref{eqn:cylinder}.

In~\autoref{fig:Class III right} \textbf{(c)} and \textbf{(d)}, we plot the shell energy density $\rho$ and pressure $p$, both of which may be positive for negative. For $\rho,~p<0$, the energy density and pressure are maximised at the pinch at $\phi=0$. For $\rho,~p>0$, the energy density and pressure are maximised at $\phi=\pm\pi$. As $\wt{M}\to\wt{M}_{\mrm{max}}$, the energy density becomes evenly distributed across the shell, as anticipated by~\eqref{eqn:large M SE}. The energy density and pressure increase with $M$ and, as before, we find both physical ($\rho>0$) and unphysical ($\rho<0$)  shells over the range of $M$. There is  a critical exterior mass parameter $M_{\mrm{crit}}$ 
\begin{equation}\label{eqn:III crit mass}
    M_{\mrm{crit}}~\coloneq~\frac{m^2}{\alpha^2}(1-m^2\mc{A}^2\ell^2)~>~0\,,
\end{equation} 
for which the stress energy of the shell vanishes, notably similar to  the critical masses in Class $\mrm{I}_{\mrm{C}}$~\eqref{eqn:IC crit mass} and Class $\mrm{II}$~\eqref{eqn:II crit mass}. Only for $M>M_{\mrm{crit}}$ does the stress energy satisfy the NEC, WEC, and SEC. We interpret this solution with $M=M_{\mrm{crit}}$ as a spacetime consisting of an accelerated black hole, pushed by a finite-length strut at $\phi=\pm\pi$ and pulled by a finite-length string at $\phi=0$. At the ends of the respective strut and string lies a point particle with respective mass $\mu_\pi$~\eqref{eqn:defect mass} and $\mu_0$~\eqref{eqn:defect mass zero}.

In~\autoref{fig:Class III right} \textbf{(e)} and \textbf{(f)}, we plot the shell speed of sound and ratio of pressure to energy density. For $M<M_{\mrm{crit}}$, both quantities are maximised at the pinch at $\phi=0$. For $M>M_{\mrm{crit}}$, they are maximised at $\phi=\pm\pi$. As in the Class I$_{\mrm{C}}$ and Class II black-hole solutions, positive stress energy violates causality and the DEC.

 In~\autoref{fig:Class III right pi mu}, we plot the mass of the point particle found at $\phi=\pm\pi$ and at $\phi=0$ as functions of the exterior mass parameter $M$. We find that the point particle at $\phi=\pm\pi$ has positive mass, whereas the one at $\phi=0$ has negative mass, respectively indicating an angular deficit and excess. As in Class II, the sign of the point particles' mass is opposite to the corresponding string/strut's tension. As $\wt{M}\to\wt{M}_{\mrm{max}}$ ($M\to\infty$), it is straightforward to show that $\mu_\pi,~\mu_0\to0$.

\subsection{Class $\text{III}_{\text{left}}$}

\begin{figure}[t]
        \centering
        \begin{tikzpicture}
        \draw (0, 0) node[inner sep=0] {\includegraphics[width=0.4\textwidth]{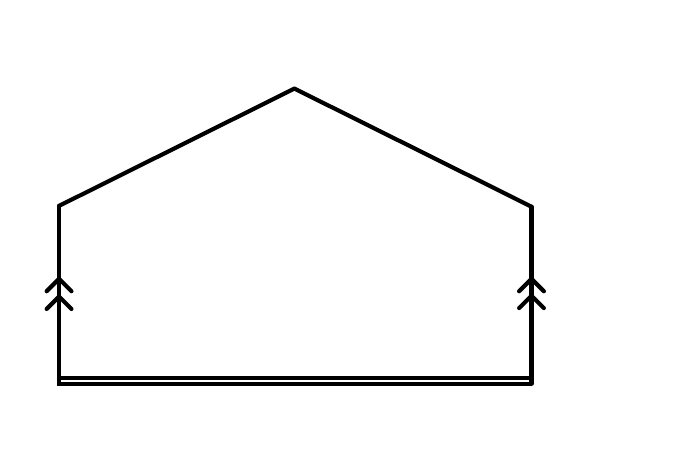}};
        \draw (2.7, -1.3) node {\large$r=r_h$};
        \draw (-2.4, -1.85) node {\large$\phi=-\pi$};
        \draw (1.8, -1.85) node {\large$\phi=\pi$};
        \node[rotate=28] at (-1.6,1.1) {\large $r=r_{\mrm{conf}}$};
        \node at (-0.45,1.46) [circle,fill,inner sep=1.5pt]{};
        \end{tikzpicture}
    \caption{\justifying Constant time slice of the Class $\mrm{III}_{\mrm{left}}$ solution. Double lines represent a Killing horizon. Dot at $\phi=0$, corresponds to $\rc=\infty$.}
    \label{fig:Class III left ranges}
\end{figure}

We find a second solution under the transformation $\mc{A}\to-\mc{A}$, the Class $\mrm{III}_{\mrm{left}}$ spacetime. This geometry is described by the metric
\begin{subequations}
    \begin{align}
        \dd s^2&~=~\frac{1}{\Omega^2(r,\phi)}\lr{-\frac{f(r)}{\alpha^2}\dd\sigma^2+\frac{\dd r^2}{f(r)}+r^2\dd\phi^2}\,,\\
        f(r)&~=~\frac{r^2}{\ell^2}-m^2\lr{1+\mc{A}^2r^2}\,,\\
        \Omega(r,\phi)&~=~1-\mc{A}r|\sinh(m\phi)|
    \end{align}
\end{subequations}
corresponding to an accelerating black-hole being pulled by a string at $\phi=\pi$ and pushed by a strut at $\phi=0$. The corresponding strut/string tensions are
\begin{subequations}
    \begin{align}
        \uptau_\pi&~=~\frac{m\mc{A}}{4\pi}\cosh(m\pi)\,,\\
        \uptau_0&~=~-\frac{m\mc{A}}{4\pi}\,.
    \end{align}
\end{subequations}
Positivity of the metric function $f(r)$ requires $m\mc{A}\ell<1$. The conformal boundary is found at
\begin{equation}
    \rc~=~\frac{1}{\mc{A}|\sinh(m\phi)|}\,.
\end{equation}There is a single Killing horizon at
\begin{equation}
     r_h~=~\frac{m\ell}{\sqrt{1-m^2\mc{A}^2\ell^2}}\,.
\end{equation}To ensure $r_h<\rc$, we require
\begin{equation}\label{eqn:III left conf order}
    m\mc{A}\ell\cosh(m\pi)~<~1\,.
\end{equation}Therefore, there is no equivalent rapid phase of acceleration in the $\mrm{III}_{\mrm{left}}$ solution. The geometry of this spacetime is depicted in~\autoref{fig:Class III left ranges}.

The minimum value of the exterior mass parameter is given by
\begin{subequations}
    \begin{align}
        \Mmin&~=~\frac{f(r_0)}{\alpha^2}\frac{\lr{m^2\mc{A}^2\ell^2\cosh^2(m\phi_*)-1}}{\lr{1-\mc{A}r_0\sinh(m\phi_*)}^2}\,,\\
        \phi_*&~=~\begin{cases}
            0&\text{for~}m\leq\frac{1}{\mc{A}\ell}\sqrt{\frac{\mc{A}r_0\lr{2-\mc{A}r_0\sinh(m\pi)}}{2\mc{A}r_0+(1-\mc{A}^2r_0^2)\sinh(m\pi)}}\,,\\
            \pi\,&\text{for~}m>\frac{1}{\mc{A}\ell}\sqrt{\frac{\mc{A}r_0\lr{2-\mc{A}r_0\sinh(m\pi)}}{2\mc{A}r_0+(1-\mc{A}^2r_0^2)\sinh(m\pi)}}\,.
        \end{cases}
    \end{align}
\end{subequations}We note that since $m\mc{A}\ell\cosh(m\pi)<1$~\eqref{eqn:III left conf order}, we have $\Mmin<0$.

\begin{figure}[t!]
    \centering
    \begin{tikzpicture}
    \draw (0, 0) node[inner sep=0] {\includegraphics[width=\textwidth]{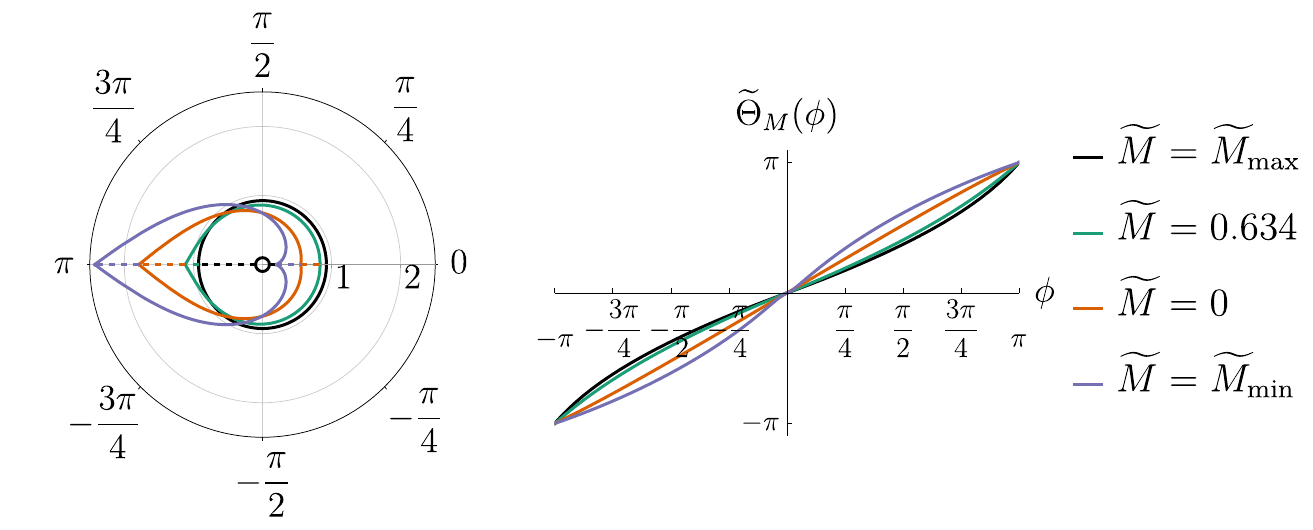}};
     \draw (-8, 2.4) node {\textbf{(a)}};
    \draw (-1.6, 2.4) node {\textbf{(b)}};
\end{tikzpicture}\vspace{-1em} 
    \begin{tikzpicture}
    \draw (0, 0) node[inner sep=0] {\includegraphics[width=\textwidth]{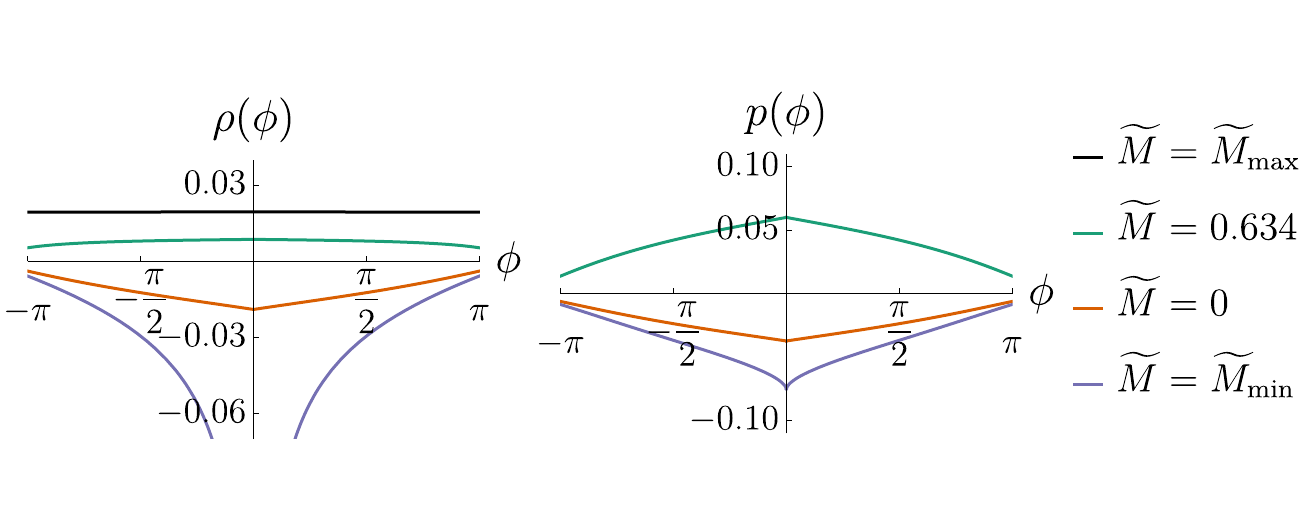}};
    \draw (-8, 2.4) node {\textbf{(c)}};
    \draw (-1.6, 2.4) node {\textbf{(d)}};
\end{tikzpicture}\vspace{-1em} 
\begin{tikzpicture}
    \draw (0, 0) node[inner sep=0] {\includegraphics[width=\textwidth]{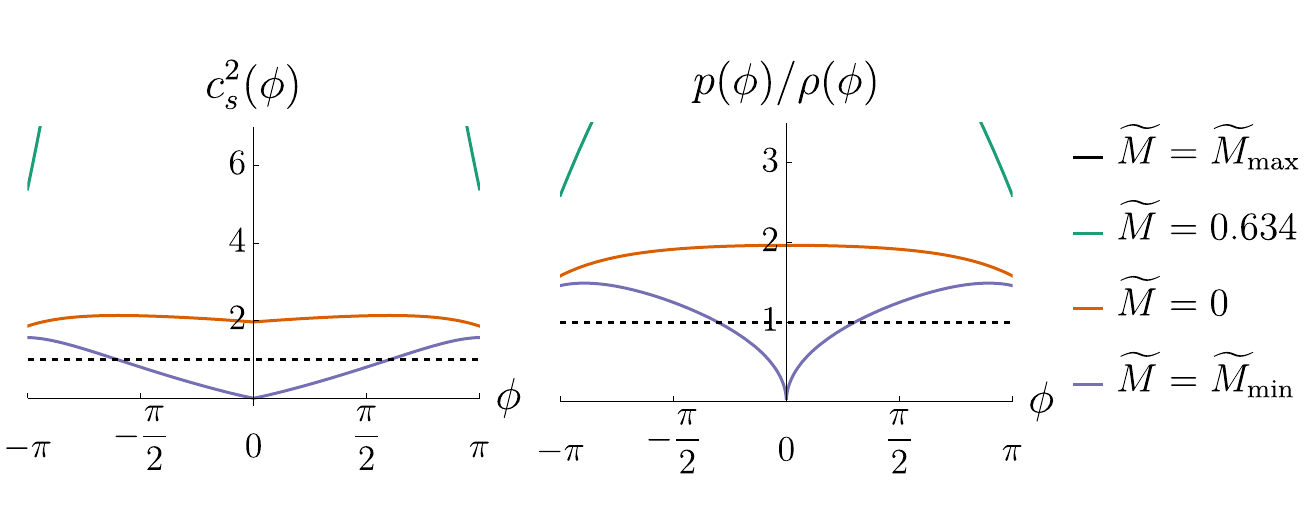}};
    \draw (-8, 2.4) node {\textbf{(e)}};
    \draw (-1.6, 2.4) node {\textbf{(f)}};
\end{tikzpicture}
    \caption{\justifying Class $\mrm{III}_{\mrm{left}}$ for $r_0=0.6$, $m=0.5$, $\mc{A}=0.5$, $\alpha=0.25$, and $\ell=1$, where $r_h\approx0.516$, $\wt{M}_{\mrm{min}}\approx-0.582$, and $\wt{M}_{\mrm{max}}\approx0.856$. \textbf{(a)} Polar plot of shell radius $\wt{R}_M$~\eqref{eqn:ADM shell coords} over $\phi\in(-\pi,\pi)$ for varying values of $\wt{M}$; dashed lines at $\phi=0,\pm\pi$ indicate the strut and string respectively and black ring represents the black hole. \textbf{(b)} Shell angular coordinate $\wt{\Theta}_M$~\eqref{eqn:ADM shell coords} for varying values of $\wt{M}$. \textbf{(c)} Energy density $\rho$ of the shell for varying values of $\wt{M}$. Black dashed line at asymptotic value as $\wt{M}\to\wt{M}_{\mrm{max}}$. \textbf{(d)} Pressure $p$ of the shell for varying values of $\wt{M}$. \textbf{(e)} Squared speed of sound $c_s^2$ of matter in the shell for varying values of $\wt{M}$; dashed line at $c_s=1$. \textbf{(f)} Ratio of shell pressure to shell energy density for varying values of $\wt{M}$; dashed line at $p/\rho=1$.
    }\label{fig:Class III left}
\end{figure}

In the case $\phi_*=0$, we have
\begin{equation}\label{eqn:III left zero diverge}
    \Mmin~=~-\frac{(1-m^2\mc{A}^2\ell^2)f(r_0)}{\alpha^2}\,,
\end{equation}at which value $\rho(\phi)\to-\infty$ as $\phi\to0$, a feature shared by the Class $\mrm{I}_{\mrm{C}}$ and Class $\mrm{II}_{\mrm{left}}$ solutions. However, unlike the Class $\mrm{I}_{\mrm{C}}$ and Class $\mrm{II}_{\mrm{left}}$ solutions, $\wt{M}_{\mrm{min}}$ remains finite.

 In~\autoref{fig:Class III left} \textbf{(a)} and \textbf{(b)}, we plot the shell radial and angular coordinates as functions of the shell's intrinsic coordinate $\phi$. As $M\to\Mmin$, the shell resembles a cardioid, pinched where the string meets the shell at $\phi=\pi$ and with a cuspoidal shape where the strut meets the shell at $\phi=0$. Although in~\autoref{fig:Class III left} \textbf{(a)} it appears graphically that the shell meets the horizon, 
 this does not actually happen: in the interior spacetime, the shell is placed at $r=r_0>r_h$, which is smoothly joined to the exterior spacetime coordinates.

 As $\wt{M}\to\wt{M}_{\mrm{max}}$, the shell becomes less deformed and tends to a perfect circle. The angular coordinate $\wt{\Theta}_M$ is strongly nonlinear for small values of $M$ and becomes increasingly linear as $\wt{M}\to0$,  returning to nonlinearity as $\wt{M}\to\wt{M}_{\mrm{max}}$.  We remark that $\wt{M}_{\mrm{min}}=-\infty$; hence, between the curves depicting the shell with $\wt{M}=\wt{M}_{\mrm{min}}$ and $\wt{M}=0$ there exists a shell whose exterior is global AdS, with $\wt{M}=-1$.

\begin{figure}[t!]
    \centering
    \begin{tikzpicture}
    \draw (0, 0) node[inner sep=0] {\includegraphics[width=0.9\textwidth]{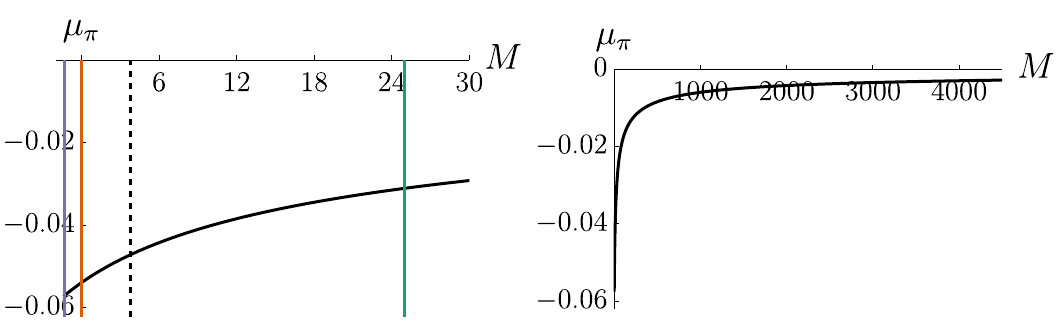}};
     \draw (-7.5, 2.4) node {\textbf{(a)}};
    \draw (0.2, 2.4) node {\textbf{(b)}};
\end{tikzpicture}
    \begin{tikzpicture}
    \draw (0, 0) node[inner sep=0] {\includegraphics[width=0.9\textwidth]{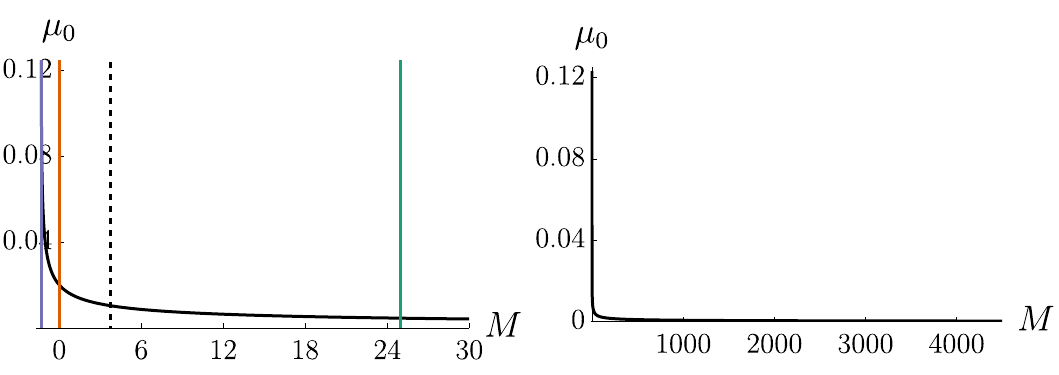}};
    \draw (-7.5, 2.6) node {\textbf{(c)}};
    \draw (0.2, 2.6) node {\textbf{(d)}};
\end{tikzpicture}
    \caption{\justifying Class $\mrm{III}_{\mrm{left}}$ for $r_0=0.6$, $m=0.5$, $\mc{A}=0.5$, $\alpha=0.25$, and $\ell=1$, where $r_h\approx0.516$, $\wt{M}_{\mrm{min}}\approx-0.582$, and $\wt{M}_{\mrm{max}}\approx0.856$. \textbf{(a)} Plot of point particle mass $\mu_\pi$ at $\phi=\pm\pi$ for varying values of $M$. Black vertical dashed line corresponds to $M=M_{\mrm{crit}}$~\eqref{eqn:III crit mass}. Coloured vertical lines correspond to the shells plotted in~\autoref{fig:Class III left}. \textbf{(b)} Plot of point particle mass $\mu_\pi$ for larger range of $M$ values, demonstrating the large-$M$ behaviour of $\mu_\pi$. \textbf{(c)} Plot of point particle mass $\mu_0$ at $\phi=0$ for varying values of $M$. The black vertical dashed line corresponds to $M=M_{\mrm{crit}}$~\eqref{eqn:III crit mass}. Coloured vertical lines correspond to the shells plotted in~\autoref{fig:Class III left}. \textbf{(d)} Plot of point particle mass $\mu_0$ for larger range of $M$ values, demonstrating the large-$M$ behaviour of $\mu_0$.
    }\label{fig:Class III left pi mu}
\end{figure}

We depict plots of the shell energy density $\rho$ and pressure $p$ in~\autoref{fig:Class III left} \textbf{(c)} and \textbf{(d)}. The energy density and pressure may both be positive or negative. For $\rho,~p<0$, the energy density and pressure are maximised where the string meets the shell at $\phi=\pi$. For $\rho,~p > 0$, the energy density and pressure are maximised at $\phi=0$. Both the energy density and pressure increase with $M$ and are positive for sufficiently large $M$, as expected by~\eqref{eqn:large M SE}. In~\autoref{fig:Class III left} \textbf{(c)}, the graph of the energy density for $\wt{M}=\wt{M}_{\mrm{min}}$ is cut off to see the other curves clearly; as $\Mmin$ is given by~\eqref{eqn:III left zero diverge} for the plotted parameters, the energy density diverges as $\phi\to0$. As in the Class $\mrm{III}_{\mrm{right}}$ solution, we find both physical ($\rho>0$) and unphysical ($\rho<0$) shells. Between these lies a solution with vanishing stress energy, there is no shell. As the Class $\mrm{III}_{\mrm{right}}$ and Class $\mrm{III}_{\mrm{left}}$ solutions are identified through the map $\mc{A}\to-\mc{A}$, the critical exterior mass parameter is again given by~\eqref{eqn:III crit mass}. Only for $M>M_{\mrm{crit}}$~\eqref{eqn:III crit mass} does the stress energy satisfy the NEC, WEC, and SEC.

In~\autoref{fig:Class III left} \textbf{(e)} and \textbf{(f)}, we plot the shell speed of sound and ratio of pressure to energy density. As in the Class I$_{\mrm{C}}$ and Class II black-hole solutions, positive stress energy violates causality and the DEC, aligning with the intuition of Section~\ref{sec: A zero limit} that positive stress energy surrounding a black-hole solution must be superluminal.

 We interpret this solution with $M=M_{\mrm{crit}}$ as a spacetime consisting of an accelerated black hole, pushed by a finite-length strut at $\phi=\pm\pi$ and pulled by a finite-length string at $\phi=0$ and at the respective ends of which  lies a point particle with  masses $\mu_\pi$~\eqref{eqn:defect mass} and $\mu_0$~\eqref{eqn:defect mass zero}.

 In~\autoref{fig:Class III left pi mu}, we plot the masses of the point particles found at $\phi=\pm\pi$ and at $\phi=0$ as functions of the exterior mass parameter $M$. We find that the point particle at $\phi=\pm\pi$ has negative mass, whereas the one at $\phi=0$ has positive mass, indicating an angular deficit and excess respectively. As in Class II, the sign of the point particles' mass is opposite to the corresponding string/strut's tension. As $\wt{M}\to\wt{M}_{\mrm{max}}$ ($M\to\infty$), we see $\mu_\pi,~\mu_0\to0$.

\section{Conclusions}\label{sec:: conclusions}
\label{Sec8}

We have carried out a comprehensive study of the C-metric in (2+1) dimensions, placing it inside a shell of stress energy and matching it to 
an exterior vacuum AdS metric. In contrast to nearly all studies of shell construction, the shells we obtain do not have circular symmetry.
We have outlined the general construction in Section~\ref{Sec2}, where we find that the angular coordinate outside the shell is a nonlinear monotonically increasing function of the interior angular coordinate matched at the shell. This has the effect of giving the shell the shape of either a teardrop or a cardioid, depending on the parameters of the interior metric.

The Class I solutions respect the weak, null, and strong energy conditions, regardless of whether they are pushed or pulled and may either respect or violate the dominant energy condition or causality (or both) depending on the choice of parameters. Both the density and the pressure tend to be concentrated on the part of the shell farthest from
the direction of acceleration: either away from the domain wall if pulled and near the domain wall if pushed. 

An exception to this is the Class I$_{\mrm{C}}$ solutions, which corresponds to 
a rapid acceleration phase that is disconnected from the ${\mathcal A}= 0$
BTZ black hole. These shells will violate the energy conditions for a sufficiently small mass parameter $M<M_{\textrm{crit}}$.  For $M=M_{\textrm{crit}}$ there is no shell: the black hole is pushed or pulled by a domain wall of finite length with a point particle at its end.

Class II shells likewise violate the energy conditions for sufficiently small mass.
 As with Class I$_{\mrm{C}}$, 
the energy conditions are violated if the mass parameter $M<M_{\textrm{crit}}$; if the interior is that of an accelerating point mass $M<0$ then the energy conditions are always violated.  If the exterior
mass has the critical value then
the solution describes  an accelerated black hole pulled by
a finite-length string at whose end lies a point particle.

We also constructed two explicit classes of Class III shells, right and left, with the right exhibiting a slow and rapid phase of acceleration. These spacetimes consist of 
two patches that are glued together across $r=\pm\infty$ in standard coordinates. For this class, the  shells are restricted as to their size: only shells with $0<r<+\infty$ can be constructed. 

We find that whilst all black-hole solutions can be enclosed by shells
that respect the weak, null, and strong energy conditions for an exterior mass parameter above a critical value, the stress energy satisfying these energy conditions must violate the dominant energy condition. This result is echoed in the $\mc{A}\to0$ limit in Section~\ref{sec: A zero limit}, in which analytic results are tractable.

Summarising, we have employed the thin-shell formalism to find not only cardioid and teardrop shaped shells matching interior C-metric solutions to vacuum AdS solutions but to also find new solutions to Einstein's field equations that have no shells. These  fall into three categories: an accelerated black hole pulled by a finite-length string with a point particle at the other end (Class $\text{I}_{\text{C}}$, Class $\text{II}_{\text{left}}$), an accelerated black hole pushed by a finite-length strut with a point particle at the other end (Class $\text{II}_{\text{right}}$), and an accelerated black hole pushed from one side by a finite-length strut and pulled from the other by a finite-length string, each with a point particle at the other end (Class $\text{III}_{\text{right}}$, Class $\text{III}_{\text{left}}$).

Our results should prove useful for several different further investigations. First, as noted in the introduction, our work can be extended to the rotating case. By 
choosing a sufficiently small shell, we can construct ($2+1$)-dimensional rotating C-metric 
solutions in the interior that do not have superluminal domain walls. Second, our results can be extended to $(3+1)$ dimensions, surrounding the black hole and a portion of the string/strut with a cardioid/teardrop 
shell having two spatial dimensions. A third problem would be that of gravitational collapse of the shells we have constructed, to see the resultant fate of the spacetime.

\ack
We thank Jorma Louko for helpful discussions on aspects of this project. The work of CRDB during the writing of this manuscript was supported by ESPRC (EP/W524402/1). This work was supported in part by the Natural Sciences and Engineering Research Council of Canada.

\section*{References}
\bibliographystyle{iopart-num} 
\bibliography{zzz_bibliography}

\providecommand{\newblock}{}
\begin{thebibliography}{10}
\expandafter\ifx\csname url\endcsname\relax
  \def\url#1{{\tt #1}}\fi
\expandafter\ifx\csname urlprefix\endcsname\relax\def\urlprefix{URL }\fi
\providecommand{\eprint}[2][]{\url{#2}}

\bibitem{LeviCivita:1917}
Levi-Civita T 1917 {\em Atti Accad. Naz. Lincei, Cl. Sci. Fis., Mat. Nat., Rend.\/} {\bf 26} 307

\bibitem{Weyl:1919}
Weyl H 1919 {\em Ann. Phys. (Leipzig)\/} {\bf 59} 185--188

\bibitem{KinnersleyWalker:1970}
Kinnersley W and Walker M 1970 {\em Phys. Rev. D\/} {\bf 2} 1359--1370

\bibitem{Bonnor:1982}
Bonnor W~B 1983 {\em Gen. Rel. Grav.\/} {\bf 15} 535--551

\bibitem{InterpretingCMetric}
Griffiths J~B, Krtous P and Podolsky J 2006 {\em Class. Quant. Grav.\/} {\bf 23} 6745--6766 (\textit{Preprint} \eprint{gr-qc/0609056})

\bibitem{PlebanskiDemianski:1976}
Pleba\'{n}ski J~F and Demia\'{n}ski M 1976 {\em Ann. Phys.\/} {\bf 98} 98--127

\bibitem{MannRoss:1995}
Mann R~B and Ross S~F 1995 {\em Phys. Rev. D\/} {\bf 52} 2254 (\textit{Preprint} \eprint{gr-qc/9504015})

\bibitem{Podolsky:2002}
Podolsk\'{y} J 2002 {\em Czech. J. Phys.\/} {\bf 52} 1--10 (\textit{Preprint} \eprint{gr-qc/0202033})

\bibitem{Dias:2003xp}
Dias O~J~C and Lemos J~P~S 2003 {\em Phys. Rev. D\/} {\bf 67} 084018 (\textit{Preprint} \eprint{hep-th/0301046})

\bibitem{Batic:2021tjh}
Batic D, Kittaneh H~A and Nowakowski M 2021 {\em Phys. Rev. D\/} {\bf 104} 124029 (\textit{Preprint} \eprint{2112.06163})

\bibitem{BHinElectricField}
Bičák J 1980 {\em Proc. Roy. Soc. Lond. A\/} {\bf 371}(1746)

\bibitem{Dowker:1993bt}
Dowker F, Gauntlett J~P, Kastor D~A and Traschen J~H 1994 {\em Phys. Rev. D\/} {\bf 49} 2909--2917 (\textit{Preprint} \eprint{hep-th/9309075})

\bibitem{Pravda:2000zm}
Pravda V and Pravdova A 2001 {\em Class. Quant. Grav.\/} {\bf 18} 1205--1216 (\textit{Preprint} \eprint{gr-qc/0010051})

\bibitem{Bini:2005qyt}
Bini D, Cherubini C, Filippi S and Geralico A 2005 {\em Class. Quant. Grav.\/} {\bf 22} 5157--5168 (\textit{Preprint} \eprint{1408.4277})

\bibitem{Bini:2007kvf}
Bini D, Cherubini C, Geralico A and Jantzen R~T 2007 {\em Int. J. Mod. Phys. D\/} {\bf 16} 1813--1828 (\textit{Preprint} \eprint{1408.4591})

\bibitem{Grenzebach:2015oea}
Grenzebach A, Perlick V and L\"ammerzahl C 2015 {\em Int. J. Mod. Phys. D\/} {\bf 24} 1542024 (\textit{Preprint} \eprint{1503.03036})

\bibitem{Alawadi:2020qdz}
Alawadi M~A, Batic D and Nowakowski M 2021 {\em Class. Quant. Grav.\/} {\bf 38} 045003 (\textit{Preprint} \eprint{2006.03376})

\bibitem{Anabalon:2018ydc}
Anabal\'on A, Appels M, Gregory R, Kubiz\v{n}\'ak D, Mann R~B and Ovg\"un A 2018 {\em Phys. Rev. D\/} {\bf 98} 104038 (\textit{Preprint} \eprint{1805.02687})

\bibitem{Anabalon:2018qfv}
Anabal\'on A, Gray F, Gregory R, Kubiz\v{n}\'ak D and Mann R~B 2019 {\em JHEP\/} {\bf 04} 096 (\textit{Preprint} \eprint{1811.04936})

\bibitem{Gregory:2019dtq}
Gregory R and Scoins A 2019 {\em Phys. Lett. B\/} {\bf 796} 191--195 (\textit{Preprint} \eprint{1904.09660})

\bibitem{CmetricThermoSuSY}
Cassani D, Gauntlett J~P, Martelli D and Sparks J 2021 {\em Phys. Rev. D\/} {\bf 104} 086005 (\textit{Preprint} \eprint{2106.05571})

\bibitem{CmetricThermoPhaseSpace}
Kim H, Kim N, Lee Y and Poole A 2023 {\em Eur. Phys. J. C\/} {\bf 83} 1095 (\textit{Preprint} \eprint{2306.16187})

\bibitem{Abbasvandi:2018vsh}
Abbasvandi N, Cong W, Kubiznak D and Mann R~B 2019 {\em Class. Quant. Grav.\/} {\bf 36} 104001 (\textit{Preprint} \eprint{1812.00384})

\bibitem{Abbasvandi:2019vfz}
Abbasvandi N, Ahmed W, Cong W, Kubiz\v{n}\'ak D and Mann R~B 2019 {\em Phys. Rev. D\/} {\bf 100} 064027 (\textit{Preprint} \eprint{1906.03379})

\bibitem{Ahmed:2019yci}
Ahmed W, Chen H~Z, Gesteau E, Gregory R and Scoins A 2019 {\em Class. Quant. Grav.\/} {\bf 36} 214001 (\textit{Preprint} \eprint{1906.10289})

\bibitem{Gregory:2020mmi}
Gregory R, Lim Z~L and Scoins A 2021 {\em Front. in Phys.\/} {\bf 9} 187 (\textit{Preprint} \eprint{2012.15561})

\bibitem{Ball:2020vzo}
Ball A and Miller N 2021 {\em Class. Quant. Grav.\/} {\bf 38} 145031 (\textit{Preprint} \eprint{2008.03682})

\bibitem{Emparan:1999wa}
Emparan R, Horowitz G~T and Myers R~C 2000 {\em JHEP\/} {\bf 01} 007 (\textit{Preprint} \eprint{hep-th/9911043})

\bibitem{Emparan:1999fd}
Emparan R, Horowitz G~T and Myers R~C 2000 {\em JHEP\/} {\bf 01} 021 (\textit{Preprint} \eprint{hep-th/9912135})

\bibitem{Kofron:2015gli}
Kofro\v{n} D 2015 {\em Phys. Rev. D\/} {\bf 92} 124064 (\textit{Preprint} \eprint{1603.01451})

\bibitem{Emparan:2020znc}
Emparan R, Frassino A~M and Way B 2020 {\em JHEP\/} {\bf 11} 137 (\textit{Preprint} \eprint{2007.15999})

\bibitem{Feng:2024uia}
Feng Y, Ma H, Mann R~B, Xue Y and Zhang M 2024 {\em JHEP\/} {\bf 08} 184 (\textit{Preprint} \eprint{2404.07192})

\bibitem{Climent:2024nuj}
Climent A, Emparan R and Hennigar R~A 2024 {\em JHEP\/} {\bf 08} 150 (\textit{Preprint} \eprint{2404.15148})

\bibitem{CMetricOriginalAstorino}
Astorino M 2011 {\em JHEP\/} {\bf 01} 114 (\textit{Preprint} \eprint{1101.2616})

\bibitem{Accin3D}
Arenas-Henriquez G, Gregory R and Scoins A 2022 {\em JHEP\/} {\bf 05} 063 (\textit{Preprint} \eprint{2202.08823})

\bibitem{RuthAccBH}
Arenas-Henriquez G, Cisterna A, Diaz F and Gregory R 2023 {\em JHEP\/} {\bf 09} 122 (\textit{Preprint} \eprint{2308.00613})

\bibitem{Oppenheimer_Snyder}
Oppenheimer J~R and Snyder H 1939 {\em Phys. Rev.\/} {\bf 56}(5) 455--459 \urlprefix\url{https://link.aps.org/doi/10.1103/PhysRev.56.455}

\bibitem{Ross:1992ba}
Ross S~F and Mann R~B 1993 {\em Phys. Rev. D\/} {\bf 47} 3319--3322 (\textit{Preprint} \eprint{hep-th/9208036})

\bibitem{Mann:2006yu}
Mann R~B and Oh J~J 2006 {\em Phys. Rev. D\/} {\bf 74} 124016 [Erratum: Phys.Rev.D 77, 129902 (2008)] (\textit{Preprint} \eprint{gr-qc/0609094})

\bibitem{CisternaMannHairyBTZ}
Cisterna A, Diaz F, Mann R~B and Oliva J 2023 {\em JHEP\/} {\bf 11} 073 (\textit{Preprint} \eprint{2309.05559})

\bibitem{BTZoriginal}
Banados M, Teitelboim C and Zanelli J 1992 {\em Phys. Rev. Lett.\/} {\bf 69} 1849--1851 (\textit{Preprint} \eprint{hep-th/9204099})

\bibitem{BTZgeometry}
Banados M, Henneaux M, Teitelboim C and Zanelli J 1993 {\em Phys. Rev. D\/} {\bf 48} 1506--1525 [Erratum: Phys.Rev.D 88, 069902 (2013)] (\textit{Preprint} \eprint{gr-qc/9302012})

\bibitem{CarlipBTZ}
Carlip S 1995 {\em Class. Quant. Grav.\/} {\bf 12} 2853--2880 (\textit{Preprint} \eprint{gr-qc/9506079})

\bibitem{QuantumCorrectedBTZ}
Casals M, Fabbri A, Mart\'\i{}nez C and Zanelli J 2019 {\em Phys. Rev. D\/} {\bf 99} 104023 (\textit{Preprint} \eprint{1902.01583})

\bibitem{Israel_thin_shell}
Israel W 1966 {\em Nuovo Cim. B\/} {\bf 44S10} 1 [Erratum: Nuovo Cim.B 48, 463 (1967)]

\bibitem{Deser_massdefect}
Deser S, Jackiw R and 't~Hooft G 1984 {\em Annals Phys.\/} {\bf 152} 220

\bibitem{Ashtekar_massdefect}
Ashtekar A and Varadarajan M 1994 {\em Phys. Rev. D\/} {\bf 50}(8) 4944--4956 \urlprefix\url{https://link.aps.org/doi/10.1103/PhysRevD.50.4944}

\bibitem{HiscockDefectMass}
Hiscock W~A 1985 {\em Phys. Rev. D\/} {\bf 31} 3288--3290

\bibitem{UnruhDefectMass}
Frolov V~P, Israel W and Unruh W~G 1989 {\em Phys. Rev. D\/} {\bf 39} 1084--1096

\bibitem{PelegGravCollapsePhaseTrans}
Peleg Y and Steif A~R 1995 {\em Phys. Rev. D\/} {\bf 51} 3992--3996 (\textit{Preprint} \eprint{gr-qc/9412023})

\bibitem{LemosThinShellEntropy}
Lemos J~P~S and Quinta G~M 2014 {\em Phys. Rev. D\/} {\bf 89} 084051 (\textit{Preprint} \eprint{1403.0579})

\bibitem{LemosThinShellRotating}
Lemos J~P~S, Lopes F~J and Minamitsuji M 2015 {\em Int. J. Mod. Phys. D\/} {\bf 24} 1542022 (\textit{Preprint} \eprint{1506.05454})

\bibitem{BTZNegativeSpectrum}
Miskovic O and Zanelli J 2009 {\em Phys. Rev. D\/} {\bf 79} 105011 (\textit{Preprint} \eprint{0904.0475})

\bibitem{Cruz:1994ar}
Cruz N and Zanelli J 1995 {\em Class. Quant. Grav.\/} {\bf 12} 975--982 (\textit{Preprint} \eprint{gr-qc/9411032})

\bibitem{BlackDroplet}
Hubeny V~E, Marolf D and Rangamani M 2010 {\em Class. Quant. Grav.\/} {\bf 27} 025001 (\textit{Preprint} \eprint{0909.0005})

\bibitem{BlackDroplet1}
Hubeny V~E, Marolf D and Rangamani M 2010 {\em Class. Quant. Grav.\/} {\bf 27} 095015 (\textit{Preprint} \eprint{0908.2270})

\end{thebibliography}

\end{document}